\documentclass[preprint,amssymb,aps,tightenlines,floatfix,nofootinbib,superscriptaddress]{revtex4-2}
\usepackage{graphicx}
\usepackage{subcaption}
\usepackage{slashed}
\usepackage{xcolor}
\usepackage{dcolumn}
\usepackage{bm}
\usepackage{amsmath}
\usepackage[super]{nth}
\usepackage{amsfonts}
\usepackage{bm}
\usepackage{physics}
\usepackage{mathtools}

\allowdisplaybreaks
\newlength{\mymathboxwidth}             
\newcommand{\Lx}{L_x}

\newcommand{\LD}{L_\Delta}
\newcommand{\LT}{L_t}

\begin{document}

\title{Radiative Corrections to Elastic Lepton-Proton Scattering with Focus on Two-Photon-Exchange Diagrams}

\author{Daniel Crowe}
\email{daniel.crowe@shu.edu}
\affiliation{Department of Physics, University at Buffalo,
The State University of New York, Buffalo, NY 14260-1500, U.S.A.}
\author{Syed Mehedi Hasan}
\email{syed-mehedi.hasan@physik.uni-regensburg.de}
\affiliation{Institute for Theoretical Physics, University of Regensburg, 93040 Regensburg, Germany}
\author{Doreen Wackeroth}
\email{dw24@buffalo.edu}
\affiliation{Department of Physics, University at Buffalo,
The State University of New York, Buffalo, NY 14260-1500, U.S.A.}

\date{\today}
\begin{abstract}
 Lepton (electron and muon) scattering experiments are excellent tools to gain insight into the nucleon structure. Elastic electron-proton scattering probes the spatial distribution of charge and magnetization inside the proton, and comparing electron-proton and  muon-proton scattering data tests lepton universality. The availability of a plethora of scattering data with increased precision and observed discrepancies such as the proton form factor puzzle and the proton radius puzzle motivated a renewed effort to improve the theoretical framework. Realizing that the one-photon-exchange approximation (OPE), i.~e. the Born approximation, is not sufficient, radiative corrections in QED, especially the two-photon-exchange (TPE) diagrams, are under investigation. The TPE diagrams are of special interest among the radiative corrections, since their hadronic-structure-dependent contribution generates an additional dependence of the cross section on the virtual-photon polarization parameter beyond the Born approximation that is not captured by standard model-independent radiative correction procedures, and that directly impacts the extraction of proton form factors.
In this work, we present a complete calculation of QED radiative corrections to elastic electron-proton and muon-proton scattering at next-to-leading order, taking into account loop-momentum-dependent form factors.  In the discussion of their numerical impact on lepton-proton scattering cross sections, we pay special attention to the elastic nucleon contribution to the TPE diagrams and compare them with existing theoretical predictions and lepton-proton scattering data. 
\end{abstract}

\maketitle
\newpage

\section{Introduction}
\label{sec:introduction}

Elastic lepton--nucleon scattering has long been one of the most powerful experimental
tools to investigate the internal structure of hadrons.
In particular, elastic electron--proton ($ep$) scattering provides direct access to the spatial
distributions of charge and magnetization in the proton through its electromagnetic form
factors.  In the one-photon-exchange (OPE) approximation, the electromagnetic interaction
is parameterized in terms of the Sachs electric and magnetic form factors,
$G_E(Q^2)$ and $G_M(Q^2)$, which depend on the spacelike four-momentum transfer
$Q^2=-q^2$ \cite{PhysRev.126.2256}.
Precision determinations of these form factors are essential for understanding how
static nucleon properties emerge from non-perturbative QCD dynamics, and they also enter
as crucial inputs in the modeling of low-energy processes in nuclear and particle physics.

\begin{figure}[htp]
    \centering
    \includegraphics[width=0.4\linewidth]{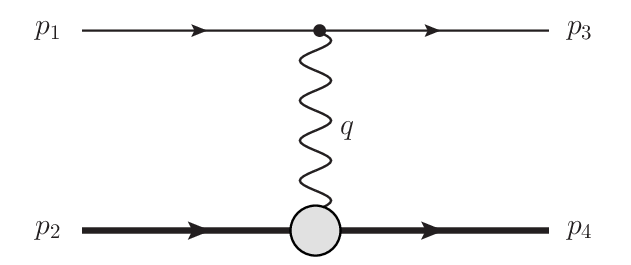}
    \caption{Elastic scattering of a lepton off a nucleon via one-photon exchange (OPE) in the Born approximation.}\label{fig:bornep}
\end{figure}

Experimentally, the two standard techniques used to extract Sachs form factors are
the Rosenbluth separation (RS) method based on unpolarized $ep$ scattering cross sections and the
polarization transfer (PT) method based on recoil polarization observables.
In the Born approximation, i.e.\ based on the one-photon exchange (OPE) diagram shown in Fig.~\ref{fig:bornep}, one introduces the reduced cross section
($\tau=Q^2/(4M^2)$, with $M$ the proton mass and the electron mass neglected)
\begin{equation}
\sigma_R
= \frac{G_E^2+\tau G_M^2}{1+\tau}
+ 2\tau G_M^2 \tan^2\!\left(\frac{\theta_e}{2}\right),
\label{eq:sigmaR}
\end{equation}
and the virtual-photon polarization parameter
\begin{equation}
\epsilon=\left[1+2(1+\tau)\tan^2\!\left(\frac{\theta_e}{2}\right)\right]^{-1},
\label{eq:eps}
\end{equation}
where $\theta_e$ is the electron scattering angle in the laboratory frame.
In the PT method, the ratio of the transverse ($P_t$) to longitudinal ($P_l$) polarization components of
the recoiling proton is related to $G_E/G_M$ as
\begin{equation}
\frac{P_t}{P_\ell}
=
-\frac{G_E}{G_M}\,
\sqrt{\frac{2\epsilon}{\tau(1+\epsilon)}}\, .
\label{eq:pt}
\end{equation}

With steadily improving experimental precision, elastic $ep$ scattering has revealed
some tensions that sharpen the need for accurate theory input.
A prominent example is the ``proton form factor puzzle'', namely the discrepancy between
RS and PT determinations of $G_E/G_M$ at multi-GeV momentum transfer
(see, e.g., \cite{Qattan:2024pco} and references therein).
Another example is the ``proton radius puzzle'', triggered by the extraction of a smaller
charge radius from muonic hydrogen spectroscopy \cite{Pohl:2010zza}
and the ensuing scrutiny of radius determinations from electron scattering and atomic
hydrogen \cite{Gao:2021sml,Lee:2015jqa}.
At very low $Q^2$, high-precision scattering programs such as A1 at MAMI \cite{A1:2013fsc},
initial-state-radiation (ISR) measurements at MAMI \cite{Mihovilovic:2016rkr,Mihovilovic:2019jiz},
and PRad at Jefferson Lab \cite{Xiong:2019umf}
push the experimental uncertainties to the percent and sub-percent level.
In parallel, low-energy elastic lepton--proton scattering also serves as a precision probe of
electroweak physics, e.g.\ via the proton weak charge as measured by Qweak \cite{Qweak:2018tjf}
and targeted by P2 \cite{Becker:2018ggl}.  Moreover, the MUSE program at PSI
\cite{MUSE:2013uhu,MUSE:2017dod} directly compares electron-proton and muon-proton scattering to
address lepton-species systematics relevant to the radius puzzle.
Muon-proton scattering is also part of the physics program proposed by the COMPASS++/AMBER collaboration for the M2 beam line of the CERN SPS, where a high-energy muon beam will be used to measure the proton charge radius~\cite{Adams:2018pwt}. More recently, the LUNE experiment has been proposed at the High-Intensity Heavy-ion Accelerator Facility (HIAF) to study elastic $\mu^\pm p$ scattering with muon momenta between 0.5 and 7.5~GeV~\cite{An:2026hqd}.

Matching this experimental precision requires a theoretically consistent inclusion of
radiative corrections (RC), including both virtual effects and real photon emission.
RC are indispensable because the extraction of physical observables depends sensitively on
how radiative events are defined and treated in the experimental analysis (see, e.~g., Ref.~\cite{Afanasev:2023gev} for a recent review).
Historically, many analyses rely on next-to-leading order (NLO) QED RC with
approximations following the pioneering work of Tsai \cite{Tsai:1961zz} and Mo--Tsai
\cite{Mo:1968cg}, while refinements and assessments of commonly used approximations
have been developed over time \cite{Maximon2000,Bystritskiy:2006ju,Kuraev:2013dra,Gerasimov:2015aoa,Akushevich:2015toa}.
Dedicated event generators implementing RC are also an integral part of modern
experimental workflows (see, e.~g., Table~2 of Ref.~\cite{Afanasev:2023gev} and Ref.~\cite{Isaacson:2022cwh}).
At the same time, recent progress in perturbative methods has enabled a range of NNLO
QED calculations for pointlike scattering processes and related building blocks,
including NNLO leptonic corrections in lepton--proton scattering
\cite{Bucoveanu:2019hxz,Banerjee:2020rww}
and fully differential NNLO treatments for pointlike four-fermion scattering amplitudes
\cite{Bonciani:2021okt}, with dedicated applications to MUSE kinematics
\cite{Engel:2023arz}.
These developments highlight the broader trend toward systematically improving the
RC framework as experiments enter the sub-percent regime. Complementary to, and beyond, fixed-order calculations, YFS exponentiation has been employed to resum soft logarithms~\cite{Yennie:1961ad} and effective field theory 
(EFT) methods have been applied systematically to elastic $\ell p$ 
scattering. Soft-Collinear Effective Theory (SCET) has been used 
to resum large QED logarithms at $Q^2\gg m_e^2$~\cite{Hill:2016gdf} 
and to organize the hard two-photon-exchange amplitude at large 
momentum transfer within QCD factorization~\cite{PhysRevLett.103.092004,Kivel:2012vs}, 
while Heavy-Baryon Chiral Perturbation Theory 
(HBChPT)~\cite{Talukdar:2019dko,PhysRevD.104.053001,Choudhary:2023rsz} 
and QED combined with Non-Relativistic QED 
(QED-NRQED)~\cite{Dye:2016uep,Dye:2018rgg} have been developed 
for the low-energy regime relevant to muon-proton scattering.

Among NLO radiative effects, two-photon exchange (TPE) corrections are of special
interest because they are sensitive to proton structure and can impact the interpretation of
form-factor extractions and cross-section ratios.
In the commonly used Mo-Tsai (MoT)~\cite{Mo:1968cg} and Maximon--Tjon (MTj)~\cite{Maximon2000} treatments, the TPE contribution is typically handled in a
soft-photon approximation.
Subsequent work incorporated loop momentum in the form factors for the TPE diagrams, in some cases
in restricted models (e.~g.\ monopole form factors)~\cite{Blunden:2003sp,Blunden:2005ew,Borisyuk:2006fh,PhysRevC.78.015205}.
Modern dispersive treatments systematically include elastic as well as $\pi N$ and resonance
intermediate states~\cite{Tomalak:2014dja,Tomalak:2016vbf,Tomalak:2015hva,Blunden:2017nby,Ahmed:2020uso},
while partonic approaches address the large-$Q^2$ regime via generalized parton
distributions~\cite{Chen:2004tw,Afanasev:2005mp}.
Extensive experimental and theoretical efforts have aimed at quantifying the size and kinematic
dependence of TPE effects (see, e.~g., Refs.~\cite{Afanasev:2023gev,AFANASEV2017245,Carlson:2007sp}
for recent reviews).

The goal of this work is twofold:
(i)~to provide an independent and complete calculation of NLO QED radiative corrections to
elastic lepton--proton scattering with particular emphasis on the elastic TPE contribution, and
(ii)~to present complete analytic results in a form suitable for easy implementation in Monte-Carlo event
generators used in experimental analyses.
A key feature of our calculation is the systematic inclusion of \emph{loop-momentum-dependent}
proton form factors in the virtual corrections, including in the elastic TPE contribution.
After presenting the theoretical framework and calculational setup, we quantify the numerical
impact of the infrared-finite, model-dependent part of the elastic TPE correction, compare to common
approximations used in data analysis and to results in the literature, and confront our results with a selection of available elastic
electron-proton scattering data in representative kinematics.

\section{Theoretical framework}
\label{sec:framework}

\begin{figure}[htbp]
\centering
  \begin{tabular}{lccr}
    \includegraphics[scale=0.4]{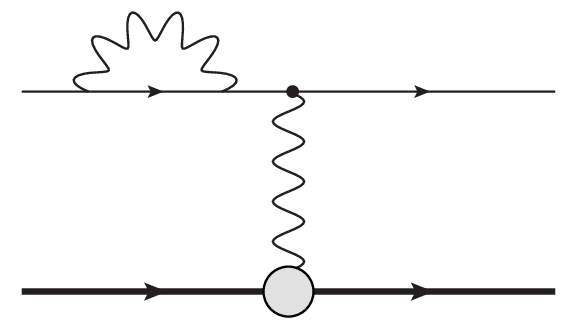} &
    \includegraphics[scale=0.4]{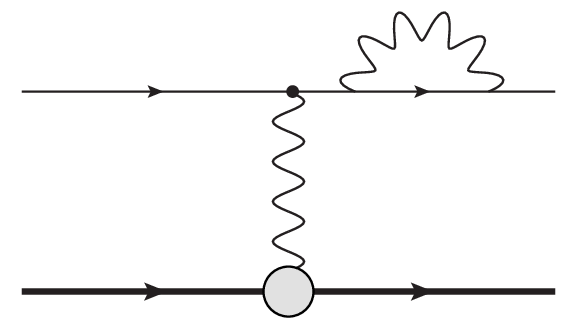} &
    \includegraphics[scale=0.4]{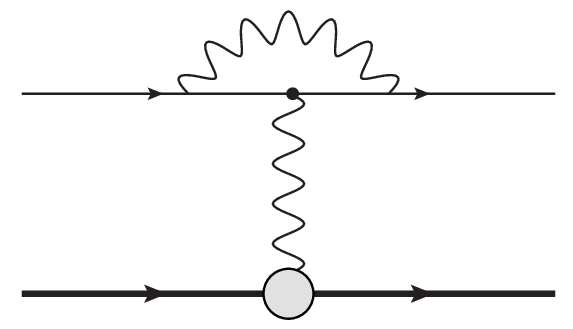} &
    \includegraphics[scale=0.4]{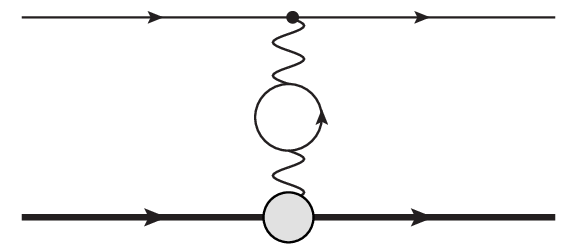} \\
    \includegraphics[scale=0.4]{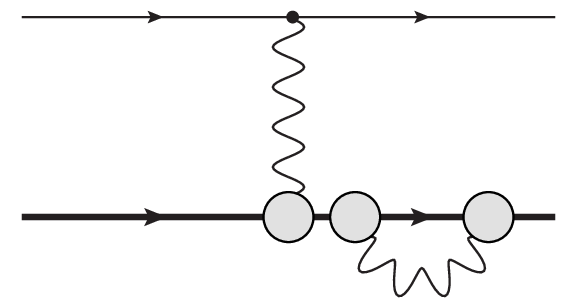} &
    \includegraphics[scale=0.4]{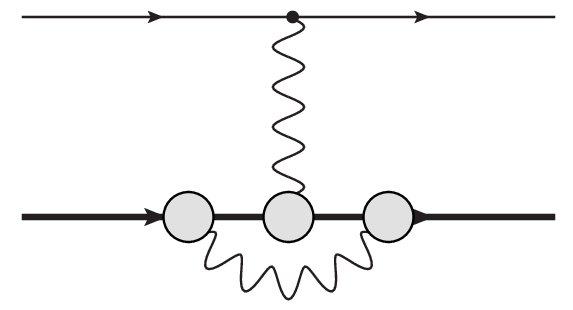}&
    \includegraphics[scale=0.4]{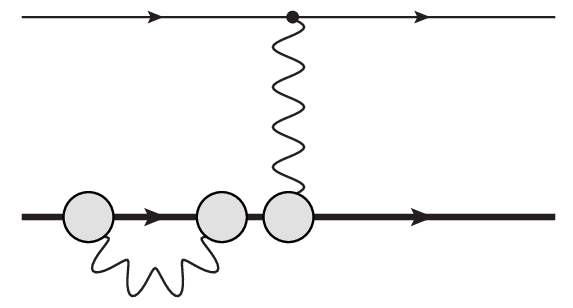} &\\
    \includegraphics[scale=0.4]{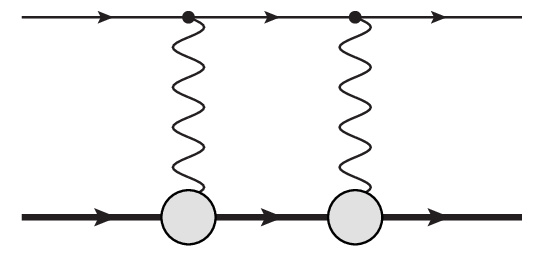} &
    \includegraphics[scale=0.4]{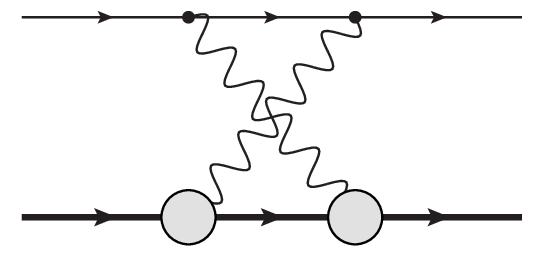}& & \\
    \includegraphics[scale=0.45]{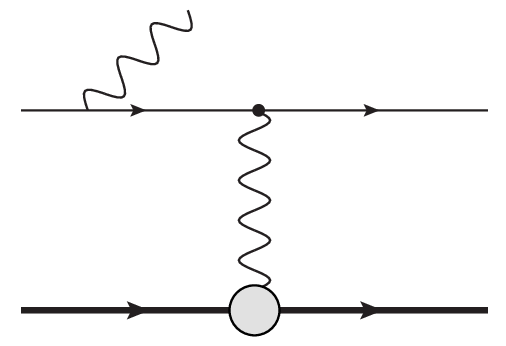} &
    \includegraphics[scale=0.4]{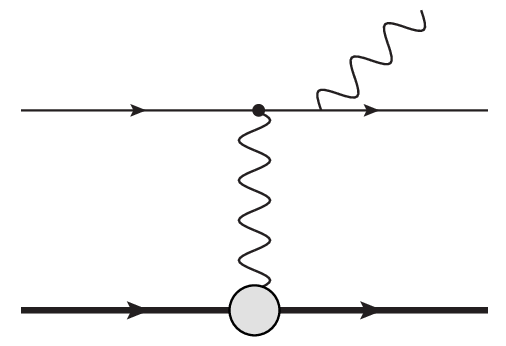}& 
    \includegraphics[scale=0.4]{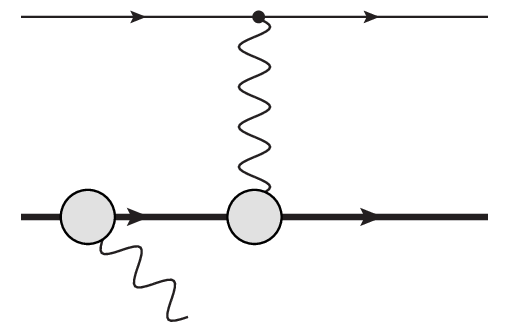} &
    \includegraphics[scale=0.4]{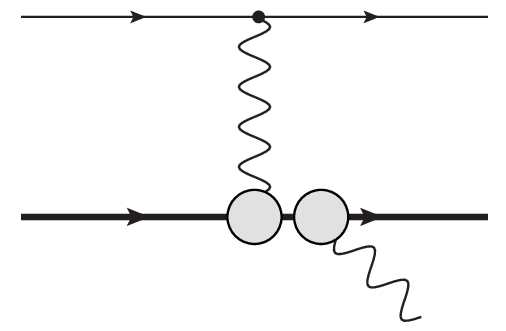}
  \end{tabular}
  \caption{Feynman diagrams entering into the NLO QED radiative corrections to lepton-proton scattering: Lepton self energies, vertex correction, and photon vacuum polarization (first row), proton self energies and vertex correction (second row), TPE diagrams (third row), Bremsstrahlung diagrams for lepton and proton legs (fourth row).}
  \label{fig:feyndiags}
\end{figure}

The NLO corrections in QED to lepton-proton scattering include one-loop self-energy and vertex corrections to all legs and internal lines of the OPE diagram and the TPE diagrams, also called box (and crossed box) diagrams, as well as real photon radiation off the leptons and protons. The corresponding Feynman diagrams are shown in Fig.~\ref{fig:feyndiags}. Note that the $W$ and $Z$ exchange diagrams (as well as weak one-loop corrections) are not included because they are negligible at the small momentum transfers usually available in elastic lepton-proton scattering experiments. Also note that we only consider the elastic proton intermediate state. A discussion of the impact of inelastic contributions, e.~g. $\Delta$ intermediate states and nucleon resonances, on the TPE amplitude can be found in Ref.~\cite{Arrington_2011,Ahmed:2020uso} for instance.
In this paper, we focus on the calculation of the virtual corrections shown in Fig.~\ref{fig:feyndiags}. Real photon emission is included in the soft approximation, i.~e. we only consider photons with energy $k_0<\Delta E_{cut}$ where $\Delta E_{cut}$ is a cut-off parameter. The contribution from soft photon emission is well-known and can be found in Ref.~\cite{Maximon2000}, for instance. We verified that we obtain the same result when using the phase space integrals in the soft approximation provided in Refs.~\cite{tHooft:1978jhc,denner07ewradcorr}.

We begin the presentation of our calculation of lepton-proton scattering cross sections by introducing kinematic conventions used throughout this work. We have $p_1$($p_3$) and $p_2$($p_4$) as the incoming (outgoing) four momenta of the lepton and proton, respectively, in the center-of-mass frame (CM). The virtuality of the exchanged photon is given by $q = p_1-p_3 = p_4-p_2$, satisfying $q^2 = -Q^2 = t < 0$.
The Mandelstam invariants $s,t,u$ are defined as follows:
\[s=(p_1+p_2)^2=(p_3+p_4)^2 \; , \; t=q^2 \;, \; u=(p_1-p_4)^2=(p_2-p_3)^2\]
where $s+t+u=2 m^2+2 M^2$. $M$ and $m$ denote the nucleon and lepton mass, respectively.
In the rest frame of the nucleon (=lab frame) the four-momenta read as follows
\begin{equation}
p_1^* = (E_1, \textbf{p}_1) \; ,\;
p_2^* =  (M, 0) \; ,\;
p_3^* = (E_3, \textbf{p}_3) \; ,\;
p_4^* = (E_4, \textbf{p}_4) = (M+\Delta E , \textbf{p}_1 - \textbf{p}_3)
\end{equation}
with $\Delta E=E_1-E_3=-t/2M$.
The CM energy squared $s$ in terms of the energy of the incident lepton in the lab frame, $E_1$, reads
\[
s=(p_1^* + p_2^*)^2 = M^2 + m^2 + 2 M E_1
\]
and $t$ in terms of the lab frame scattering angle of the lepton $\theta_l$ reads
\[t=2 m^2-2 E_1 E_3+2 \sqrt{(E_1^2-m^2)( E_3^2-m^2)} \cos\theta_l\]
The elastic scattering cross section can be defined in terms of $Q^2=-t$ and $E_1$. In the case of $ep$ scattering, instead of $E_1$ often the dimensionless quantity $\epsilon$ of Eq.~\ref{eq:eps} is used which can be written in terms of Mandelstam invariants as follows (assuming $m=0$):
\begin{equation}\label{eq:epsilon2}
    \epsilon=\frac{2 (M^4-su)}{s^2+u^2-2 M^4}
\end{equation}
For a lepton with non-zero mass, the generalization of $\epsilon$ introduced in Refs.~\cite{Borisyuk:2008es,Tomalak:2014dja,Tomalak:2018jak}, which we denote by $\epsilon_m$, reads in terms of Mandelstam invariants
\begin{equation}\label{eq:epsilonm}
    \epsilon_m=\frac{(s-u)^2-t(t-4M^2)}{(s-u)^2-t(t-4M^2)+2(t-4M^2)(t+2m^2)}\, ,
\end{equation}
which reduces to $\epsilon$ of Eq.~\ref{eq:epsilon2} for $m=0$. Note that $\epsilon_m$ varies between $2m^2/Q^2$ and 1, so that $\epsilon_m>1$ for $Q^2<2m^2$, and the interpretation as the degree of longitudinal polarization of the virtual photon only holds for $\epsilon$ in the massless case. In $\mu p$ scattering in the kinematic range of the MUSE experiment, $Q^2$ is of the same order as $2m_\mu^2\simeq 0.022~{\rm GeV}^2$, and we therefore present our results for $\mu p$ scattering in Section~\ref{sec:results} in dependence of $Q^2$ for fixed beam momenta.
The differential cross section for the $lp \to lp$ ($l=e,\mu$) scattering process including virtual and soft NLO QED corrections reads
\begin{equation}\label{eq:sig}
\frac{d\sigma^{(0+1)}_{vs}}{dt}=\frac{1}{16 \pi \lambda(s,m,M)} \frac{1}{4} \Big[{\sum} |M_0|^2(s,t) (1+\delta_s)+2 Re {\sum} (M_{virt} \cdot M_0^*)(s,t)\Big]
\end{equation}
where $\delta_s$ denotes the relative soft correction and $\lambda(s,m,M)=(s-(m-M)^2)(s-(m+M)^2)$ is the K\"all\'en function. In this work we assume unpolarized incident leptons and that the outgoing proton is not observed. In this case the matrix element squared is averaged/summed over initial/final state spin degrees of freedom denoted by $\frac{1}{4}{\sum}$. 
We keep the lepton and proton masses non-zero throughout, so that the calculation is free of collinear divergences. To regulate the infra-red (IR) divergences we use a fictitious photon mass, $\lambda_\gamma$. As is to be expected, in the sum of the virtual and soft contributions the dependence on $\lambda_\gamma$ cancels. On the lepton side (first row in Fig.~\ref{fig:feyndiags}), ultra-violet (UV) divergences are regulated by dimensional regularization and are canceled in the sum of the lepton self energies and the $\gamma ll$ vertex correction and by renormalization of the photon vacuum polarization which is performed in the on-shell renormalization scheme~\cite{Bohm:1986rj,denner07ewradcorr}. On the proton side (second row in Fig.~\ref{fig:feyndiags}), the virtual contributions can be shown to be UV finite from simple power counting arguments.

The Born matrix element for lepton-nucleon scattering for one-photon exchange shown in Fig.~\ref{fig:bornep} is given by
\begin{equation}
\mathcal{M}_0=
    \bar{u}(p_3,s_3) (- i e Q_l\gamma_{\mu})u(p_1,s_1)\Bigl(\frac{-i g^{\mu\nu}}{q^2+i \delta} \Bigr)
    \bar{u}(p_4,s_4) (-i Z e \Gamma_{\nu}(q))u(p_2,s_2)
\label{eq:born}        
\end{equation}
where $Z$ is the nucleon charge ($Z=1$ for a proton) and $Q_l=-1, \,(l=e,\mu)$. 
Since we work in the regime wherein the quark substructure of the proton is not resolved, however the distribution of the electromagnetic structure of the proton is not simply that of a point particle, we parametrize the nucleon-photon interaction vertex with the usual Dirac ($F_1$) and Pauli ($F_2$) form factors:
\begin{equation}
\Gamma_{\mu}(q^2) = F_1 (q^2) \gamma_{\mu} + \frac{i}{2M} F_2(q^2) \sigma_{\mu \nu} q^{\nu}
\label{eq:pvertex}
\end{equation}
with $i\sigma_{\mu \nu}=-\frac{1}{2} (\gamma_\mu \gamma_\nu-\gamma_\nu\gamma_\mu)$. The functional form of the form factors is chosen to be
\begin{equation}\label{eq:formff}
F_1(q^2) = \frac{F_2(q^2)}{\kappa} = \left(\frac{-\Lambda^2}{q^2 - \Lambda^2} \right)^n, 
\qquad 
n = \left\{ 
\begin{matrix}
1 & \text{monopole} \\
2 & \text{dipole}
\end{matrix}
\right.
\end{equation}
$\kappa=\mu_p-1$ is the anomalous magnetic moment of the proton and
$\Lambda$ is a constant parameter of the order of $1 \text{ GeV}$. 
Inserting the definition of the proton vertex function and taking the spin average/sum of the square of the Born matrix element we obtain (with $e^2=4 \pi \alpha$)
\begin{eqnarray}\label{eq:born2}
\sum |\mathcal{M}_0|^2 &=& \frac{(4 \pi  \alpha Z Q_l)^2 \Lambda^{4n}}{t^2 \Bigl(\Lambda^2-t\Bigr)^{2n}} 
    \times  \nonumber \\
     & &8\Big\{2 m^2 (m^2+2M^2-2 s)+2(M^2-s)^2+t(2s+t)+2 \kappa t (2m^2+t) \nonumber \\
     &+&  \kappa^2 \frac{t}{2 M^2} [m^2 (2M^2+2s+t-m^2)+(s+t)(2M^2-s)-M^4] \Big\}\, .
\end{eqnarray}

The calculation of the virtual NLO QED corrections has been performed independently in three different ways: 1) Performing Passarino--Veltman (PV)~\cite{Passarino:1978jh} reductions to scalar loop integrals using the \textsc{Mathematica} package \textsc{Tracer.m}~\cite{Jamin:1991dp} and in-house routines for the evaluation of the scalar loop integrals based on~\cite{tHooft:1978jhc,Bohm:1986rj,Beenakker:1988jr,Denner:1991qq,Denner:2010tr}, 2) also PV based using \textsc{FeynCalc}~\cite{MERTIG1991345,Shtabovenko_2016,Shtabovenko_2020} and \textsc{Package-X}~\cite{Patel:2015tea,Patel:2016fam}, and 3) based on Integration-By-Parts (IBP) identities and using \textsc{LiteRed}~\cite{Lee:2013mka}, \textsc{Package-X}~\cite{Patel:2015tea,Patel:2016fam},
and \textsc{Collier}~\cite{Denner:2014gla,Denner:2002ii,Denner:2005nn,Denner:2010tr} as described in Section~\ref{subsec:TPE}. The results obtained with these three different calculations agree, which provides a powerful cross-check. The following presentation of the complete analytic result is based on the first approach, and a stand-alone \textsc{Fortran} code for the virtual and soft NLO QED corrections can be obtained from the authors upon request.

\section{Virtual NLO QED corrections}\label{sec:virt}

We will separately discuss the gauge-invariant contributions of the virtual NLO QED corrections to the differential cross section of Eq.~\ref{eq:sig} and decompose $M_{virt}$ in powers of the proton charge $Z$:
\begin{equation}\label{eq:matv}
    M_{virt}=Z^0  M_l+ Z^2 M_p+Z^1 (M_{box}+M_{cbox})\, .
\end{equation}
$M_l \,, (l=e,\mu)$  comprises the corrections shown in the first row of Fig.~\ref{fig:feyndiags}, i.e. the lepton self energy corrections, corrections to the $\gamma ll$ vertex, and the renormalized photon vacuum polarization. $M_p$ describes the corrections shown in the second row of Fig.~\ref{fig:feyndiags}, i.~e. proton self energy and vertex corrections. $M_{box}$ and $M_{cbox}$ denote the contributions from the direct and crossed box, respectively, as shown in the third row of Fig.~\ref{fig:feyndiags}. Structurally the expressions of the one-loop diagrams for low energy scattering are similar to that of high energy scattering of leptons off point-like partons. The only modification comes from the Feynman rule for the $\gamma pp$ vertex of Eq.~\ref{eq:pvertex}. As we will see, the form factors of Eq.~\ref{eq:formff} can be loop-momentum dependent, so that they enter into the propagator structure of the one-loop amplitude.

\subsection{The $Z^0$ contribution}
The $Z^0$ contribution is straightforward to calculate and can be found in the literature (see, e.~g., Ref.~\cite{Beenakker:1991ca}), but for completeness we also provide the result here in terms of form factors $F_{V,E}^{l}$ which contain the loop integrals and the coefficient $C_{E}^{l}$:
\begin{equation}
 \sum (M_{l} \cdot M_0^*) 
 = \frac{\alpha}{2\pi}\left({\sum} |M_0|^2 F_V^{l} + C_E^{l}  F_E^{l}\right) 
\end{equation}
with
\begin{eqnarray}
C_E^l &=& \frac{8 m(4 \pi  \alpha Z Q_l)^2 \Lambda^{4n}}{t^2 \Bigl(\Lambda^2-t\Bigr)^{2n}} 
 \Bigl\{4 (m^2 (m^2+2 (M^2-s))+(M^2-s) (M^2-s-t)) \nonumber \\
 &+&2 \kappa  t (4 m^2-t)-\kappa^2 \frac{t}{4 M^2} 
\left(4 m^2 (m^2-(2 M^2+2 s+t)) \right.\nonumber \\
&+& \left. 4 M^2 (M^2-2 s)+(2 s+t)^2\right)\Bigr\} 
\end{eqnarray}
and 
\begin{eqnarray}
    F_V^{l} &=& \frac{Q_l^2}{2} \left[2 (2 m^2-t) \text{C}_0(t,m,\lambda_\gamma,m)-3 \text{B}_0(t,m,m)+3\Delta_m-2 \ln\Bigl(\frac{\lambda_\gamma^2}{m^2}\Bigr)+2\right]\nonumber  \\
    &+&\frac{\hat\Pi_T(q^2)}{q^2}
    \\
   F_E^{l} &=& \frac{m Q_l^2}{(4 m^2-t)} \left[\text{B}_0(t,m,m)-\Delta_m-2\right]\, .
\end{eqnarray}
with $\Delta_m=\frac{2}{4-d}-\gamma_E-\ln(\frac{m^2}{4\pi \mu^2})$ denoting the UV-divergent part in $d$ dimensions.
The contribution from the transverse part of the renormalized photon vacuum polarization reads as follows: 
\begin{eqnarray}\label{eq:vacpol}
\frac{\hat\Pi_T(q^2)}{q^2} &=& \frac{2}{9} \sum_{l=e,\mu,\tau}Q_l^2\Big[1-3 (1+\frac{2 m_l^2}{t}) \left(\text{B}_0(t,m_l,m_l)-\Delta_{m_l}\right)\Big]
\,.\end{eqnarray}
We only include the leptonic contributions to the photon vacuum polarization in Eq.~\ref{eq:vacpol}, as in Ref.~\cite{Maximon2000}, since the hadronic contribution $\Delta\alpha_{had}(t)$ is not calculable in perturbative QED but is obtained from a dispersion integral over the measured $e^+e^-\to$ hadrons cross section. Being a correction to the photon propagator, it does not depend on the lepton flavor, on the nucleon model, or on the treatment of real photon emission, and it is routinely included in experimental analyses by means of parametrizations of $\Delta\alpha_{had}(t)$~\cite{Burkhardt:2005se,Jegerlehner:2008rs}. It can be added to the results presented here by adding $2\,{\rm Re}\,\Delta\alpha_{had}(t)$ to the relative correction $\delta_{vs}$ of Eq.~\ref{eq:dvs}. For $Q^2=0.01, 0.1, 1$ and $6~{\rm GeV}^2$ it amounts to $0.02\%$, $0.16\%$, $0.74\%$ and $1.5\%$ of the Born cross section, respectively, i.~e. it is comparable to the muon contribution at $Q^2 \approx 0.1~{\rm GeV}^2$ and exceeds it at larger momentum transfer, while the $\tau$ contribution is negligible in the whole range. Since the vacuum polarization is even under $e^+ \leftrightarrow e^-$, it only contributes to $\delta_{even}$ in Eq.~\ref{eq:rpm} and therefore does not enter the ratio $R_{2\gamma}$ at the order considered here, while the hadronic contribution shifts the relative correction $\delta_{vs}$ shown in Fig.~\ref{fig:dvsZ0Z1Z2} by the amounts quoted above, which at $Q^2$ of a few ${\rm GeV}^2$ are of the same size as the hard elastic TPE correction.

With $\beta_l=\sqrt{1-4m^2/t+i\varepsilon}$ and $x_t=(\beta_l-1)/(\beta_l+1)$ the scalar functions \cite{denner07ewradcorr,Beenakker:1988jr} can be written as:
\begin{eqnarray}\label{eq:b0c0l}
B_0(t,m,m)&=&\Delta_m+2+\beta_l \ln(x_t)  \\
C_0(t,m,\lambda_\gamma,m)&=&-\frac{1}{\beta_l t} \Big\{\ln(x_t) \ln(\frac{m^2}{\lambda_\gamma^2})
+\ln(x_t) \Big(-\frac{1}{2} \ln(x_t)+2\ln(1-x_t^2)\Big) \nonumber\\
&-&\frac{\pi^2}{6} + {\rm Li}_2(x_t^2)+
2 {\rm Li}_2(1-x_t)\Big\}
\end{eqnarray}

\subsection{The $Z^2$ contribution}

The QED $\gamma pp$ vertex and proton self energy corrections can be shown to still satisfy a Ward-Takahashi identity even with the form factor dependence on the loop momentum. We still calculate both corrections directly to have an additional check.
The vertex and self energy loops contain only the nucleon as intermediate state. For the photon-proton couplings we use the on-shell vertex of Eq.~\ref{eq:pvertex} either evaluated at the physical momentum transfer $q^2=-Q^2$ or the loop momentum $k^2$. Note that the latter constitutes a model for the off-shell electromagnetic structure of the proton inside the loop, and should be distinguished from the on-shell Sachs form factors $G_E(Q^2)$ and $G_M(Q^2)$ that parametrize the Born amplitude  and are the physical observables extracted from elastic scattering data. Inelastic excitations such as the $\Delta(1232)$ resonance are not included. This is consistent with the treatment of the box diagrams. The $Z^2$ contribution consists of two parts. The IR-divergent part does not depend on the form factors and is needed to cancel the IR divergence of the soft photon emission. The finite remainder depends on $\Lambda$. This remainder is not an observable by itself. Form factors extracted from elastic scattering data are corrected only for the IR-divergent part and therefore absorb the finite remainder. We nevertheless keep it, as in Ref.~\cite{Maximon2000}, to quantify its size and model dependence. We have also verified that the expressions of Appendix~\ref{sec:pvertex} satisfy charge non-renormalization, i.~e. $F_V^{p,\Lambda}+2M F_E^{p,\Lambda}+F_V^{p,\Sigma}\to 0$ for $t\to 0$, as required by the Ward-Takahashi identity.
Again, we write the $Z^2$ contribution in terms of form factors and coefficients, separately for the proton self energy and vertex correction:
\begin{equation}
 \sum (M_{p} \cdot M_0^*) 
 = \frac{\alpha}{2\pi}\left({\sum} |M_0|^2 F_V^{p,\Sigma} + C_V^{p,\Lambda}  F_V^{p,\Lambda}+C_E^{p,\Lambda}  F_E^{p,\Lambda}\right) 
\end{equation}
Note that the proton self energy only contributes to $F_V^p$. The coefficients are
\begin{eqnarray}
C_V^{p,\Lambda} &=&
(4 \pi  \alpha Z Q_l)^2 
\frac{8}{t^2\left(1 - \frac{t}{\Lambda^{2}}\right)^{2n}}
\times \\ \nonumber
&&\left[
\kappa\, t\left(2 m^{2} + t\right)
+
2 m^{4}
+ 2 M^{4}
+ 4 m^{2}(M^{2} - s)
- 4 M^{2}s
+ 2 s^{2}
+ 2 s t
+ t^{2}
\right]\\
C_E^{p,\Lambda} &=&
(4 \pi  \alpha Z Q_l)^2 
\frac{8 M}{t^2\left(1 - \frac{t}{\Lambda^{2}}\right)^{2n}}
\times \\ \nonumber
&&\left(4 + \frac{\kappa t}{M^2}\right)
\left[
m^{4}
+ M^{4}
- 2 M^{2} s
+ m^{2}(2 M^{2} - 2 s - t)
+ s(s+t)
\right] 
\end{eqnarray}
The proton self energy contribution is given in terms of derivatives of the proton self energy
\begin{equation}
\frac{\alpha}{2\pi} Z^2 F_V^{p,\Sigma}=\frac{1}{2}\frac{\partial \Sigma(\slashed p_2)}{\partial \slashed p_2} \Bigr|_{\slashed p_2 \to M}+
\frac{1}{2}\frac{\partial \Sigma(\slashed p_4)}{\partial \slashed p_4} \Bigr|_{\slashed p_4 \to M}
\end{equation}
where  $-i\Sigma(\slashed p)$ (with $p=p_{2,4}$) describes the correction shown in Fig.~\ref{fig:feyndiags} (1. and 3. diagram in the second row):  
\begin{equation}
\label{ref:intpself}
-i\Sigma(\slashed p) = -e^2 Z^2 (\Lambda^2)^{2n} \int \frac{d^4 k}{(2\pi)^4}\tilde{\Gamma}_{\mu}(k)
\frac{1}{(\slashed p - \slashed k - M)} \tilde{\Gamma}^{\mu}(-k)
\frac{1}{(k^2-\lambda_\gamma^2)}\frac{1}{(k^2-\Lambda^2)^{2n}}
\end{equation}
where we used the loop-momentum dependent vertex function
\begin{align}
\Gamma_{\mu}(k) = \left( \frac{-\Lambda^2}{k^2 - \Lambda^2} \right)^n \left( \gamma_{\mu} + \frac{i \kappa}{2 M} \sigma_{\mu \nu} k^{\nu} \right) =  \left( \frac{-\Lambda^2}{k^2 - \Lambda^2} \right)^n \tilde \Gamma_{\;\mu}(k).
\end{align}
We follow~\cite{Maximon2000} and use
\begin{equation}\label{eq:decomp}
\frac{1}{(k^{2}-\Lambda^{2})^{2n}(k^{2}-\lambda_\gamma^{2})}
=
\frac{1}{(2n-1)!}\,\frac{\partial^{2n-1}}{\partial (\Lambda^2)^{2n-1}} \frac{1}{(\Lambda^2-\lambda_\gamma^2)}\left[\frac{1}{(k^2-\Lambda^2)} -\frac{1}{(k^2-\lambda_\gamma^2)}\right]\, .
\end{equation}
This allows us to perform the standard PV reduction to scalar integrals and express $F_V^{p,\Sigma}$ in terms of (at most) two-point functions and their derivatives with respect to $\slashed p$ and $\Lambda^2$. The 
final result then reads for $n=2$:
\begin{equation}
F_V^{p,\Sigma}=  2 \ln \left( \frac{\Lambda}{\lambda_\gamma} \right) - \frac{3x}{8 \beta_c^7} \mathcal{C}_{d,1} \ln \left( x + \beta_c \right) + \frac{1}{24 \beta_c^6} \mathcal{C}_{d,2} 
\end{equation}
where $x = \frac{\Lambda}{2 M}$, $\beta_c = \sqrt{x^2 - 1+i \varepsilon}$, and the coefficients $\mathcal{C}_{d,(1,2)}$ are
\begin{align*}
\mathcal{C}_{d,1} &= \kappa^2 \left(2 x^2-7 x^4\right)+\kappa \left(4 x^6-22 x^4+8
    x^2\right)+8 x^6-28 x^4+20 x^2-5 \\
\mathcal{C}_{d,2} &= \kappa^2 \left(-6 x^8+3 x^6-66 x^4+24 x^2\right)+\kappa \left(6x^6-168 x^4+72 x^2\right)\\
&-32 x^6+30 x^4-87 x^2+44
\end{align*}
For $n=1$ we find:
\begin{equation}
F_V^{p,\Sigma} = 2 \ln \left( \frac{\Lambda}{\lambda_\gamma} \right) + \frac{1}{\beta_c^2} \mathcal{C}_{m,1} + \frac{3 x}{\beta_c^3} \mathcal{C}_{m,2} \ln \left(x+\beta_c \right) +  12 x^4 \mathcal{C}_{m,3} \ln \left( 2 x \right)  
\end{equation}
with
\begin{align*}
\mathcal{C}_{m,1} &=  \kappa^2 \left(3 x^2-6 x^6\right)+\kappa \left(9 x^2-15 x^4\right)-6
    x^4+2 x^2+1 \\
\mathcal{C}_{m,2} &=\kappa^2 \left(-8 x^8+8 x^6+3 x^4-2 x^2\right)+\kappa \left(-20
    x^6+30 x^4-8 x^2\right)-8 x^6+12 x^4-4 x^2+1
\\
\mathcal{C}_{m,3} &= 
\kappa^2 \left(2 x^2+1\right)+5 \kappa+2
\end{align*}
The proton vertex one-loop correction shown in Fig.~\ref{fig:feyndiags} (2. diagram in the second row) reads
\begin{eqnarray}\label{eq:pvertcorr}
-ieZ\Lambda_{\mu} (p_4,p_2) &=& 
-e^3 Z^3 (\Lambda^2)^{2n} \int \frac{d^4 k}{(2\pi)^4}
\frac{1}{(k^2 - \lambda_\gamma^2)} \frac{1}{(k^2-\Lambda^2)^{2n}} \nonumber \\
&&\quad \times\;
    \tilde{\Gamma}_{\sigma}(-k)
    \frac{1}{\slashed p_4 + \slashed k - M}
    \Gamma_{\mu}(q)
    \frac{1}{(\slashed p_2 + \slashed k - M)} \tilde{\Gamma}^{\sigma}(k) \nonumber \\
    & =:& (-i e Z) \frac{\alpha}{4\pi} Z^2 [\gamma_\mu F_{V}^{p,\Lambda}+(p_2+p_4)_\mu F_E^{p,\Lambda}+ (p_2-p_4)_\mu F_M^{p,\Lambda}] .
\end{eqnarray}
Note that $F_M^{p,\Lambda}$ does not contribute to the cross section. Using Eq.~\ref{eq:decomp} and performing again PV reduction to scalar integrals we can express $F_{V,E}^{p,\Lambda}$ in terms of (at most) three-point functions and their derivatives with respect to $\Lambda^2$. Analytic results for $F_{V,E}^{p,\Lambda}$ for $n=1,2$ are provided in Appendix~\ref{sec:pvertex}.

\subsection{The $Z^1$ contribution: Two-Photon Exchange (TPE) diagrams}\label{subsec:TPE}

\begin{figure}[htbp]
\centering
\includegraphics[scale=.55]{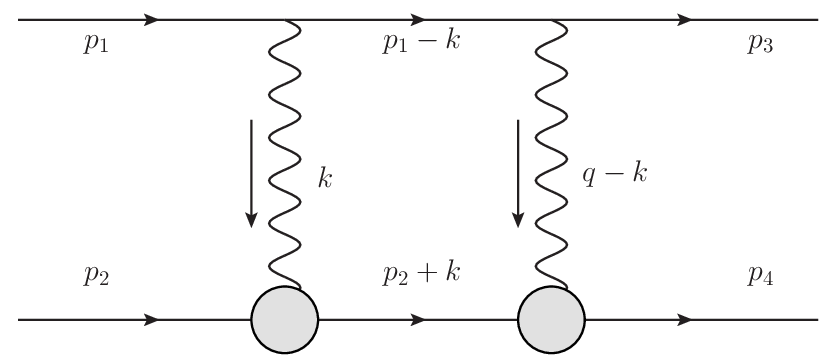}
\caption{Box diagram for lepton proton scattering}
\label{fig:tpe}
\end{figure}

Throughout this work, the box and crossed box diagrams are evaluated with a nucleon intermediate state only. We refer to their contribution as the elastic TPE contribution. It corresponds to the Born TPE contribution in the terminology of Refs.~\cite{Tomalak:2014dja,Tomalak:2014sva,Tomalak:2018jak}.
The two-photon exchange diagram shown in Fig.~\ref{fig:tpe} leads to the
following expression for $\mathcal{M}_{\rm box}$ (with $u_i\equiv u(p_i,s_i)$):
\begin{eqnarray}
\mathcal{M}_{\rm box} &=& (4\pi\alpha)^2 Q_l^2 \, Z \, (\Lambda^2)^{2n} \times \nonumber \\
&&\int \frac{d^4 k}{(2\pi)^4}
\frac{\bar{u}_4\,\tilde\Gamma^\nu(q-k)\,(\slashed{p}_2+\slashed{k}+M)\,
       \tilde\Gamma^\mu(k)\,u_2\;
       \bar{u}_3\,\gamma_\nu\,(\slashed{p}_1-\slashed{k}+m)\,\gamma_\mu\,u_1}
      {D_0^n\,D_1^n\,D_2\,D_3\,D_4\,D_5}\,,
\label{eq:Mbox}
\end{eqnarray}
with propagator denominators
\begin{align}\label{eq:boxprop}
D_0 &= k^2-\Lambda^2\,,       &  D_1 &= (q-k)^2-\Lambda^2\,, \nonumber \\
D_2 &= k^2-\lambda_\gamma^2\,,       &  D_3 &= (q-k)^2-\lambda_\gamma^2\,,  \nonumber\\
D_4 &= (p_1-k)^2-m^2\,,       &  D_5 &= (p_2+k)^2-M^2\,.
\end{align}
The fictitious photon mass $\lambda_\gamma$ acts as an IR regulator, while
$\Lambda$ is the form-factor cut-off scale.  Because the loop-momentum-dependent
vertex $\tilde\Gamma^\mu(k)$ introduces form-factor propagators $D_{0,1}$ raised
to the power $n$ ($n=1$ for a monopole, $n=2$ for a dipole), the denominator
structure of Eq.~\eqref{eq:Mbox} is non-standard and requires special treatment
before any reduction to scalar integrals can be applied. We address this by first applying a partial-fraction decomposition that separates
the form-factor denominators from the remaining propagators: 
\begin{equation}\label{eq:partfrac}
\frac{(\Lambda^2)^{2n}}{D_0^{\,n} D_1^{\,n} D_2 D_3}
=
\frac{1}{D_2 D_3} 
- \sum_{j=1}^{n} \frac{(-\Lambda^2)^{j-1}}{D_3 D_0^j} 
- \sum_{m=1}^{n} \frac{(-\Lambda^2)^{m-1}}{D_2 D_1^m} \\
+ \sum_{j=1}^{n} \sum_{m=1}^{n} \frac{(-\Lambda^2)^{j+m-2}}{D_0^j D_1^m}
\end{equation}
While for the monopole case ($n=1$) Eq.~\eqref{eq:Mbox} immediately reduces to
a standard form
which is directly amenable to PV reduction to at most
four-point scalar functions, the dipole case contains higher powers of $D_0$ and $D_1$,
which would generate scalar integrals beyond the standard four-point
basis.  We eliminate these higher powers by differentiation with respect to $\Lambda^2$ similar to our treatment of the proton vertex corrections and find: 
\begin{eqnarray}\label{eq:masterderiv}
\frac{(\Lambda^2)^{2n}}{D_0^{\,n} D_1^{\,n} D_2 D_3}
&=&
\frac{1}{D_2 D_3} 
- \sum_{j=1}^{n} \frac{(-\Lambda^2)^{j-1}}{(j-1)!}
\left(\frac{\partial}{\partial \Lambda^2}\right)^{j-1} 
\frac{1}{D_3 D_0(\Lambda)} \nonumber \\
&-& \sum_{m=1}^{n} \frac{(-\Lambda^2)^{m-1}}{(m-1)!}
\left(\frac{\partial}{\partial \Lambda^2}\right)^{m-1} 
\frac{1}{D_2 D_1(\Lambda)} 
\nonumber \\
&+& \sum_{j=1}^{n} \sum_{m=1}^{n} \frac{(-\Lambda^2)^{j+m-2}}{(j-1)!(m-1)!} 
\left(\frac{\partial}{\partial \Lambda_1^2}\right)^{j-1} 
\left(\frac{\partial}{\partial \Lambda_2^2}\right)^{m-1} \frac{1}{D_0(\Lambda_1) D_1(\Lambda_2)}\Bigg|_{\Lambda_1=\Lambda_2=\Lambda}
\end{eqnarray}
After these manipulations, every term can be written in a form suitable for
standard PV reduction, and $\sum(\mathcal{M}_{\rm box}\cdot\mathcal{M}_0^*)$ is
expressed as a combination of (at most) scalar four-point functions and for $n=2$ their first- and second-order derivatives with respect to $\Lambda^2$.  The result
for the crossed-box diagram follows from crossing symmetry,
\begin{equation}
\sum(\mathcal{M}_{\rm cbox}\cdot\mathcal{M}_0^*)(s,u)
= -\left.\sum(\mathcal{M}_{\rm box}\cdot\mathcal{M}_0^*)(s,t)\right|_{s\to u}\,.
\label{eq:crossing}
\end{equation}

Integration-by-Parts (IBP) identities provide an independent and complementary
route to the same scalar basis.  The key idea behind IBP reduction is that, in
dimensional regularization, any total derivative of a Feynman integrand
vanishes upon integration over the loop momentum.  This generates linear
relations among loop integrals with different propagator powers, making it
possible to express every member of a given \emph{integral family} as a linear
combination of a small set of irreducible \emph{master integrals}. In practice
the IBP approach requires that all denominators belong to the same family, i.~e.\
that every scalar product involving loop momenta can be written in terms of the
denominators of that family.

\begin{figure}[htp]
    \centering
    \includegraphics[width=1\linewidth]{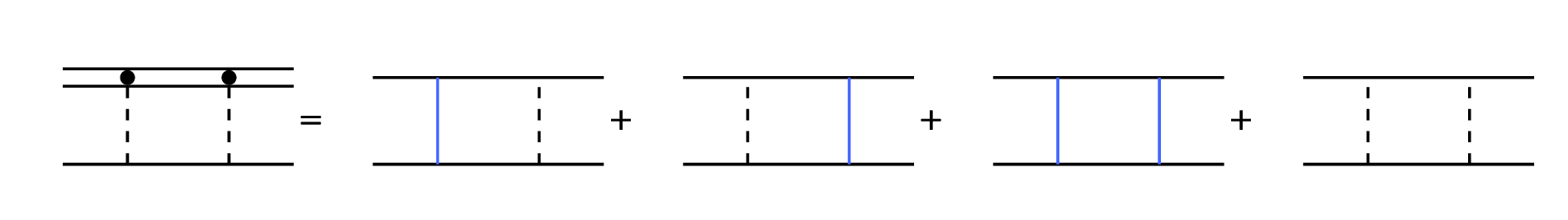}
    \caption{Schematic decomposition of TPE diagrams after partial fraction, where the double
line on left hand side represents proton, the blue lines are massive propagators and dashed
ones are massless propagators.}\label{fig:partial}
\end{figure}

After the partial-fraction decomposition in Eq.~\eqref{eq:partfrac}, every
resulting denominator structure maps onto one of the following four families:
\begin{align*}
\text{Family 1:}&\quad D_0,\,D_1,\,D_4,\,D_5\,, \\
\text{Family 2:}&\quad D_0,\,D_3,\,D_4,\,D_5\,, \\
\text{Family 3:}&\quad D_1,\,D_2,\,D_4,\,D_5\,, \\
\text{Family 4:}&\quad D_2,\,D_3,\,D_4,\,D_5\,.
\end{align*}
Schematically the partial decomposition can be represented by Fig.~\ref{fig:partial}, where the double line on left hand side represents proton, the blue lines are massive propagators and dashed ones are massless propagators.
Using the compact notation $J(n_0',n_1',n_2',n_3',n_4',n_5')$, where $n_i'$ is
the power of denominator $D_i$, any term of Eq.~\eqref{eq:partfrac} is first
identified with its family and rewritten accordingly, e.~g., the entry
$J(0,2,1,0,1,1)$ belongs to Family~3 and is denoted
$J_3(2,1,1,1)$.  The tensor numerator of Eq.~\eqref{eq:Mbox}, after
multiplication by the complex conjugate of the Born amplitude, becomes a set of
scalar products that can be expressed in terms of the denominators of the chosen
family, yielding a representation of the form $C\times J_k(n_1,n_2,n_3,n_4)$
with $C$ a rational function of the Mandelstam variables.  We then applied IBP
reduction via \textsc{LiteRed}~\cite{Lee:2013mka} to decompose all such integrals
into 23 master integrals, comprising four one-point, eight two-point, eight
three-point, and three four-point scalar functions~\cite{tHooft:1978jhc}. Analytic
expressions for these master integrals were taken from \textsc{Package-X}~\cite{Patel:2015tea,Patel:2016fam},
which was linked to \textsc{Collier}~\cite{Denner:2014gla,Denner:2002ii,Denner:2005nn,Denner:2010tr} for numerical
evaluation.  All UV divergences in the master integrals cancel
identically, consistent with simple power-counting arguments for the box
topology, leaving only the expected IR singularities captured by
$\ln\lambda_\gamma^2$.

The PV-based and IBP-based calculations were performed independently and yield
identical numerical results, providing a stringent consistency check on the
entire computation.  We note that one may equivalently regulate the IR
divergences with dimensional regularisation ($\lambda_\gamma\to 0$,
$\ln \lambda_\gamma^2 \to 1/\varepsilon -\gamma +\ln{4\pi}$~\cite{Dittmaier:2003bc}), provided the finite-part mismatch arising from
the $\mathcal{O}(\varepsilon)$ terms of the $d$-dimensional Born amplitude is
properly accounted for when combining virtual and real-radiation contributions. We computed the IR part in both mass regularization and dimensional regularization approach and found agreement.

The elastic TPE contribution has also been evaluated within dispersion relations, for $ep$ scattering in Ref.~\cite{Tomalak:2014sva} and for $\mu p$ scattering in Ref.~\cite{Tomalak:2018jak}. In this approach the imaginary parts of the TPE amplitudes are obtained from unitarity, with the on-shell form factors as input, and the real parts follow from dispersion integrals. The relation to the box diagram calculation with on-shell vertices has been studied in detail in these works. For the $F_1F_1$ and $F_1F_2$ vertex structures the unsubtracted dispersion relations reproduce the box diagram result. For the $F_2F_2$ structure one of the amplitudes does not satisfy an unsubtracted dispersion relation, and agreement with the box diagram model is obtained only after one subtraction. The subtraction function is then fixed to a model estimate or to data. Our calculation thus coincides with the dispersive elastic contribution for the $F_1F_1$ and $F_1F_2$ terms. The difference between the two approaches resides in the terms of order $\kappa^2$. In the MUSE kinematics the subtracted dispersive result agrees within $10\%$ with the model calculation of Ref.~\cite{Tomalak:2015hva}, and the difference is expected to grow with $Q^2$. The dispersive approach with a subtraction function constrained by data is the method of choice when such data are available. The box diagram calculation presented here provides closed analytic expressions for the dependence on the lepton mass, on the form factor parametrization and on the kinematics.

\section{Numerical results}
\label{sec:results}

\begin{figure}[htbp]
  \begin{subfigure}[b]{0.49\textwidth}
    \includegraphics[trim={4cm 14cm 4cm 4cm}, clip,width=\textwidth]{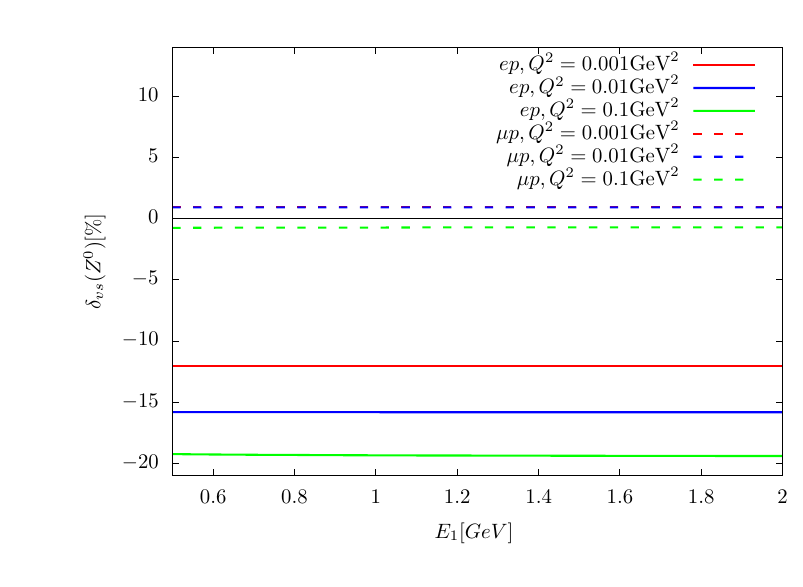}
  \end{subfigure} \hfill
  \begin{subfigure}[b]{0.49\textwidth}
    \includegraphics[trim={4cm 14cm 4cm 4cm}, clip,width=\textwidth]{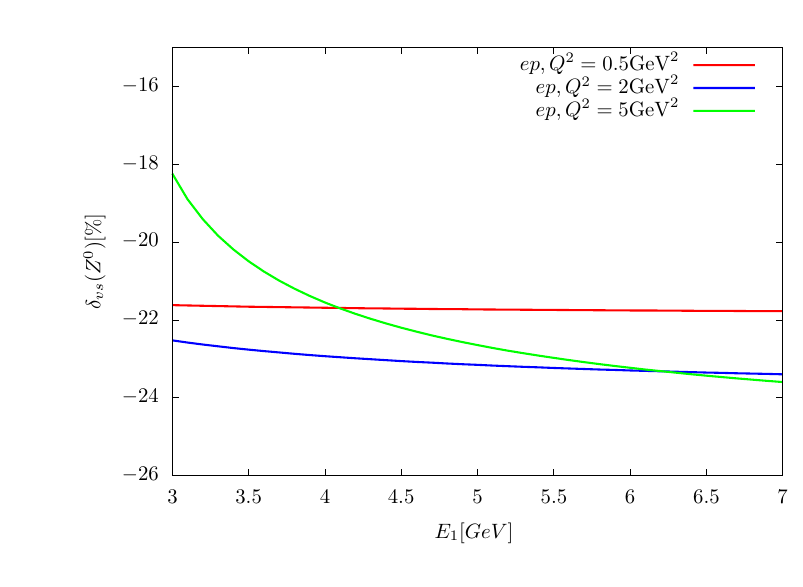}
  \end{subfigure}
  \begin{subfigure}[b]{0.49\textwidth}
    \includegraphics[trim={4cm 14cm 4cm 4cm}, clip,width=\textwidth]{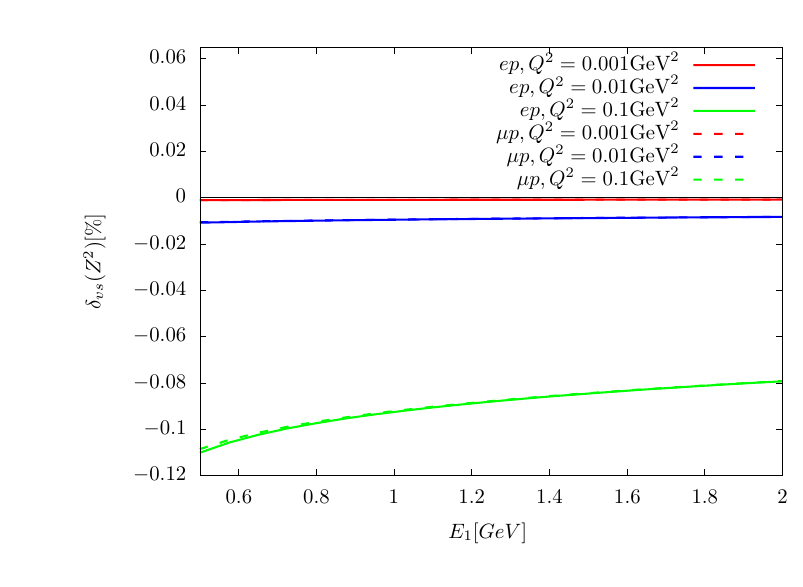}
  \end{subfigure} \hfill
  \begin{subfigure}[b]{0.49\textwidth}
    \includegraphics[trim={4cm 14cm 4cm 4cm}, clip,width=\textwidth]{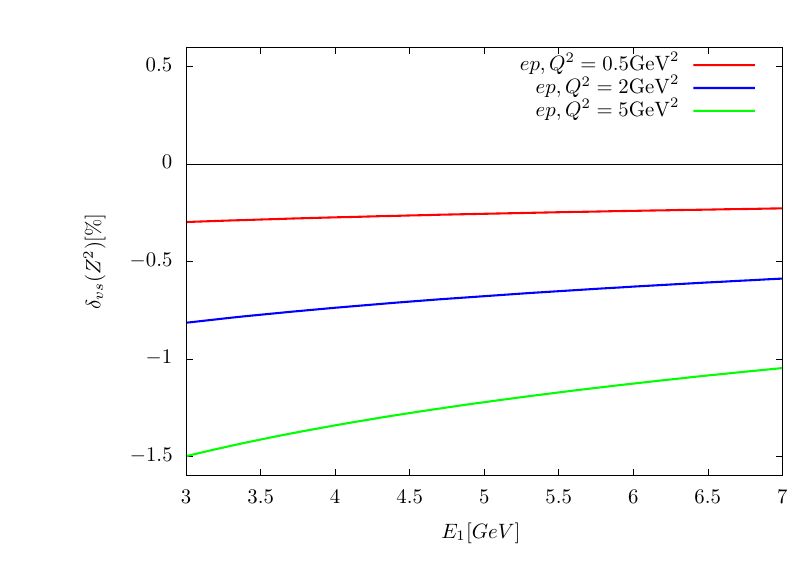}
  \end{subfigure}
  \begin{subfigure}[b]{0.49\textwidth}
    \includegraphics[trim={4cm 14cm 4cm 4cm}, clip,width=\textwidth]{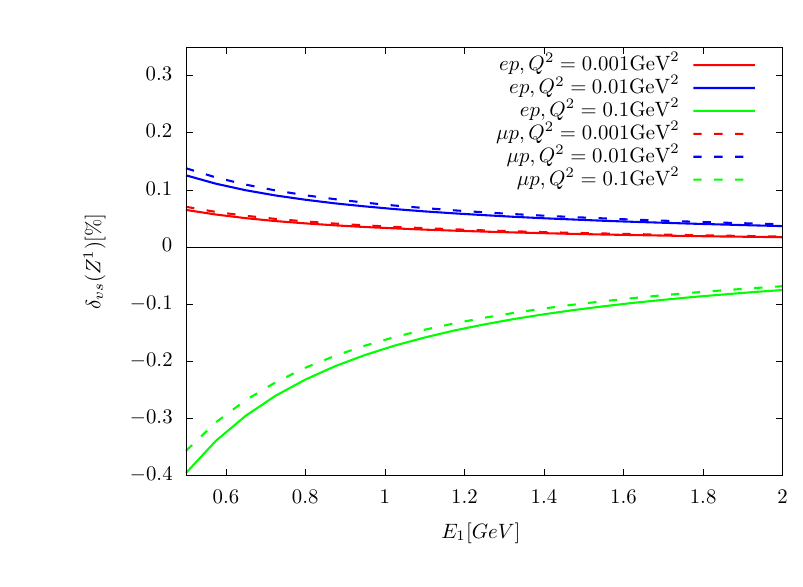}
  \end{subfigure} \hfill
  \begin{subfigure}[b]{0.49\textwidth}
    \includegraphics[trim={4cm 14cm 4cm 4cm}, clip,width=\textwidth]{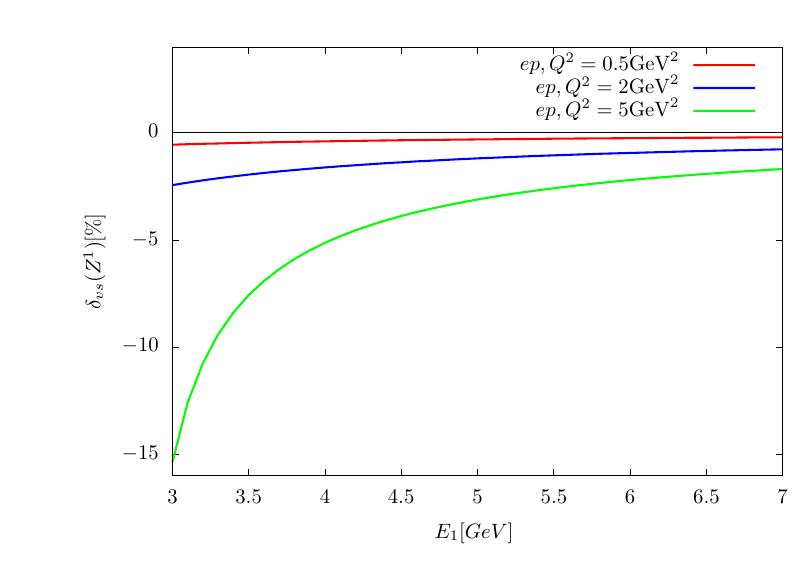}
  \end{subfigure}
\caption{The $Z^0$ (first row), $Z^2$ (middle row), and $Z^1$ (last row) contributions to the relative correction $\delta_{vs}$ for $e^-p$ scattering in dependence of the beam energy $E_1$ for different low and moderate values of $Q^2$. Results for $\mu^-p$ scattering are also provided for low values of $Q^2$.}\label{fig:dvsZ0Z1Z2}
\end{figure}

If not noted otherwise, we use the following values for the input parameters when presenting numerical results which are taken from Ref.~\cite{ParticleDataGroup:2024cfk} (2025 update):
\[\alpha(0)=1/137.035999177 \; ,\; M=0.93827208816~\rm{GeV} \; , \; m_e=0.51099895~ \rm{MeV} \; ,\]
\[m_\mu=105.6583755~ \rm{MeV} \; , m_\tau=1.77693~\rm{GeV}\; ,\kappa=\mu_p-1=1.79284734463 \; .\]
We choose $\Lambda=0.7~\rm{GeV}$ and use dipole form factors ($n=2$). The cut on the photon energy in the lab frame, $\Delta E_{cut}$, to define the soft photon region is taken to be $1\%$ of the beam energy: $\Delta E_{cut}=0.01 \, E_1$.
We first discuss the impact of the different contributions to the relative virtual and soft ${\cal O}(\alpha)$ corrections to elastic lepton-proton scattering by providing results for the relative correction 
\begin{equation}\label{eq:dvs}
\delta_{vs}(E_1,Q^2)=\frac{d\sigma_{vs}^{(0+1)}-d\sigma^{(0)}}{d\sigma^{(0)}} 
\end{equation}
with $d\sigma_{v+s}^{(0+1)}$ of Eq.~\ref{eq:sig}.
Figure~\ref{fig:dvsZ0Z1Z2} shows the impact of the $Z^0$ (leptonic diagrams), $Z^2$ (proton vertex and self energy diagrams), and $Z^1$ (box and crossed box diagrams) contributions in percent of the Born cross section
for different values of the lepton beam energy $E_1$ and the momentum transfer squared $Q^2$.
Note that, since $\delta_{vs}$ includes the soft photon contribution, the $Z^1$ part of $\delta_{vs}$ contains both the interference of the box and crossed box diagrams with the Born amplitude and the interference term of the soft photon emission off the lepton and the proton, and it depends on the cut-off $\Delta E_{cut}$. It should be distinguished from the IR-finite elastic TPE contribution $\delta_{\gamma\gamma}$ introduced in Eq.~\ref{eq:dggir} below.

The leptonic $Z^0$ contribution shown in the first row of Fig.~\ref{fig:dvsZ0Z1Z2}
dominates the virtual-plus-soft correction and reaches the
$\mathcal{O}(20\%)$ level for $e^-p$ scattering, while the corresponding
$\mu^-p$ correction is significantly smaller. This pattern is governed
by the collinear logarithm $L_c = \ln(Q^2/m_\ell^2)$, which is regulated
by the lepton mass. For typical kinematics one has
$L_c^{(e)} \simeq \mathcal{O}(13)$ for the electron at
$Q^2 \sim 0.1\,\text{GeV}^2$, but only $L_c^{(\mu)} \simeq \mathcal{O}(2)$
for the muon, accounting for the order-of-magnitude difference between
the two channels.

After the Bloch--Nordsieck~\cite{Bloch:1937pw} cancellation of $\ln\lambda_\gamma^2$ between
the virtual loop and the soft real-emission contribution, a soft
remnant $L_s = \ln(\Delta E_{cut}^2/E_1^2)$ survives and combines with $L_c$
into a Sudakov-type double logarithm $\sim L_c\, L_s$, which is
responsible for the negative sign and the bulk of the magnitude of
$\delta_{vs}(Z^0)$. The collinear logarithm itself is not cancelled in
the sum of virtual and soft real corrections: the
Kinoshita--Lee--Nauenberg theorem~\cite{Kinoshita:1962ur,Lee:1964is}
guarantees the cancellation of mass singularities only for an
observable that is fully inclusive over collinear photon emission, but
an elastic measurement defined by an upper cutoff $\Delta E_{cut}$ on the
emitted photon energy explicitly rejects hard collinear radiation along
the lepton direction. The numerical result for $\delta_{vs}(Z^0)$
therefore depends on $\Delta E_{cut}$, and a quantitative comparison with
experimental data requires $\Delta E_{cut}$ to be matched to the actual
elastic-event selection used in the analysis (detector resolution,
missing-mass window, or coincidence cuts on the recoil
proton)~\cite{Afanasev:2023gev}. The values shown in Fig.~\ref{fig:dvsZ0Z1Z2}
correspond to the canonical choice $\Delta E_{cut} = 0.01\,E_1$ adopted throughout this work and should be interpreted accordingly. The $Z^1$ and $Z^2$ contributions to $\delta_{vs}$ do not suffer from large collinear logarithms, but are governed by the remainder of the soft logarithms after cancellation of the IR singularity. The virtual-plus-soft $Z^1$ contribution can reach $-15\%$ for a beam energy of $E_1=3$~GeV and a relatively large momentum transfer of $Q^2=5~{\rm GeV}^2$. For a more detailed study of the impact of experimental selection cuts on the NLO QED total and differential cross sections, the virtual and soft contributions should be combined with the hard photon radiation ($E_\gamma>\Delta E_{cut}$) in a Monte Carlo event generator (see, e.~g., Ref.~\cite{Kuraev:2013dra}), which is work in progress and left to a future publication.    

We first compare with the results presented by Maximon and Tjon (MTj)~\cite{Maximon2000} and adjust our input parameters to the MTj values ($M=0.938$~GeV, $\alpha=1/137.036$, $\kappa=1.79$). And we only include the electron contribution to the photon vacuum polarization of Eq.~\ref{eq:vacpol}. In Table~\ref{tab:MTjcomp} we show the comparison with the results presented in Table~I of Ref.~\cite{Maximon2000} where available.
\begin{table}[htbp]
\begin{tabular}{c|cc|cc|cc}
        & \multicolumn{2}{|c|}{$E_1 = 4.4 $ GeV} & 
        \multicolumn{2}{|c|}{$E_1 = 12 $ GeV} &
\multicolumn{2}{|c}{$E_1 = 21.5 $ GeV} \\      
 &        \multicolumn{2}{|c|}{$Q^2 = 6 \; (\text{GeV})^2$} & 
		 \multicolumn{2}{|c|}{$Q^2 = 16 \; (\text{GeV})^2$}
		&  \multicolumn{2}{|c}{$Q^2 = 31.3 \; (\text{GeV})^2$} \\[.2 cm] \hline
                        & this work & MTj          & this work & MTj         & this work & MTj          \\ \hline
$Z^0$                   & $-0.2187$ & $-0.2187$  & $-0.2330$ & $-0.2330$ & $-0.2323$ & $-0.2323$  \\ \hline
$Z^1$(soft)                   & $-0.0569$ & $-0.0569$  & $-0.0517$ & $-0.0517$ & $-0.0625$ & $-0.0625$  \\ \hline
$Z^1$                   & $-0.0598$ & --& $-0.0327$ & --& $-0.0304$ & -- \\ \hline
$Z^2$(soft) & $-0.0242$ & $-0.0242$  & $-0.0359$ & $-0.0359$ & $-0.0452$ & $-0.0452$  \\ \hline
$\delta^{(1)}_{el}$     & $+0.0096$ & $+0.0068$  & $+0.0194$ & $+0.0116$ & $+0.0279$ & $+0.0185$  \\ \hline
$\delta_{vs}$                & $-0.2931$ & $-0.2930$  & $-0.2822$ & $-0.3090$ & $-0.2800$ & $-0.3214$ 
\end{tabular}
\caption{Comparison of the different contributions to $\delta_{vs}$ in electron proton scattering provided by \cite{Maximon2000} in Table~I to our work for different values of the electron beam energy $E_1$ and the momentum transfer $Q^2$.}
\label{tab:MTjcomp}
\end{table}
As expected, there is agreement on the leptonic corrections and when a soft approximation is applied indicated 'soft' in the table. We differ in $\delta_{el}^{(1)}$ which is the remaining part of the $Z^2$ contribution after $Z^2$(soft) is subtracted as described in Ref.~\cite{Maximon2000}. We note that $\delta_{el}^{(1)}$ is a model-dependent quantity. As discussed in Section~\ref{sec:virt}, it is absorbed in the empirical form factors. Both calculations agree that it is small compared to the other contributions. Since the focus of this paper is the elastic TPE contribution, we concentrate on the latter in the remaining discussion. 

There are two approximations for the elastic TPE contribution commonly used in the experimental analysis, based on Mo and Tsai (MoT)~\cite{Mo:1968cg} and MTj~\cite{Maximon2000}, which both use a soft approximation in the calculation of the direct and crossed box diagrams. The MTj approximation neglects the loop momentum in the form factors and in the numerator of the loop integral shown in Eq.~\ref{eq:Mbox}, so that the direct box contribution to the relative correction reads:
\begin{equation}\label{eq:boxMTj}
\delta_{box}(MTj)= \frac{2 Re \sum (M_{box} (MTj)\cdot M_0^*)}{\sum |M_0|^2}=\frac{Z Q_l \alpha}{\pi} t (s-m^2-M^2) Re(D_0^{(2345)})
\end{equation}
where the IR-divergent four-point function is provided in~\cite{Beenakker:1988jr} (for $s \ne (m-M)^2$ and $x_s=K(s,m,M)$ of Eq.~\ref{eq:kfunction}):
\begin{equation}
D_0^{(2345)}=-\frac{2}{t\sqrt{\lambda(s,m,M)}}  \ln[-x_s] \ln[\frac{\lambda_\gamma^2}{-t-i \varepsilon}]  \; .
\end{equation}
The superscript $(2345)$ of the scalar function denotes the contributing propagator denominators of Eq.~\ref{eq:boxprop}.
Adding the crossed box contribution the relative elastic TPE contribution then reads:
\begin{eqnarray}\label{eq:mtjfullm}
\delta_{IR}(MTj)&=&\delta_{box}(MTj)-\delta_{box}(MTj)(s \to u) \nonumber \\
&=&\frac{2Z Q_l\alpha}{\pi} Re\left[-\frac{(s-m^2-M^2)}{\sqrt{\lambda(s,m,M)}}  \ln[-x_s]-(s\to u) \right] \ln[\frac{\lambda_\gamma^2}{-t -i\varepsilon}] 
\end{eqnarray}
In Table~\ref{tab:MTjcomp} we show the result for $\delta_{IR}(MTj)$ (indicated as $Z^1$ (soft)) and our complete result for $\delta_{vs}(Z^1)$. As can be seen, the hard elastic TPE contribution increases with $E_1$ and $Q^2$ and can reach $3\%$ of the  Born cross section for $E_1=21.5$~GeV and $Q^2=31.3~{\rm GeV}^2$.

In the MoT approximation, the soft approximation is also applied to one of the photon propagators so that the box contribution (see also ~\cite{AFANASEV2017245,Arrington_2011}):
\begin{equation}\label{eq:boxMoT}
\delta_{box}(MoT)= \frac{2 Re \sum (M_{box} (MoT)\cdot M_0^*)}{\sum |M_0|^2}=\frac{2Z Q_l \alpha}{\pi} (s-m^2-M^2) Re(C_0^{(245)})
\end{equation}
where the IR-divergent three-point function is also provided in~\cite{Beenakker:1988jr} (for $s \ne (m-M)^2$)
\begin{eqnarray}
C_0^{(245)}&=&-\frac{1}{\sqrt{\lambda(s,m,M)}} \ln[-x_s] \ln[\frac{\lambda_\gamma^2}{M m}]+f(s)
\end{eqnarray}
with
\begin{eqnarray}
f(s)&=&\frac{1}{\sqrt{\lambda(s,m,M)}} \Big\{\ln[-x_s] \Big[-\frac{1}{2} \ln[-x_s]+2\ln[1-x_s^2]\Big]-\frac{\pi^2}{6}+
Li_2[x_s^2] \nonumber \\ 
&+&\frac{1}{2} \ln^2[\frac{m}{M}]+
Li_2[1+x_s \frac{m}{M}]+Li_2[1+x_s 
\frac{M}{m}]\Big\}
\end{eqnarray}
denoting the IR-finite part of $C_0$.
Adding the crossed box contribution and making one more change in the finite part of $C_0$ in the box contribution ($s\to s'=2m^2+2M^2-s$) as described in Refs.~\cite{AFANASEV2017245,Arrington_2011}, one finds for the relative elastic TPE contribution
\begin{eqnarray}\label{eq:motfullm}
\delta_{IR}(MoT)&=&\delta_{box}(MoT)-\delta_{box}(MoT)(s \to u) \nonumber \\
&=&\frac{2Z Q_l\alpha}{\pi} Re\Big\{\left[-\frac{(s-m^2-M^2)}{\sqrt{\lambda(s,m,M)}}  \ln[-x_s]-(s\to u) \right] \ln[\frac{\lambda_\gamma^2}{Mm}] 
\nonumber \\
&+&(s'-m^2-M^2)f(s')-(u-m^2-M^2) f(u)\Big\}
\end{eqnarray}
In the limit $m\ll s,|t|,|u|$ these approximations simplify to
\begin{eqnarray}
\delta_{IR}(MTj)&=&\frac{2ZQ_l\alpha}{\pi}
Re \ln[\frac{M^2-u}{M^2-s}] \ln[\frac{\lambda_\gamma^2}{(-t+i\varepsilon)}]=\frac{2ZQ_l\alpha}{\pi}  \ln[\frac{E_1}{E_3}] \ln[\frac{-t+i\varepsilon}{\lambda_\gamma^2}] \\
\delta_{IR}(MoT)&=&
\frac{2ZQ_l\alpha}{\pi}
Re \Big\{\ln[\frac{M^2-u}{M^2-s}] \ln[\frac{\lambda_\gamma^2}{\sqrt{(s-M^2)(M^2-u)}}]
\nonumber \\
 &-& Li_2[1+\frac{M^2}{M^2-s}]+ Li_2[\frac{u}{u-M^2}]\Big\}
\nonumber \\
&=&\frac{2ZQ_l\alpha}{\pi} \left[ \ln[\frac{E_1}{E_3}] \ln[\frac{2 M \sqrt{E_1E_3}}{\lambda_\gamma^2}]-Li_2[1-\frac{M}{2E_1}]+Li_2[1-\frac{M}{2E_3}]\right] \, ,
\end{eqnarray}
which agree with the results provided in Refs.~\cite{AFANASEV2017245,Arrington_2011}. 
 
\begin{figure}[!htbp]
\includegraphics[width=0.9\linewidth,clip=true, trim=5cm 14.5cm 2cm 4cm]{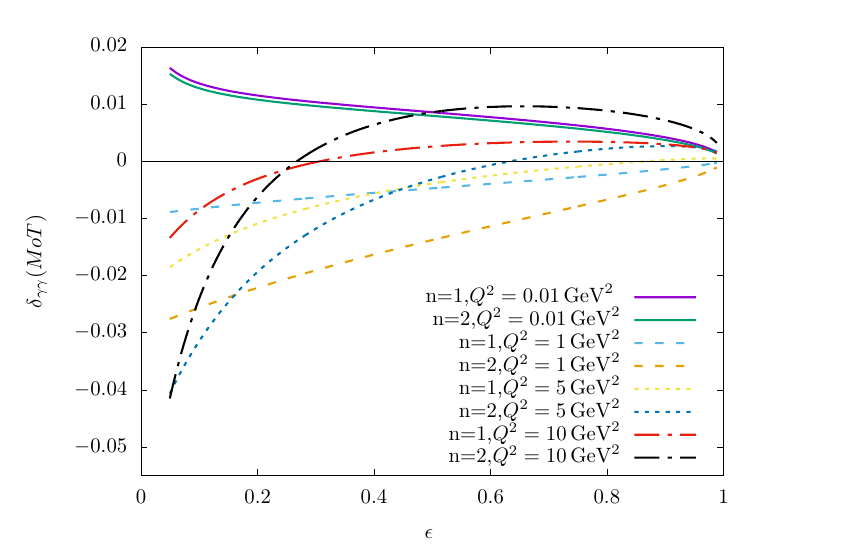}
\caption{Our results for $\delta_{\gamma\gamma}(MoT)$ using monopole and dipole form factors. }
\label{fig:monodipole}
\end{figure}

As discussed for instance in  Ref.~\cite{AFANASEV2017245}, lepton scattering data are often already corrected for the model-independent IR-divergent contribution either using the MoT or MTj approximation. Thus, we will discuss
\begin{equation}\label{eq:dggir}
\delta_{\gamma\gamma}(MoT,MTj)=\delta_{virt}(Z^1)-\delta_{IR}(MoT,MTj)
\end{equation}
This corresponds to the hard IR-finite and model-dependent part of the virtual elastic TPE contribution which is usually not taken into account in the experimental analysis. In Fig.~\ref{fig:monodipole}
we show our results for $\delta_{\gamma\gamma}(MoT)$ for $e^-p$ scattering in dependence of the virtual-photon polarization parameter $\epsilon$ of Eq.~\ref{eq:eps} for different values of $Q^2$ for monopole ($n=1$) and dipole ($n=2$) form factors.  
At small momentum transfer $Q^2 \ll \Lambda^2$ the monopole and
dipole results are essentially indistinguishable. This is expected from the model-independent low-$Q^2$ expansion of the elastic TPE contribution~\cite{Brown:1970te,Tomalak:2015hva,Peset:2021iul}: the leading term of order $Q$, the so-called Feshbach term which arises from the coupling of two Coulomb photons to the proton, and the terms of order $Q^2 \ln Q^2$ do not depend on the proton structure, while the form factors, as well as inelastic intermediate states, only enter at order $Q^2$. As $Q^2$ approaches and exceeds $\Lambda^2$, the loop samples
the form factors at hadronic virtualities where the monopole and dipole
shapes differ markedly and the sensitivity of $\delta_{\gamma\gamma}$ to
the choice of $n$ is correspondingly enhanced, becoming clearly visible
at $Q^2 = 5,10\,\text{GeV}^2$ (see also Fig.~3a of \cite{Blunden:2005ew}).

In the following, we compare our results for the model-dependent, hard part of the relative elastic TPE corrections ($\delta_{\gamma\gamma}(MoT,MTj)$) to $e^-p, \mu^-p$ scattering with results provided in the literature. While we follow MTj~\cite{Maximon2000} and adopt the
vector-dominance parametrization of Eq.~\ref{eq:formff}, in which the
Dirac and Pauli form factors share a common functional form, the results we will use for comparison are based on the
Blunden--Melnitchouk--Tjon (BMT) convention~\cite{Blunden:2003sp}, in which the
Sachs form factors are parametrized as
\begin{equation}\label{eq:gegm}
G_E(q^2) = \frac{G_M(q^2)}{\mu_p}
        = \left(1 - \frac{q^2}{\Lambda_{\rm BMT}^2}\right)^{-n} ,
\end{equation}
and $F_1, F_2$ are obtained from the standard Sachs relations
\begin{equation}
F_1 = \frac{G_E + \tau G_M}{1+\tau}, \qquad
F_2 = \frac{G_M - G_E}{1+\tau},
\end{equation}
with $\tau = Q^2/(4M^2)$. The conversion makes $F_1$ and $F_2$ different
functions of $Q^2$ and introduces a $\tau$-dependent ratio $F_2/F_1$
that is absent in the MTj ansatz. The two parametrizations closely agree for $Q^2 \ll 4 M^2$ for Born kinematics, but inside the TPE
loop the form factors are sampled at virtualities up to the hadronic
scale $\sim \Lambda^2$, where the two ans\"atze differ in shape.  We therefore expect to find differences in the predictions for $\delta_{\gamma\gamma}$ which are more pronounced for larger values of $Q^2$. At small $Q^2$, on the other hand, these differences are suppressed, since the form factors only enter the elastic TPE contribution at order $Q^2$, as discussed above.

\begin{figure}[!htbp]
\includegraphics[width=0.9\linewidth,clip=true, trim=5cm 14.5cm 2cm 4cm]{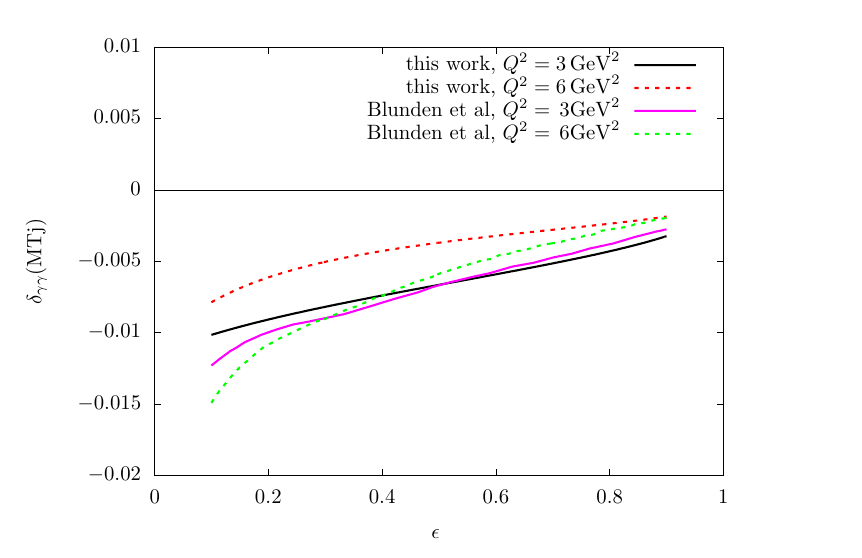}
\caption{Comparison of our results (with $n=1$ and $\Lambda=0.84~\rm{GeV}$) for $\delta_{\gamma\gamma}(MTj)$ with those presented in Ref.~\cite{Blunden:2003sp} (Fig.~3). }
\label{fig:Blunden}
\end{figure}

\begin{figure}[!htbp]
\centering
\includegraphics[width=0.8\linewidth,clip=true,trim=5cm 14.5cm 4cm 4cm]{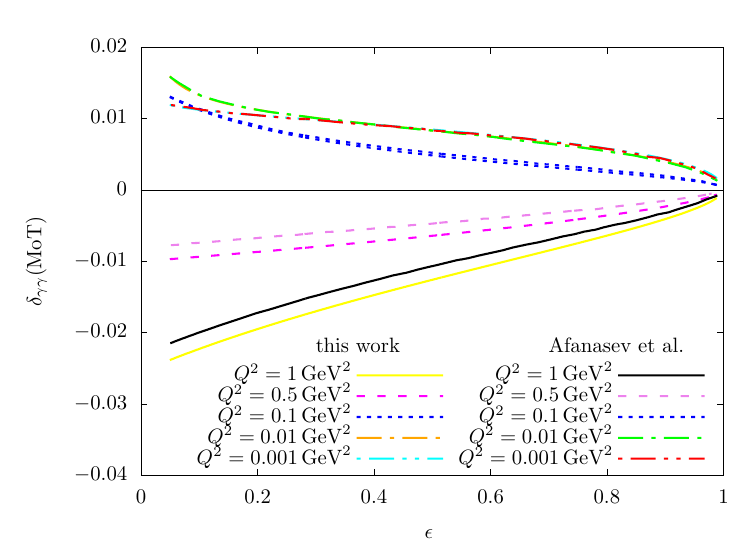}
\caption{Comparison of our results for $\delta_{\gamma\gamma}(MoT)$ with those presented in Fig.~2.3 (left) of Ref.~\cite{AFANASEV2017245} for the range of $Q^2$ from 0.001 GeV$^2$ to 1 GeV$^2$.}
\label{fig:fig13a}
\end{figure}

\begin{figure}[!htbp]
\centering
\includegraphics[width=0.8\linewidth,clip=true,trim=5cm 14.5cm 4cm 4cm]{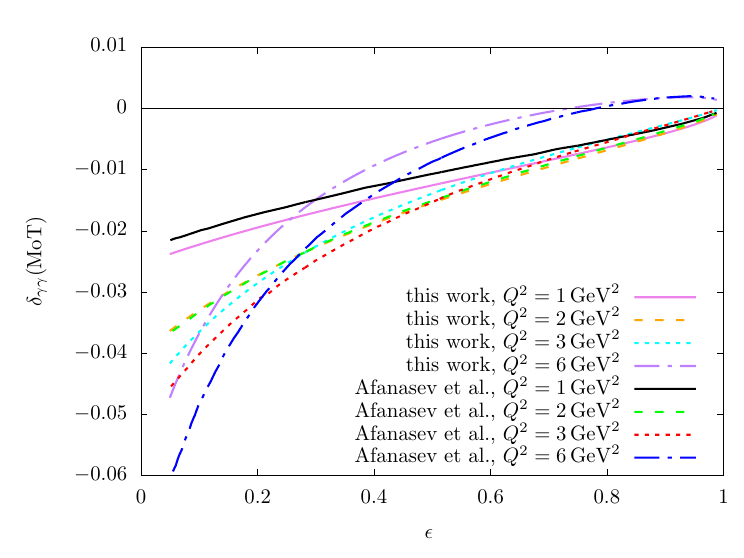}
\caption{Comparison of our results for $\delta_{\gamma\gamma}(MoT)$ with those presented in Fig.~2.3 (right) of Ref.~\cite{AFANASEV2017245} for the range of $Q^2$ from 1 GeV$^2$ to 6 GeV$^2$.}
\label{fig:fig13b}
\end{figure}

In Fig.~\ref{fig:Blunden} we compare $\delta_{\gamma\gamma}(MTj)$ of Fig.~3 in Ref.~\cite{Blunden:2003sp} which is obtained with $n=1$ in Eq.~\ref{eq:gegm}, with our results for $e^-p$ scattering obtained with monopole form factors ($n=1$) and $\Lambda=0.84~\rm{GeV}$ for different values of $\epsilon$ and $Q^2$. 
In Figs.~\ref{fig:fig13a} and \ref{fig:fig13b} 
we compare $\delta_{\gamma\gamma}(MoT)$ of Fig.~2.3 of Ref.~\cite{AFANASEV2017245} 
which is also obtained with $n=1$ in Eq.~\ref{eq:gegm}, 
with our results for $e^-p$ scattering obtained with dipole form factors ($n=2$) and $\Lambda=0.84~\rm{GeV}$ for different values of $\epsilon$ and $Q^2$. 

\begin{figure}[htbp]
\includegraphics[width=0.47\linewidth]{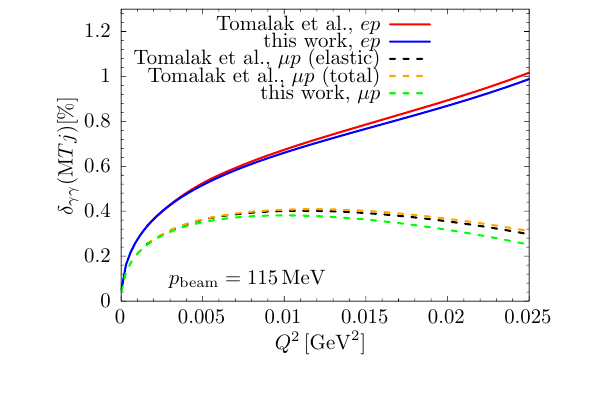}
\includegraphics[width=0.47\linewidth]{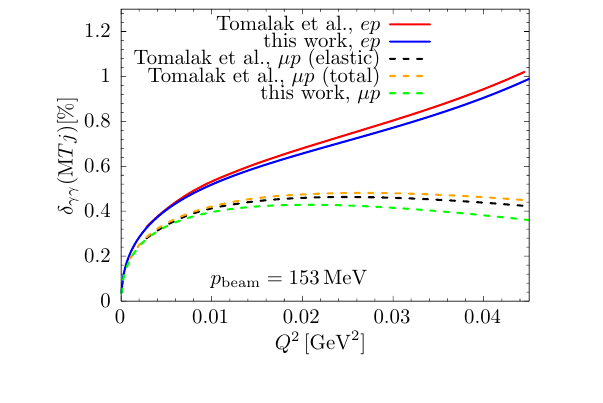}
\includegraphics[width=0.47\linewidth]{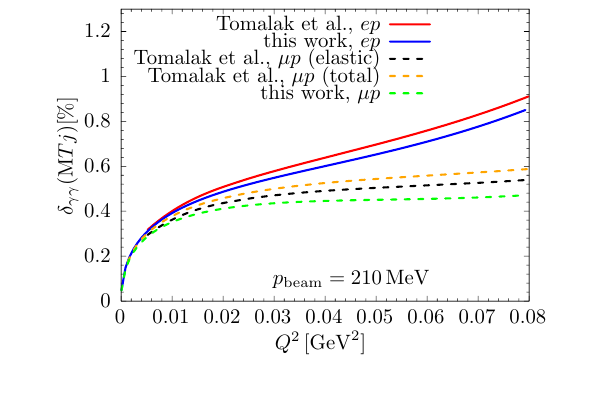}
\caption{Comparison of our results for $\delta_{\gamma\gamma}(MTj)$ for $e^- p$ (solid lines) and $\mu^-p$ (dashed lines) scattering with those presented in Fig.~5 of Ref.~\cite{Tomalak:2014dja} for three different beam momenta and $\Lambda^2=0.71 \, \rm{GeV}^2$. Also shown is the total TPE correction for $\mu^- p$ scattering (orange dashed line) presented in Fig.~14 of Ref.~\cite{Tomalak:2015hva}. The IR subtraction is based on the MTj prescription as described in the text. }\label{fig:Tomalak}
\end{figure}

\begin{figure}[htbp]
 \centering
  \begin{subfigure}[b]{0.49\textwidth}
    \includegraphics[width=\textwidth]{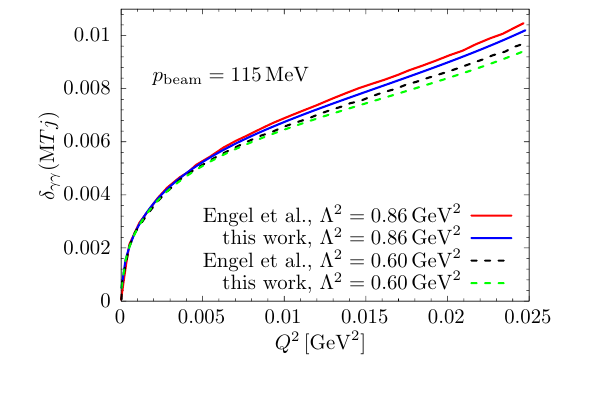}
  \end{subfigure} \hfill
  \begin{subfigure}[b]{0.49\textwidth}
    \includegraphics[width=\textwidth]{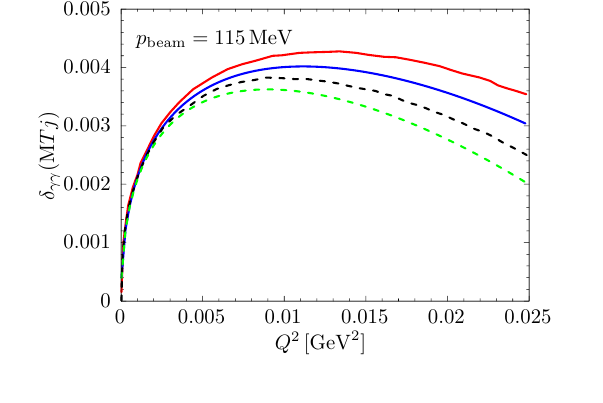}
  \end{subfigure}

  \vspace{10pt} 

  \begin{subfigure}[b]{0.49\textwidth}
    \includegraphics[width=\textwidth]{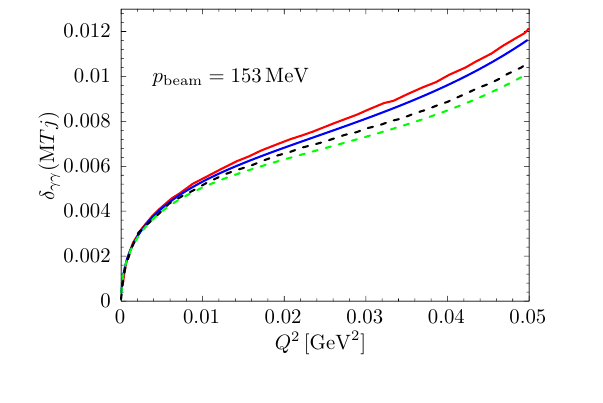}
  \end{subfigure} \hfill
  \begin{subfigure}[b]{0.49\textwidth}
    \includegraphics[width=\textwidth]{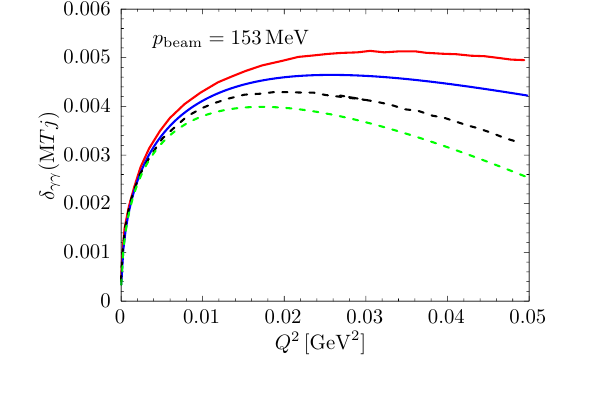}
  \end{subfigure}

  \vspace{10pt}

  \begin{subfigure}[b]{0.49\textwidth}
    \includegraphics[width=\textwidth]{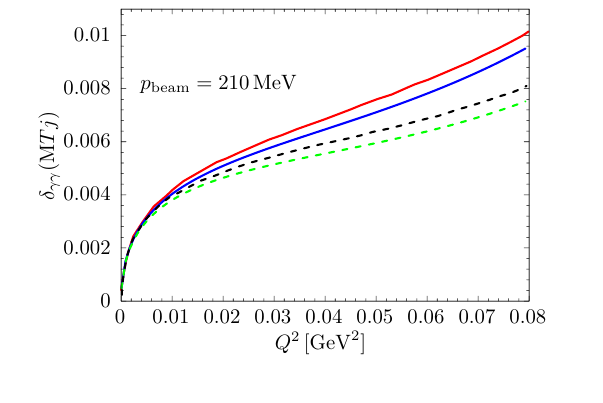}
  \end{subfigure} \hfill
  \begin{subfigure}[b]{0.49\textwidth}
    \includegraphics[width=\textwidth]{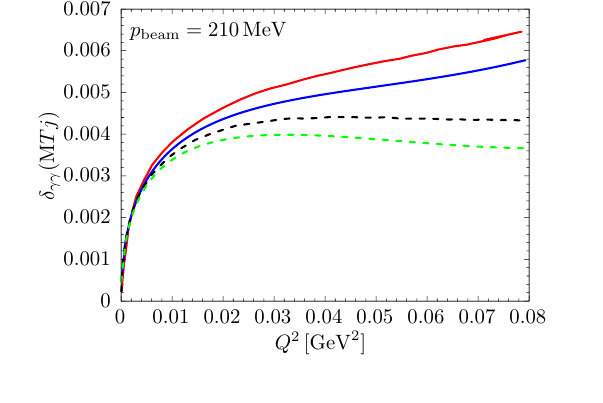}
  \end{subfigure}
\caption{The elastic TPE corrections $\delta_{\gamma\gamma}(MTj)$ to $e^- p$ (left column) and $\mu^-p$ (right column) scattering for three different beam momenta and for two different values of $\Lambda^2$. The IR subtraction is based on the MTj prescription as described in the text. We show our results for $\Lambda^2=0.86,0.60 \, \rm{GeV}^2$ (blue solid and green dashed lines) compared to the predictions presented in Fig.~2 of Ref.~\cite{Engel:2023arz} (red solid and black dashed lines).}
  \label{fig:d2gcomp}
\end{figure}

In Fig.~\ref{fig:Tomalak} and Fig.~\ref{fig:d2gcomp} we compare our results for the $Q^2$ dependence of
$\delta_{\gamma\gamma}(MTj)$ for $e^- p$ and $\mu^-p$ scattering with those presented in Fig.~5 of Ref.~\cite{Tomalak:2014dja} and Fig.~2 of Ref.~\cite{Engel:2023arz}, respectively, which both use $n=2$ in Eq.~\ref{eq:gegm}. The chosen $Q^2$ range and the three beam three-momenta $p_{beam}$ represent the kinematic regime of the MUon-Scattering Experiment MUSE~\cite{MUSE:2013uhu,MUSE:2017dod}, which uses both a beam of electrons and muons ($e^\pm$ and $\mu^\pm$). We adjust our input parameters to the ones provided in Refs.~\cite{Tomalak:2014dja,Engel:2023arz} and use dipole form factors. As expected from our discussion of the difference in the form factor parametrization,
we observe an offset between our prediction and those of Refs.~\cite{Tomalak:2014dja,Engel:2023arz}, small for $e^-p$ and more pronounced for $\mu^-p$, which increases with $Q^2$, where
the loop kinematics are most sensitive to the form-factor shape at
hadronic virtualities. The qualitative $Q^2$-dependence and the
$\Lambda^2$-dependence of $\delta_{\gamma\gamma}$ agree between the three
calculations. In Fig.~\ref{fig:Tomalak} we also show the total TPE correction for $\mu^- p$ scattering presented in Fig.~14 of Ref.~\cite{Tomalak:2015hva}, which in addition accounts for inelastic intermediate states. As discussed in detail in Ref.~\cite{Tomalak:2015hva}, the difference between the total and the elastic contribution is small in the MUSE kinematic range, consistent with the expectation that inelastic contributions are suppressed at low $Q^2$, and it increases with the beam momentum and with $Q^2$. The elastic TPE contribution thus captures the bulk of the total TPE correction in this kinematic regime. The difference between the two form factor parametrizations is of comparable size as the inelastic contribution, so that at MUSE the choice of the form factor model matters as much as the inclusion of inelastic intermediate states.

\begin{figure}[htbp]
\includegraphics[width=0.47\linewidth]{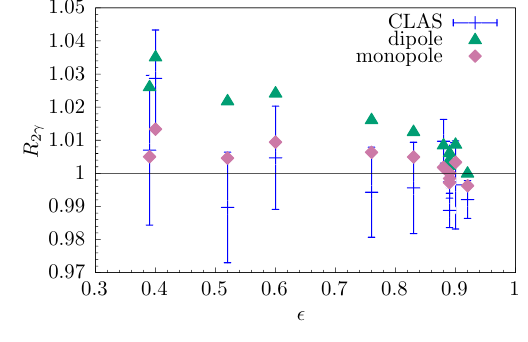}
\includegraphics[width=0.47\linewidth]{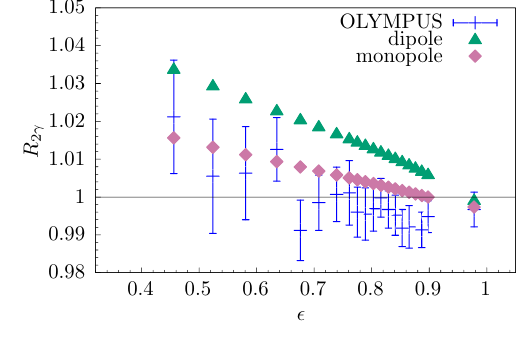}
\includegraphics[width=0.47\linewidth]{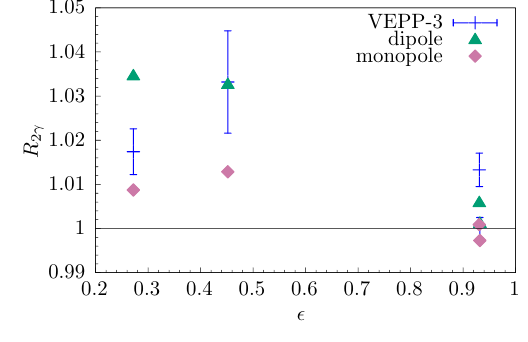}
\caption{CLAS (upper left), OLYMPUS (upper right), and VEPP-3 (bottom) results for $R_{2\gamma}$ for different values of $\epsilon$ and $Q^2$ as provided in Tab.~3.3 of Ref.~\cite{AFANASEV2017245} and our results when using dipole and monopole form factors with $\Lambda^2=0.71 \, \rm{GeV}^2$. Both data and theory are corrected for MoT-type radiative corrections as described in the text.}\label{fig:CLASOlympus}
\end{figure}

One way to study the TPE corrections in unpolarized elastic
lepton-proton scattering is through the measurement of the ratio of electron proton ($e^- p$) to positron proton ($e^+p$) scattering cross sections~\cite{Arrington:2003ck,Arrington:2009qd,CLAS:2013mza} 
\begin{equation}\label{eq:rpm}
    R^{\pm}
    = \frac{\sigma(e^+p)}{\sigma(e^-p)} \approx
    \frac{1+\delta_{even}+\delta_{odd}^{e^+ p}}{1+\delta_{even}+\delta_{odd}^{e^-p}}=
    \frac{1+\delta_{even}+\delta_{\gamma\gamma}^{e^+ p} +\delta_{b}^{e^+p}}{1+\delta_{even}+\delta_{\gamma\gamma}^{e^-p}+\delta_{b}^{e^-p}}
    \approx
    1-\frac{2 (\delta_{\gamma\gamma}^{e^-p}+\delta_{b}^{e^-p})}{1+\delta_{even}}
\end{equation}
where $\delta$ denote the corrections relative to the Born cross section. Here $\delta_{odd}^{e^\pm p}$ refers to relative corrections that change sign under the inversion of $e^+ \to e^-$ consisting of the $Z^1$ contributions, i.~e. the interference bremsstrahlung ($\delta_{b}^{e^\pm p}$) and TPE diagrams: $\delta_{odd}^{e^\pm p} = \delta_{\gamma \gamma}^{e^\pm p}+\delta_{b}^{e^\pm p}$. $\delta_{even}$ comprises the remaining contributions. After a suitable subtraction of the bremsstrahlung contribution, $R_{2\gamma}=1-2\delta_{\gamma\gamma}^{e^-p}$ is usually used for comparison with experimental data. In Fig.~\ref{fig:CLASOlympus} we show our results for $R_{2\gamma}$ obtained with $\delta_{\gamma\gamma}(MoT)$ of Eq.~\ref{eq:dggir} and using dipole and monopole form factors together with the CLAS~\cite{CLAS:2003umf}, OLYMPUS~\cite{OLYMPUS:2013lem}, and VEPP-3~\cite{Rachek:2014fam} measurements, as provided in Tab.~3.3 of Ref.~\cite{AFANASEV2017245}. We observe an at most $3.5\%$ effect of the elastic TPE in the dipole case and a better agreement with data in the monopole case. However, a more complete comparison should also include other hadronic intermediate states, such as the $\Delta(1232)$ resonance, as done for instance in Ref.~\cite{Ahmed:2020uso}. There it was found that in the $\epsilon$ and $Q^2$ range of the OLYMPUS experiment the inclusion of this resonance can reduce the effect of the nucleon elastic contribution to the TPE by up to about $2\%$.   

\section{Conclusions}
\label{sec:conclusions}
We have presented a complete calculation of the NLO QED radiative
corrections to elastic lepton-proton scattering, with particular
attention to the elastic two-photon-exchange (TPE) contribution. Loop momentum dependent
proton form factors are retained throughout the calculation of the virtual one-loop corrections,
including the box and crossed-box topologies. The resulting
non-standard denominator structure is reduced to the Passarino-Veltman
basis along two independent routes, i.~e. partial fractioning combined with
differentiation in the cut-off scale $\Lambda$, and integration-by-parts (IBP) identities
applied family by family, which yield identical results and provide an
internal check of the calculation. Lepton and proton masses are kept
finite, and the infrared sector is regulated equivalently by a
fictitious photon mass and by using dimensional regularization.

Numerically, the leptonic $Z^{0}$ contribution dominates the
virtual-plus-soft correction, the proton-side $Z^{2}$ correction is
comparatively small, and the structure-dependent $Z^{1}$ elastic TPE contribution exhibits
the $\epsilon$- and $Q^{2}$-dependence relevant to form-factor
extractions. Comparisons with a selection of existing
calculations agree at the qualitative level, with residual differences possibly
traceable to the choice of form-factor parametrization. We find our results for $R_{2\gamma}$ to be consistent with CLAS and OLYMPUS measurements within their experimental uncertainties when using monopole form factors, but observe less agreement with VEPP-3 data.

The framework presented here is restricted to the elastic proton
intermediate state. Inelastic contributions, such as $\pi N$ and nucleon resonances at moderate momentum transfer, and the partonic regime at
large $Q^{2}$ lie outside its scope and remain natural directions for
further work. With MUSE, the precision programs at MAMI and Jefferson
Lab, and the ongoing concerted efforts of both experiment and theory in reducing uncertainties in the extraction of $e^{+}p/e^{-}p$ observables from data approaching the level at
which structure-dependent corrections such as the TPE contribution become quantitatively visible, we
hope the results reported here to be of use in the corresponding future analyses.

\begin{acknowledgments}
This work was initiated through discussions within the Fermilab NPC Joint Theory-Experiment Working Group and we thank their organizers and members, in particular Minerba Betancourt and Adi Ashkenazi, for helpful comments and feedback in the early stages of this work. We are grateful to Wolf Wackeroth for providing cross-checks of the results presented in Appendix~\ref{sec:BCDderiv}. DW is supported in part by the U.S.\ National Science Foundation
under award PHY-2309085,
and DC was supported in part by the U.S.\ National Science Foundation under award PHY-2014021. Part of SMH’s work was conducted during a research visit to the University at Buffalo. SMH would like to thank Taushif Ahmed and Andreas Rapakoulias for useful discussions. SMH and DW are grateful to the Galileo Galilei Institute for hospitality and support
during the scientific program on “Theory Challenges in the Precision Era of
the Large Hadron Collider”, where part of this work was done.
\end{acknowledgments}

\appendix

\section{Proton Vertex Form Factors}\label{sec:pvertex}

The form factors $F_{V,E}^{(p,\Lambda)}$ describing the proton vertex correction of Eq.~\ref{eq:pvertcorr} can be written in terms of the IR-divergent 3-point scalar function $C_0$ of Eq.~\ref{eq:b0c0l}  (with the replacement $m\to M$), the finite function $C_{0,\Lambda}\equiv C_0(t,M,\Lambda,M)$, and its derivatives $C_{0,\Lambda}^{(i)}=\frac{\partial^i C_{0,\Lambda}}{\partial L^i}$ ($i=1,2,3$ for the dipole case and $i=1$ for the monopole case and $L=\Lambda^2$). We define
\begin{equation}\label{eq:betas}
\beta=\sqrt{1-\frac{4M^2}{t}} \; , \; \beta_\Lambda=\sqrt{1-\frac{4M^2}{L}+i\varepsilon} \; , \; L_\Lambda=\ln\Bigl(\frac{1-\beta_\Lambda}{1+\beta_\Lambda}\Bigr)  \; . 
\end{equation}
Using $C_{0,\Lambda}$ brought in a concise form~\cite{tHooft:1978jhc,Maximon2000}
\begin{equation}
    C_{0,\Lambda} = \frac{1}{\beta t}\left(Li_2(1-\frac{1}{rx})-Li_2(1-\frac{r}{x})-2 \ln(r) \ln(1+\frac{1}{x})\right),
\end{equation}
where $x = (1+\beta_\Lambda)/(1-\beta_\Lambda)$ and $r=(\beta+1)/(\beta-1)$, 
we can easily derive the \nth{1}, \nth{2} and \nth{3} derivatives $C_{0,\Lambda}^{(i)}$ with respect to $L$, noting that $x' = dx/dL = x/k$, with $k=L \beta_\Lambda$. A dot represents differentiation with respect to $x$:
\begin{align*}
C_{0,\Lambda}^{(1)} &=  x'\dot{C}_{0,\Lambda} = \frac{G}{\beta t k} \;  \mbox{with}  \;
G = \frac{\ln(rx)}{rx-1}-\frac{\ln(x/r)}{x/r-1}+2\frac{\ln(r)}{x+1} \\
C_{0,\Lambda}^{(2)} &= \frac{1}{\beta t k^2} \left(-\frac{(k^2+2 M^2 L)}{Lk}G+x \dot{G}\right) \; \mbox{with} \\ 
\dot{G} &= \frac{1}{r-x}+\frac{r}{rx-1}-2\frac{\ln r}{(x+1)^2}
+\frac{r \ln(x/r)}{(r-x)^2}-\frac{r\ln(rx)}{(rx-1)^2} \\
C_{0,\Lambda}^{(3)} &= \frac{1}{\beta t k^3} \left(\frac{2(k^2 +6 M^4)}{k^{2}} G 
+x\left(1-\frac{3 (k^2+2 M^2L)}{kL}\right)\dot{G}
+x^2\ddot{G} \right) \; \mbox{with} \\
\ddot{G} &= \frac{1+r/x}{(r-x)^2}-\frac{r(r+1/x)}{(rx-1)^2}+4\frac{\ln r}{(x+1)^3}
+2\frac{r\ln(x/r)}{(r-x)^3}+2\frac{r^2\ln(rx)}{(rx-1)^3}.
\end{align*}

For the dipole case ($n=2$) we find for $F_{V,E}^{(p,\Lambda)}$: 

\noindent\parbox{\mymathboxwidth}{%
  \raggedright
  \relpenalty=100
  \binoppenalty=100
 \small
  $
48 \beta^2 \beta_\Lambda^5 L^2 M^2 t\,\mathbf{F_V^{(p,\Lambda)}} = \beta_\Lambda^5 L^2 \Biggl(-4 \beta^2 t^2 \biggl(2 (\kappa +1) M^2 \Bigl(-6 \text{C}_{0,\Lambda}+6 \text{C}_0+6 \text{C}_{0,\Lambda}^{(1)} L-3 \text{C}_{0,\Lambda}^{(2)} L^2+\text{C}_{0,\Lambda}^{(3)} L^3\Bigr)+\text{C}_{0,\Lambda}^{(3)} \kappa^2 L^4\biggr)-2 t \biggl(-8 \beta^2 (\kappa +1) M^4 \Bigl(-6 \text{C}_{0,\Lambda}+6 \text{C}_0+6 \text{C}_{0,\Lambda}^{(1)} L-3 \text{C}_{0,\Lambda}^{(2)} L^2+\text{C}_{0,\Lambda}^{(3)} L^3\Bigr)+2 \beta^2 \text{C}_{0,\Lambda}^{(3)} \Bigl(\kappa  \bigl(\kappa^2+\kappa +6\bigr)+4\Bigr) L^4 M^2+3 \kappa^2 L^4 (3 \text{C}_{0,\Lambda}^{(2)}+\text{C}_{0,\Lambda}^{(3)} L)\biggr)+L^4 \biggl((\kappa -1) \kappa^2 (6 \text{C}_{0,\Lambda}^{(1)}+L (6 \text{C}_{0,\Lambda}^{(2)}+\text{C}_{0,\Lambda}^{(3)} L))-4 ((\kappa -2) (\kappa -1) \kappa +2) M^2 (3 \text{C}_{0,\Lambda}^{(2)}+\text{C}_{0,\Lambda}^{(3)} L)\biggr)\Biggr)+\beta_\Lambda L \Biggl(L \biggl(-(\kappa -1) \kappa^2 L^2+2 (\kappa  (\kappa  (3 \kappa -1)-8)-8) L M^2+8 (\kappa  (\kappa  (5 \kappa +13)+36)+42) M^4\biggr)-t \biggl(-16 \beta^2 (\kappa  (4 \kappa +11)+14) M^4+\kappa^2 (\kappa +8) L^2+4 (\kappa  (\kappa  (2 \kappa +5)+17)+20) L M^2\biggr)\Biggr)-6 \text{L}_\Lambda M^2 t \Biggl(80 \beta^2 (\kappa +2) M^4+(\kappa  (\kappa  (\kappa +6)+10)+16) L^2-4 (\kappa  (3 \kappa +13)+24) L M^2\Biggr)+12 L \text{L}_\Lambda M^4 \Biggl((\kappa  (\kappa  (\kappa +5)+16)+28) L+4 (\kappa  ((\kappa -3) \kappa -24)-46) M^2\Biggr)$
}

\noindent\parbox{\mymathboxwidth}{%
  \raggedright
  \relpenalty=100
  \binoppenalty=100
 \small
  $
96 \beta^4 \beta_\Lambda^5 L^2 M^3 t^2\,\mathbf{F_E^{(p,\Lambda)}} = 2 \beta_\Lambda L \Biggl(-t \biggl(32 \beta^2 (\kappa +1) M^6+\kappa^3 L^3+2 \kappa  (\kappa  (8 \kappa +15)-16) L^2 M^2+8 (\kappa  (3 \kappa  (2 \kappa +5)+20)-4) L M^4\biggr)+2 \kappa  t^2 \biggl(-8 \beta^2 (2 \kappa  (\kappa +3)+7) M^4+2 \kappa^2 L^2+(\kappa  (5 \kappa +21)+20) L M^2\biggr)+2 L M^2 \biggl(-(\kappa -3) \kappa^2 L^2+2 (\kappa  (3 (\kappa -7) \kappa -56)-24) L M^2+8 (\kappa  (\kappa  (5 \kappa -3)+6)-2) M^4\biggr)\Biggr)+12 \text{L}_\Lambda M^2 \Biggl(80 \beta^2 M^4 t \biggl(\kappa  t-2 (\kappa +1) M^2\biggr)+2 L M^2 \biggl(8 \Bigl(\kappa  \bigl(\kappa^2+\kappa -6\bigr)-14\Bigr) M^4-3 \kappa  \Bigl(\kappa^2+\kappa +8\Bigr) t^2+4 (\kappa  (\kappa  (4 \kappa +7)+30)+10) M^2 t\biggr)+L^2 \biggl(4 (\kappa +2) ((\kappa -9) \kappa -2) M^4-2 (\kappa  (\kappa  (8 \kappa +13)+12)+4) M^2 t+\kappa  (3 \kappa  (\kappa +1)+8) t^2\biggr)\Biggr)+\beta_\Lambda^5 L^2 \Biggl(2 \beta^2 \kappa  t^3 \biggl(4 \beta^2 M^2 \Bigl(-6 \text{C}_{0,\Lambda}+6 \text{C}_0+6 \text{C}_{0,\Lambda}^{(1)} L-3 \text{C}_{0,\Lambda}^{(2)} L^2+\text{C}_{0,\Lambda}^{(3)} L^3\Bigr)+\text{C}_{0,\Lambda}^{(3)} \kappa^2 L^4\biggr)-2 L^4 t \biggl(8 \beta^2 \text{C}_{0,\Lambda}^{(3)} (\kappa  (\kappa  (\kappa +3)+12)+4) M^4-\kappa^3 (6 \text{C}_{0,\Lambda}^{(1)}+L (6 \text{C}_{0,\Lambda}^{(2)}+\text{C}_{0,\Lambda}^{(3)} L))+2 \kappa  (\kappa  (7 \kappa +9)-8) M^2 (3 \text{C}_{0,\Lambda}^{(2)}+\text{C}_{0,\Lambda}^{(3)} L)\biggr)+\kappa  t^2 \biggl(-16 \beta^4 M^4 \Bigl(-6 \text{C}_{0,\Lambda}+6 \text{C}_0+6 \text{C}_{0,\Lambda}^{(1)} L-3 \text{C}_{0,\Lambda}^{(2)} L^2+\text{C}_{0,\Lambda}^{(3)} L^3\Bigr)-8 \beta^2 \text{C}_{0,\Lambda}^{(3)} \Bigl(\kappa^2-6\Bigr) L^4 M^2+5 \kappa^2 L^4 (3 \text{C}_{0,\Lambda}^{(2)}+\text{C}_{0,\Lambda}^{(3)} L)\biggr)+4 L^4 M^2 \biggl((\kappa -3) \kappa^2 (6 \text{C}_{0,\Lambda}^{(1)}+L (6 \text{C}_{0,\Lambda}^{(2)}+\text{C}_{0,\Lambda}^{(3)} L))-4 \Bigl(\kappa^3+14 \kappa +6\Bigr) M^2 (3 \text{C}_{0,\Lambda}^{(2)}+\text{C}_{0,\Lambda}^{(3)} L)\biggr)\Biggr)$
}

For the monopole case ($n=1$) also $\overline B_0 \equiv B_0-\Delta_M$ contributes which is the UV-finite part of $B_0(t,M,M)$ of Eq.~\ref{eq:b0c0l} (with $m\to M$) and the form factors are: 

\noindent\parbox{\mymathboxwidth}{%
  \raggedright
  \relpenalty=100
  \binoppenalty=100
 \small
  $
128 \beta^2 \beta_\Lambda M^8 t\,\mathbf{F_V^{(p,\Lambda)}} = \beta_\Lambda^3 L^2 \Biggl(t \biggl(8 \beta^2 \Bigl(-\kappa^2+\kappa +4\Bigr) M^4+\kappa^3 L^2-4 \kappa  ((\kappa -3) \kappa -1) L M^2\biggr)+2 L M^2 \biggl(\kappa^2 \Bigl(-3 \kappa  L+16 (\kappa -1) M^2+L\Bigr)+8 M^2\biggr)\Biggr)+\beta_\Lambda \Biggl(\kappa^3 L^2 \biggl(16 M^6 \Bigl(-4 M^2 \bigl(\beta^2 \text{C}_{0,\Lambda}^{(1)} t+\text{C}_{0,\Lambda}\bigr)+\overline{\text{B}}_0+2\Bigr)+8 L M^2 \Bigl(4 \text{C}_{0,\Lambda} M^4-8 \text{C}_{0,\Lambda}^{(1)} M^6-7 M^2+t\Bigr)+L^2 \Bigl(16 \text{C}_{0,\Lambda}^{(1)} M^6+6 M^2-t\Bigr)\biggr)-2 \kappa^2 L^2 M^2 \biggl(8 M^4 \Bigl(4 \beta^2 \text{C}_{0,\Lambda}^{(1)} t^2+\overline{\text{B}}_0+6 \text{C}_{0,\Lambda} t+8\Bigr)+32 M^6 \Bigl(\beta^2 \text{C}_{0,\Lambda}^{(1)} t-3 \text{C}_{0,\Lambda}\Bigr)+4 L M^2 \Bigl(4 M^2 \bigl(\text{C}_{0,\Lambda}+3 \text{C}_{0,\Lambda}^{(1)} \left(t-2 M^2\right)\bigr)-5\Bigr)+L^2 \Bigl(8 \text{C}_{0,\Lambda}^{(1)} M^4+1\Bigr)+6 L t-28 M^2 t\biggr)+4 \kappa  M^2 \biggl(32 \beta^2 M^6 t (\text{C}_{0,\Lambda}-\text{C}_0) \Bigl(t-2 M^2\Bigr)+2 L^2 M^2 \Bigl(-16 M^4 \bigl(3 \beta^2 \text{C}_{0,\Lambda}^{(1)} t+\text{C}_{0,\Lambda}\bigr)+4 M^2+t\Bigr)+16 \beta^2 L M^4 t \Bigl(4 \text{C}_{0,\Lambda}^{(1)} M^4-2 \text{C}_{0,\Lambda}^{(1)} M^2 t+1\Bigr)-L^3 \Bigl(32 \text{C}_{0,\Lambda}^{(1)} M^6+t\Bigr)\biggr)+16 M^4 \biggl(-8 \beta^2 M^4 t (\text{C}_{0,\Lambda}-\text{C}_0) \Bigl(2 M^2-t\Bigr)-2 L^2 \Bigl(4 M^4 \bigl(2 \beta^2 \text{C}_{0,\Lambda}^{(1)} t+\text{C}_{0,\Lambda}\bigr)-6 M^2+t\Bigr)+8 \beta^2 L M^2 t \Bigl(2 \text{C}_{0,\Lambda}^{(1)} M^4-\text{C}_{0,\Lambda}^{(1)} M^2 t+1\Bigr)-L^3 \Bigl(8 \text{C}_{0,\Lambda}^{(1)} M^4+1\Bigr)\biggr)\Biggr)-2 \text{L}_\Lambda t \Biggl(64 \beta^2 (\kappa +2) M^8+\kappa^3 L^4+2 \kappa  ((3-2 \kappa ) \kappa +1) L^3 M^2+4 \kappa  ((\kappa -6) \kappa -2) L^2 M^4+16 \kappa  (3 \kappa +1) L M^6\Biggr)+2 \beta_\Lambda^4 L^3 \text{L}_\Lambda \Biggl(\kappa  t \biggl(\kappa^2 L+2 (3 \kappa +1) M^2\biggr)+2 (1-3 \kappa ) \kappa^2 L M^2+4 \biggl((\kappa -3) \kappa^2+2\biggr) M^4\Biggr)+8 \beta_\Lambda L^2 M^2 \log \Biggl(\frac{L}{M^2}\Biggr) \Biggl(-\kappa  t \biggl(\kappa^2 L+(2-(\kappa -6) \kappa ) M^2\biggr)+2 \kappa^2 (3 \kappa -1) L M^2-4 \biggl((2 \kappa -3) \kappa^2+2\biggr) M^4\Biggr)+4 L \text{L}_\Lambda M^2 \Biggl(\kappa^2 (3 \kappa -1) L^3-2 \biggl((7 \kappa -5) \kappa^2+2\biggr) L^2 M^2+4 \biggl((5 \kappa -7) \kappa^2+4\biggr) L M^4-16 \biggl((\kappa -3) \kappa^2+2\biggr) M^6\Biggr)$
}

\noindent\parbox{\mymathboxwidth}{%
  \raggedright
  \relpenalty=100
  \binoppenalty=100
 \small
  $
16 \beta^4 \beta_\Lambda \Biggl(\beta_\Lambda^2-1\Biggr) M^5 t^2\,\mathbf{F_E^{(p,\Lambda)}} = L^2 \text{L}_\Lambda \Biggl(-8 \biggl(\kappa  \Bigl(\kappa^2+\kappa +18\Bigr)+10\biggr) M^4+4 \kappa^2 (3 \kappa -5) L M^2+3 \kappa^2 (\kappa +1) t^2+2 \biggl(L \kappa^2+2 (\kappa  (6-\kappa  (4 \kappa +7))+2) M^2\biggr) t\Biggr) \beta_\Lambda^4+\Biggl(L \biggl(2 \text{C}_{0,\Lambda}^{(1)} L M^2 t^3+\Bigl(-16 \text{C}_{0,\Lambda}^{(1)} L M^4+\bigl(5 L (\text{C}_{0,\Lambda}+\text{C}_{0,\Lambda}^{(1)} L)-2 \beta^2\bigr) M^2+3 L\Bigr) t^2+2 L M^2 \Bigl(8 \text{C}_{0,\Lambda}^{(1)} M^4-14 (\text{C}_{0,\Lambda}+\text{C}_{0,\Lambda}^{(1)} L) M^2+\text{C}_{0,\Lambda}^{(1)} L^2+\overline{\text{B}}_0+2 \text{C}_{0,\Lambda} L-8\Bigr) t+2 L M^2 \Bigl(32 \text{C}_{0,\Lambda}^{(1)} M^6-8 (\text{C}_{0,\Lambda}+\text{C}_{0,\Lambda}^{(1)} L) M^4+2 \bigl(\text{C}_{0,\Lambda}^{(1)} L^2+2 \text{C}_{0,\Lambda} L+\overline{\text{B}}_0-8\bigr) M^2+3 L\Bigr)\biggr) \kappa^3+L \biggl(2 L \Bigl(96 \text{C}_{0,\Lambda}^{(1)} M^6-2 \bigl(3 \text{C}_{0,\Lambda}^{(1)} L^2+6 \text{C}_{0,\Lambda} L+3 \overline{\text{B}}_0-8\bigr) M^2-5 L\Bigr) M^2+3 \Bigl(L-2 \beta^2 M^2\Bigr) t^2+L \Bigl(-48 \text{C}_{0,\Lambda}^{(1)} M^6-36 (\text{C}_{0,\Lambda}+\text{C}_{0,\Lambda}^{(1)} L) M^4-32 M^2+L\Bigr) t\biggr) \kappa^2+8 M^2 \biggl((\text{C}_{0,\Lambda}-\text{C}_0) M^2 \Bigl(2 M^2-t\Bigr) t^2 \beta^4+L t \Bigl(\bigl(\beta^2 \text{C}_{0,\Lambda}^{(1)} t \left(t-2 M^2\right)-4\bigr) M^2+t\Bigr) \beta^2+4 \text{C}_{0,\Lambda}^{(1)} L^3 M^2 \Bigl(t-7 M^2\Bigr)+L^2 \Bigl(2 \bigl(3 \text{C}_{0,\Lambda}^{(1)} \left(t-4 M^2\right)^2+2 \text{C}_{0,\Lambda} \left(t-7 M^2\right)-9\bigr) M^2+3 t\Bigr)\biggr) \kappa +8 L M^2 \biggl(\Bigl(-8 \text{C}_{0,\Lambda}^{(1)} L M^4-4 \beta^2 M^2+L\Bigr) t-2 L M^2 \Bigl(-16 \text{C}_{0,\Lambda}^{(1)} M^4+6 (\text{C}_{0,\Lambda}+\text{C}_{0,\Lambda}^{(1)} L) M^2+5\Bigr)\biggr)\Biggr) \beta_\Lambda^3+\text{L}_\Lambda \Biggl(4 L \biggl(-4 \Bigl(\kappa  \bigl(\kappa^2+\kappa +18\bigr)+10\Bigr) M^4+(\kappa  (\kappa  (5 \kappa -3)+36)+20) L M^2+(5-3 \kappa ) \kappa^2 L^2\biggr) M^2+\kappa  \biggl(16 \beta^2 M^4+6 \kappa  (\kappa +1) L M^2-3 \kappa  (\kappa +1) L^2\biggr) t^2-2 \biggl(16 \beta^2 (\kappa +1) M^6+4 (\kappa  (\kappa  (4 \kappa +7)-6)-2) L M^4+(\kappa  (12-\kappa  (8 \kappa +15))+4) L^2 M^2+\kappa^2 L^3\biggr) t\Biggr) \beta_\Lambda^2+\Biggl(16 L \biggl(-16 \text{C}_{0,\Lambda}^{(1)} L M^4+2 (3 L (\text{C}_{0,\Lambda}+\text{C}_{0,\Lambda}^{(1)} L)-14) M^2+5 L\biggr) M^4-8 \biggl(8 \Bigl(\beta^2-\text{C}_{0,\Lambda}^{(1)} L^2\Bigr) M^4-8 L M^2+L^2\biggr) t M^2+8 \kappa  \biggl(4 \text{C}_{0,\Lambda}^{(1)} M^2 \Bigl(7 M^2-t\Bigr) L^3+\Bigl(2 M^2 \bigl(-3 \text{C}_{0,\Lambda}^{(1)} \left(t-4 M^2\right)^2+2 \text{C}_{0,\Lambda} \left(7 M^2-t\right)+9\bigr)-3 t\Bigr) L^2+\Bigl(2 \bigl(\beta^4 \text{C}_{0,\Lambda}^{(1)} t^2-44\bigr) M^4+t \bigl(20-\beta^4 \text{C}_{0,\Lambda}^{(1)} t^2\bigr) M^2-t^2\Bigr) L-\beta^2 M^2 \Bigl(2 M^2-t\Bigr) t \Bigl((\text{C}_{0,\Lambda}-\text{C}_0) t \beta^2+4\Bigr)\biggr) M^2+\kappa^2 L \biggl(2 \Bigl(-96 \text{C}_{0,\Lambda}^{(1)} L M^6-16 M^4+6 L \bigl(\text{C}_{0,\Lambda}^{(1)} L^2+2 \text{C}_{0,\Lambda} L+\overline{\text{B}}_0-6\bigr) M^2+5 L^2\Bigr) M^2-3 \Bigl(L-6 M^2\Bigr) t^2+\Bigl(48 \text{C}_{0,\Lambda}^{(1)} L M^6+4 (9 L (\text{C}_{0,\Lambda}+\text{C}_{0,\Lambda}^{(1)} L)-34) M^4+36 L M^2-L^2\Bigr) t\biggr)+\kappa^3 L \biggl(-64 \text{C}_{0,\Lambda}^{(1)} L M^8+16 (L (\text{C}_{0,\Lambda}+\text{C}_{0,\Lambda}^{(1)} L)-2) M^6-4 L \Bigl(\text{C}_{0,\Lambda}^{(1)} L^2+2 \text{C}_{0,\Lambda} L+\overline{\text{B}}_0-14\Bigr) M^4-2 \text{C}_{0,\Lambda}^{(1)} L t^3 M^2-6 L^2 M^2-2 \Bigl(36 M^2+L \bigl(8 \text{C}_{0,\Lambda}^{(1)} M^4-14 (\text{C}_{0,\Lambda}+\text{C}_{0,\Lambda}^{(1)} L) M^2+\text{C}_{0,\Lambda}^{(1)} L^2+\overline{\text{B}}_0+2 \text{C}_{0,\Lambda} L-8\bigr)\Bigr) t M^2+\Bigl(16 \text{C}_{0,\Lambda}^{(1)} L M^4+(14-5 L (\text{C}_{0,\Lambda}+\text{C}_{0,\Lambda}^{(1)} L)) M^2-3 L\Bigr) t^2\biggr)\Biggr) \beta_\Lambda+\Biggl(\beta_\Lambda^2-1\Biggr) L^2 \Biggl(-16 \biggl(\kappa^3+9 \kappa +5\biggr) M^4+4 \kappa^2 (3 \kappa -5) L M^2+3 \kappa^2 (\kappa +1) t^2+2 \biggl(L \kappa^2+(\kappa  (12-\kappa  (7 \kappa +15))+4) M^2\biggr) t\Biggr) \log \Biggl(\frac{L}{M^2}\Biggr) \beta_\Lambda+2 \text{L}_\Lambda M^2 \Biggl(2 L \biggl((5-3 \kappa ) L \kappa^2+4 \Bigl(\kappa  \bigl(\kappa^2+\kappa +18\bigr)+10\Bigr) M^2\biggr) M^2-\kappa  \biggl(8 \beta^2 M^2+3 \kappa  (\kappa +1) L\biggr) t^2+\biggl(16 \beta^2 (\kappa +1) M^4+4 (\kappa  (\kappa  (4 \kappa +7)-6)-2) L M^2-\kappa^2 L^2\biggr) t\Biggr)$
}

\section{Direct Box Contribution}\label{sec:dbox}

Following Eq.~\ref{eq:masterderiv} for the monopole ($n=1$) and dipole case ($n=2$), we separately provide the contributions originating from 
the $\Lambda^0$ term, which contains no derivatives ($i=0$), 
the $\Lambda^2$ term, which contains first derivatives ($i=1$), and the $\Lambda^4$ term, which contains first and second derivatives ($i=2$) of the scalar functions:
\begin{equation}\label{eq:boxsum}
    \sum (M_{box} \cdot M_0^*)= \frac{Q_l^3 Z^2 4\pi \alpha^3}{t} \Big(\frac{\Lambda^2}{\Lambda^2-t}\Big)^n\sum_{i=0}^{2n-2} ({\rm box}_{B}(i)+{\rm box}_C(i)+{\rm box}_{D}(i))
\end{equation}
We work in $d=4$ dimensions and only keep the fictitious photon mass $\lambda_\gamma$ non-zero when it is needed as IR regulator. 
The contributions of the scalar two-point functions, $B_0^{(jk)}$, three-point functions, $C_0^{(jkl)}$, four-point functions, $D_0^{(jklm)}$, and their derivatives, can be written as follows:
\begin{eqnarray}\label{eq:bcoeffs}
{\rm box}_{B}(0)&=&  c_{1ab}(1) B_0^{(23)}+c_{1ab}(2) B_0^{(12)}+
c_{1ab}(3) B_0^{(01)}+c_{1ab}(4) B_0^{(24)}+
c_{1ab}(5) B_0^{(04)} \nonumber \\
&+&c_{1ab}(6) B_0^{(25)}+c_{1ab}(7) B_0^{(05)}
\nonumber \\ 
 {\rm box}_{B}(1)&=& 
 c_{2ab}(1) B_0^{(12)}+
c_{2ab}(2) B_0^{(01)}+c_{2ab}(3) B_0^{(24)}+
c_{2ab}(4) B_0^{(04)}+c_{2ab}(5) B_0^{(25)}
\nonumber \\
&+&c_{2ab}(6) B_0^{(05)}+
c_{2ab}(7) \frac{\partial B_0^{(04)}}{\partial \Lambda}
+c_{2ab}(8) \frac{\partial B_0^{(05)}}{\partial \Lambda}
+c_{2ab}(9) \frac{\partial B_0^{(12)}}{\partial \Lambda}
+c_{2ab}(10) \frac{\partial B_0^{(01)}}{\partial \Lambda}
 \nonumber \\
{\rm box}_{B}(2)&=& 
c_{3ab}(1) \frac{\partial B_0^{(04)}}{\partial \Lambda}
+c_{3ab}(2) \frac{\partial B_0^{(05)}}{\partial \Lambda}
+c_{3ab}(3) \frac{1}{2}\frac{\partial B_0^{(01)}}{\partial \Lambda}
+c_{3ab}(4)
\frac{\partial^2 B_0^{(01)}}{\partial \Lambda_1\partial\Lambda_2}\Bigl|_{\Lambda_1=\Lambda_2=\Lambda} 
 \\ \label{eq:ccoeffs}
{\rm box}_{C}(0)&=&  c_{1c}(1) C_0^{(245)}+c_{1c}(2) C_0^{(234)}+
c_{1c}(3) C_0^{(235)}+c_{1c}(4) C_0^{(124)}+
c_{1c}(5) C_0^{(125)} \nonumber \\
&+&c_{1c}(6) C_0^{(045)}+c_{1c}(7) C_0^{(014)}+
c_{1c}(8) C_0^{(015)}
\nonumber \\
{\rm box}_{C}(1)&=& 
c_{2c}(1) C_0^{(245)}+c_{2c}(2) C_0^{(124)}+
c_{2c}(3) C_0^{(125)} 
+c_{2c}(4) C_0^{(045)}+c_{2c}(5) C_0^{(014)}
\nonumber \\
&+&c_{2c}(6) C_0^{(015)}+
c_{2c}(7) \frac{\partial C_0^{(034)}}{\partial \Lambda}
+c_{2c}(8) \frac{\partial C_0^{(035)}}{\partial \Lambda}
\nonumber \\
&+&c_{2c}(9) \frac{\partial C_0^{(045)}}{\partial \Lambda}
+c_{2c}(10) \frac{\partial C_0^{(014)}}{\partial \Lambda}
+c_{2c}(11) \frac{\partial C_0^{(015)}}{\partial \Lambda}
\nonumber \\
{\rm box}_{C}(2)&=& c_{3c}(1) C_0^{(014)} +c_{3c}(2) C_0^{(015)} 
+c_{3c}(3) \frac{\partial C_0^{(045)}}{\partial \Lambda}
+c_{3c}(4) \frac{1}{2}\frac{\partial C_0^{(014)}}{\partial \Lambda}
\nonumber \\ 
&+&c_{3c}(5) \frac{1}{2}\frac{\partial C_0^{(015)}}{\partial \Lambda}
+c_{3c}(6) \frac{\partial^2 C_0^{(014)}}{\partial \Lambda_1\partial\Lambda_2}\Bigl|_{\Lambda_1=\Lambda_2=\Lambda}
+c_{3c}(7)
\frac{\partial^2 C_0^{(015)}}{\partial \Lambda_1\partial\Lambda_2}\Bigl|_{\Lambda_1=\Lambda_2=\Lambda}\\
{\rm box}_{D}(0)&=& c_{1d}(1) D_0^{(2345)}+ 
c_{1d}(2) D_0^{(1245)}+ c_{1d}(3) D_0^{(0145)} \nonumber \\ \label{eq:dcoeffs}
{\rm box}_{D}(1)&=& c_{2d}(1) D_0^{(1245)}+ 
c_{2d}(2) D_0^{(0145)}+ c_{2d}(3) \frac{\partial D_0^{(1245)}}{\partial \Lambda} +c_{2d}(4) \frac{\partial D_0^{(0145)}}{\partial \Lambda}
\nonumber \\
{\rm box}_{D}(2)&=& c_{3d}(1) D_0^{(0145)}+ c_{3d}(2) \frac{1}{2}\frac{ \partial  D_0^{(0145)}}{\partial \Lambda}+
c_{3d}(3) \frac{\partial^2 D_0^{(0145)}}{\partial \Lambda_1\partial\Lambda_2}\Bigl|_{\Lambda_1=\Lambda_2=\Lambda} 
\end{eqnarray}
The superscripts $(jklm) \in \{0,1,2,3,4,5\}$ of the scalar functions denote the contributing propagator denominators of Eq.~\ref{eq:boxprop}.
The scalar functions can be found, e.~g., in Refs.~\cite{tHooft:1978jhc,Denner:2014gla}, and can be easily obtained with tools such as \textsc{COLLIER}~\cite{Denner:2014gla,Denner:2002ii,Denner:2005nn,Denner:2010tr}. Their derivatives are provided in Appendix~\ref{sec:BCDderiv}.
The $c_{abi}(j),c_{ci}(j),c_{di}(j)$ coefficients are provided in Appendix~\ref{sec:bcdcoeffs}.

\section{Derivatives of $B_0, C_0$ and $D_0$ functions}\label{sec:BCDderiv}

Here we provide analytic expressions for the single and double derivatives needed for Eqs.~\ref{eq:bcoeffs},\ref{eq:ccoeffs},\ref{eq:dcoeffs}. Since the results here are obtained for  derivatives with respect to $\Lambda^2$, they have to be multiplied with $2\Lambda$ for single derivatives and $4\Lambda^2$ for double derivatives when used in Eqs.~\ref{eq:bcoeffs},\ref{eq:ccoeffs},\ref{eq:dcoeffs}. 

\subsection{Derivatives of $B_0$ and $C_0$ functions needed for the elastic TPE contribution}

The derivatives of the scalar 2-point functions of Eq.~\ref{eq:bcoeffs} read as follows ($L=\Lambda^2$):
\begin{equation}
\frac{\partial B_0^{(04)}}{\partial \Lambda^2}=
\frac{1}{m^2} \left(-\ln\left(\frac{\sqrt{L}}{m}\right)+\frac{(L-2m^2)}{\sqrt{L}\sqrt{L-4m^2}} \ln\left(\frac{(\sqrt{\sqrt{L}+2m}+\sqrt{\sqrt{L}-2m})^2}{4m}\right)\right)
          \end{equation}
\begin{equation}
\frac{\partial B_0^{(05)}}{\partial \Lambda^2}=
\frac{1}{M^2}
\left(\frac{4 (L-2 M^2)}{\sqrt{4M^2-L}} \arctan\left(\frac{\sqrt{2M-\sqrt{L}}}{\sqrt{2M+\sqrt{L}}}\right)- \sqrt{L} \ln\left(\frac{L}{M^2}\right) \right)
\end{equation}
\begin{equation}
\frac{\partial B_0^{(12)}}{\partial \Lambda^2}=
\frac{1}{t} \ln\left(1- \frac{t}{L}\right)
\end{equation}
\begin{equation}
\frac{\partial B_0^{(01)}}{\partial \Lambda^2}=
-\frac{2}{t\beta_t}\ln\left(\frac{\beta_t-1}{\beta_t+1}\right)
\end{equation}
\begin{equation}
\frac{\partial^2 B_0^{(01)}}{\partial \Lambda_1^2 \partial \Lambda_2^2}\Big|_{\Lambda_1=\Lambda_2=\Lambda}=
\frac{(\beta_t^2-1)}{4 L\beta_t^3 t} \left(2\beta_t+(\beta_t^2+1) \ln\left(\frac{\beta_t-1}{\beta_t+1}\right)\right)
\end{equation}
with $\beta_t=\sqrt{1-4 L/t}$.
For the derivatives of the scalar 3-point functions of Eq.~\ref{eq:ccoeffs} we find:
\begin{equation}
\frac{\partial C_0^{(035)}}{\partial \Lambda^2}=
\frac{2}{L\beta_M (L-t)}
\ln\left(\frac{\sqrt{L} (1+\beta_M)}{2M} \right)
\end{equation}
\begin{eqnarray}
\frac{\partial C_0^{(045)}}{\partial \Lambda^2}&=&
\frac{1}{L (\Delta+s L)}
\left(\frac{\left(m^2-M^2+s\right) }{\beta_m}
    \ln\left(\frac{(\beta_m+1) \sqrt{L}}{2
    m}\right) \right.\nonumber \\    
    &+& \left. \frac{\left(-m^2+M^2+s\right) }{\beta_M}
    \ln\left(\frac{(\beta_M+1) \sqrt{L}}{2
    M}\right)
    +\sqrt{\Delta } 
    \ln\left(\frac{\sqrt{\Delta }+m^2+M^2-s}{2 m M}\right)\right)
\end{eqnarray}
\begin{equation}
\frac{\partial C_0^{(015)}}{\partial \Lambda^2}=
-\frac{2}{L
    t \left(L^2-4 L M^2+M^2 t\right)}
 \left(\frac{L^2}{\beta_t} \ln\left(1-\frac{(\beta_t+1) t}{2 L}\right)+\frac{2 M^2 t}{\beta_M}
    \ln\left(\frac{(\beta_M+1) \sqrt{L}}{2 M}\right)\right)
\end{equation}
\begin{eqnarray}
\frac{\partial^2 C_0^{(015)}}{\partial \Lambda_1^2 \partial \Lambda_2^2}\Big|_{\Lambda_1=\Lambda_2=\Lambda}&=&
-\frac{1}{\left(\beta_t^2 M^2 t+L^2\right)^2}
\left(\frac{(2 L-t) \left(\beta_t^2 M^2 t+L^2\right)}{\beta_t^2
    t^2} \right. \nonumber \\
    &+& \left. \frac{\left(4 L^4+M^2 \left(-16 L^3+12 L^2 t-6 L t^2+t^3\right)\right)}{\beta_t^3 t^3}
\ln\left(1-\frac{(\beta_t+1) t}{2 L}\right)
\right. \nonumber \\
&-& \left. \frac{2
    M^2 \left(L-2 M^2\right)}{\beta_M L} \ln\left(\frac{(\beta_M+1) \sqrt{L}}{2
    M}\right)\right)
\end{eqnarray}
where $\beta_x=\sqrt{1-4 x^2/L}, x=m,M$ and $\Delta=\lambda(s,m,M)$.
The derivatives of $C_0^{(014)}, C_0^{(034)}$ can be obtained from the derivatives of $C_0^{(015)}, C_0^{(035)}$, respectively, with the replacement $M \to m$.

\subsection{First derivatives of $D_0^{(1245)}$, $D_0^{(0145)}$ and second derivative of $D_0^{(0145)}$}

For convenience, we introduce $L=\Lambda^2$ and write the logarithms appearing in the derivatives of the scalar 4-point functions of Eq.~\ref{eq:dcoeffs} as follows ($x=m$ or $M$)
\[
\Lx = 
\ln\!\left(
\frac{L+\sqrt{\Delta_x}}{2x\sqrt{L}}
\right) \; , \;
\LD = 
\ln\!\left(
\frac{m^{2}+M^{2}-s + \sqrt{\Delta}}{2mM}
\right) \; , \;
\LT = 
\ln\!\left(
\frac{2L-t+\sqrt{\Delta_L}}{2L}
\right),
\]
where $\Delta = \lambda(s,m,M), \, \Delta_L=\lambda(t,\sqrt{L},\sqrt{L})$, and $\Delta_x=\lambda(L,x,x)$ are K\"all\'en  functions. We also define
\begin{equation}
A_x=L^2+(t-4L)x^2 \,.
\end{equation}
Using these definitions, the first derivative of $D_0^{(0145)}$
takes the following compact form:
\begin{equation}\label{eq:dd0}
\frac{\partial D_0^{(0145)}}{\partial L}=\frac{2}{(4 s L^2-(t-4L)\Delta)} [\mathcal{F}_{\Delta} L_\Delta+ \mathcal{F}_{t} L_t+\mathcal{F}(m,M) L_m+\mathcal{F}(M,m) L_M+\mathcal{F}_{D} D_0^{(0145)}  ]
\end{equation}
The $\mathcal{F}$ coefficients read as follows
\begin{equation}
\mathcal{F}_{\Delta}=
 -\frac{(\Delta+2 Ls)\sqrt{\Delta}}{(\Delta+Ls)L}
\end{equation}
\begin{equation}
\mathcal{F}(m,M)=
\frac{(s+m^2-M^2)(t \Delta m^2+(4 s m^2-
\Delta-2Ls)L^2)}{(\Delta+Ls) A_m \sqrt{\Delta_m}}
\end{equation}
\begin{equation}
\mathcal{F}_{D}=-(\Delta+2Ls) 
\end{equation}
\begin{equation}
\mathcal{F}_{t}=
-\frac{L(2L-t) \sqrt{\Delta_L}((4L-t)(\Delta+s(m^2+M^2-s))+2sL^2)}{t(4L-t)A_m A_M}
\end{equation}
The second derivative of $D_0^{(0145)}$ can be written as
\begin{eqnarray}\label{eq:ddd0}
 \frac{\partial^2 D_0^{(0145)}}{\partial L_1 \partial L_2}\Big|_{L_1=L_2=L}&=&\mathcal{N}_{\Delta} L_\Delta+ \mathcal{N}_{t} L_t+\mathcal{N}(m,M) L_m+\mathcal{N}(M,m) L_M \nonumber \\
 &+&\mathcal{N}_{D} D_0^{(0145)} +\mathcal{N}_R 
\end{eqnarray}
where the $\mathcal{N}$ coefficients are defined as follows
\begin{eqnarray}
\mathcal{N}_{R}&=&\frac{(t-2L)^2 (2 L^2 s+(4L-t) (\Delta+s(m^2+M^2-s)))}{t(4L-t) (4 s L^2-(t-4L)\Delta) A_m A_M} \nonumber \\
\mathcal{N}_{\Delta}&=&\frac{2 \sqrt{\Delta}((2 L + t) \Delta^2 + 2 L s (L + 3 t) \Delta + 6 L^2 s^2 t)}{Lt (Ls+\Delta)(4 L^2 s+(4L-t) \Delta)^2}
\nonumber \\
\mathcal{N}_D&=& \frac{2 (2 (t - L) \Delta^2 + s (t^2 + 2 L (t - L)) 
\Delta + 2 L^2 s^2 t)}{t(4 L^2 s+(4L-t) \Delta)^2}\nonumber \\
\mathcal{N}(m,M)&=& \frac{2 (m^2-M^2+s)}{\sqrt{\Delta_m}  t  (\Delta + s L) (\Delta_m + t m^2)^2  ((4  L - t)  \Delta + 4  s  L^2)^2} \sum_{i=0}^6 L^i\eta_i
\nonumber \\
\mathcal{N}_t&=&\frac{\sqrt{t(t-4L)}}{4 s^2 t^2 (t-4L)^2 (4 s L^2-(t-4L)\Delta)^2 A_m^2 A_M^2} \times \nonumber \\
&&\sum_{i=0}^3\sum_{j=0}^3 \Delta^i (\Delta+s (m^2+M^2-s))^j A_{ij}
\end{eqnarray}
with the $\eta_i$ coefficients
\begin{eqnarray}
\eta_0&=&-3 \Delta^2 m^4 t^3 \nonumber \\
\eta_1&=&\Delta m^2 t^2 \left(16 \Delta m^2+t \left(m^4-2 m^2 \left(M^2+3 s\right)+\left(M^2-s\right)^2\right)\right) \nonumber \\
\eta_2&=&\Delta m^2 t \left(-24 m^6+6 m^4 \left(8 \left(M^2+s\right)-t\right)+4 m^2 \left(t \left(3 M^2+7 s\right)-6 \left(M^2-s\right)^2\right) \right. \nonumber \\
&+& \left. t \left(s t-6 \left(M^2-s\right)^2\right)\right) \nonumber \\
\eta_3&=&2 m^2 \left(16 \Delta^2 m^2+\Delta t \left(3 m^4-2 m^2 \left(3 M^2+s\right)+3 \left(M^2-s\right)^2\right) \right. \nonumber \\
&+& \left. s t^2 \left(-3 m^4+2 m^2 \left(3 M^2+s\right)-3 \left(M^2-s\right)^2\right)\right) \nonumber \\
\eta_4&=&t \left(m^8-2 m^6 \left(2 M^2+11 s\right)+m^4 \left(6 M^4+40 M^2 s+90 s^2\right) \right. \nonumber \\
&+& \left. 2 m^2 \left(s^2 t-\left(M^2-s\right)^2 \left(2 M^2+11 s\right)\right)+\left(M^2-s\right)^4\right) \nonumber \\
&-&16 \Delta m^2 \left(m^4-2 m^2 \left(M^2+2 s\right)+\left(M^2-s\right)^2\right) \nonumber \\
\eta_5&=&2 \left(m^8-4 m^6 \left(M^2+3 s\right)+m^4 \left(6 M^4+20 M^2 s+s (22 s+3 t)\right)\right. \nonumber \\
&+& \left. 2 m^2 \left(-s t \left(3 M^2+11 s\right)-2 \left(M^2+3 s\right) \left(M^2-s\right)^2\right)+3 s t \left(M^2-s\right)^2+\left(M^2-s\right)^4\right) \nonumber \\
\eta_6&=&2 s \left(m^4-2 m^2 \left(M^2+s\right)+\left(M^2-s\right)^2+3 s t\right)
\end{eqnarray}
and the non-zero $A_{ij}$ coefficients 
\begin{eqnarray}
A_{00}&=& 16 L^6 s^4 \left(16 L^5-64 L^4 s+8 L^4 t+16 L^3 s t-16 L^3 t^2 \right. \nonumber \\
&+& \left. 32 L^2 s t^2+3 L^2 t^3-12 L s t^3+s t^4\right) \nonumber \\
A_{01}&=&-8    L^4 s^3 (t-4 L) \left(-16 L^5-64 L^4 s+16 L^4 t-16 L^3 s t-8 L^3 t^2+72 L^2 s t^2 \right.\nonumber \\
&+&\left. 3 L^2 t^3-28 L s t^3+3 s t^4\right) \nonumber \\
A_{02}&=&-4s^2 L^4 (t-4 L)^2 \left(16 L^3+24 L^2 t-32 L t^2+7 t^3\right)\nonumber \\
A_{03}&=&-2s L^2 (t-4 L)^3 \left(16 L^3-8 L t^2+t^3\right)\nonumber \\
A_{10}&=&4  L^4 s^3 (t-4 L) \left(-80 L^5+128 L^4 s+4 L^4 t-112 L^3 s t+36 L^3 t^2 \right.\nonumber \\
&+& \left. 36 L^2 s t^2-4 L^2 t^3-16 L s t^3+3 s t^4\right) \nonumber \\
A_{11}&=&2s^2 L^2 (t-4 L)^2 \left(-16 L^5-128 L^4 s+92 L^4 t+48 L^3 s t \right. \nonumber \\
&-&\left. 84 L^3 t^2+44 L^2 s t^2+22 L^2 t^3-16 L s t^3+s t^4\right) \nonumber \\
A_{12}&=&4 s L^2 (t-4 L)^3 \left(20 L^3+3 L^2 t-13 L t^2+2 t^3\right) \nonumber \\
A_{13}&=&A_{31}  \nonumber \\
A_{20}&=&-4 s^2 L^2 (t-4 L)^2 \left(-32 L^5+16 L^4 s+20 L^4 t-24 L^3 s t-4 L^3 t^2 \right. \nonumber \\
&+&\left. 17 L^2 s t^2+3 L^2 t^3-7 L s t^3+s t^4\right) \nonumber \\
A_{21}&=& s(t-4 L)^3 \left(-32 L^5+32 L^4 s-32 L^4 t-32 L^3 s t+48 L^3 t^2 \right. \nonumber \\
&+&\left. 14 L^2 s t^2-10 L^2 t^3-6 L s t^3+s t^4\right) \nonumber \\
A_{22}&=&-2A_{31}  \nonumber \\
A_{30}&=&4 s L^2 (t-4 L)^3 \left(-4 L^3+5 L^2 t-3 L t^2+t^3\right) \nonumber \\
A_{31}&=&(4 L-t)^4 \left(8 L^3-6 L^2 t+2 L t^2-t^3\right)
\end{eqnarray}

The derivative of $D_0^{(1245)}$ can be obtained from taking the derivative with respect to $L$ of the result for $D_0^{(1245)}$ provided in Ref.~\cite{Beenakker:1988jr}. We provide the result here as well for completeness:
\begin{eqnarray}
\frac{\partial D^{(1245)}_0 }{\partial L}&=&\frac{D_0^{(1245)}}{t-\Lambda^2} +\frac{x_s}{ M m (1-x_s^2)(t-\Lambda^2)}
\left[
\frac{(\Lambda^2+t)\ln(x_s)}{\Lambda^2(\Lambda^2-t)}
+\frac{2}{x_M}d_M\ln x_M
+\frac{2}{x_m}d_m\ln x_m \right. \nonumber \\
&-& \left. \sum_{k=\pm 1}\sum_{r=\pm 1} 
\frac{
x_s\bigl(\ln(x_s)+k\ln x_M+r\ln x_m\bigr) \, x_M^{-1+k} \, x_m^{-1+r}
\bigl(k\,x_m\,d_M+r\,x_M\,d_m\bigr)}{-1+x_s x_M^{k} x_m^{r}}
\right] \nonumber \\
\end{eqnarray}
with $x_s=-K(s,m,M), x_m=-K(m^2,m,\Lambda), x_M=-K(M^2,M,\Lambda)$. The $K$ function introduced in~\cite{Beenakker:1988jr} reads
\begin{equation}\label{eq:kfunction}
K(z,m,m')=\frac{1-\sqrt{1-\frac{4mm'}{z+i\varepsilon-(m-m')^2}}}{1+\sqrt{1-\frac{4mm'}{z+i\varepsilon-(m-m')^2}}} \; \mbox{for} \; z\neq (m-m')^2
\end{equation}
and
\[d_m =
\frac{m}
{(2m-\Lambda)\Lambda\left(2m+\Lambda
+\Lambda\sqrt{1+\dfrac{4m}{-2m+\Lambda}+\mathrm{i}\epsilon}\right)}.
\]

\section{The coefficients $c_{iab}, c_{ic}$, and $c_{id}$ ($i=1,2,3)$ }\label{sec:bcdcoeffs}

The coefficients $c_{iab}(j)$ ($i=1,2,3, j=1-10$) of the scalar 2-point functions and their derivatives of Eq.~\ref{eq:bcoeffs} are: 

\noindent\parbox{\mymathboxwidth}{%
  \raggedright
  \relpenalty=100
  \binoppenalty=100
 \small
  $
2 M^2 \Biggl(t-4 m^2\Biggr)^2 \Biggl(t-4 M^2\Biggr)^2\,\mathbf{c_{1ab}(1)} = -\Biggl(\kappa^3 t \biggl(4 m^2-t\biggr) \biggl(8 m^6 \Bigl(16 M^2-t\Bigr)-2 m^4 \Bigl(-8 t \bigl(7 M^2+s\bigr)+128 M^2 \bigl(M^2+s\bigr)+15 t^2\Bigr)+4 m^2 \Bigl(2 M^2-2 s-t\Bigr) \Bigl(16 M^4-M^2 (16 s+9 t)+t (s-t)\Bigr)+t \Bigl(-160 M^6+2 M^4 (96 s+73 t)-4 M^2 (s+2 t) (8 s+5 t)+t \bigl(2 s^2+6 s t+3 t^2\bigr)\Bigr)\biggr)\Biggr)+\kappa^2 t \Biggl(2 t^4 \biggl(2 m^2-20 M^2-9 s\biggr)+2 t^3 \biggl(-91 m^4+m^2 \Bigl(94 M^2+70 s\Bigr)+45 M^4+38 M^2 s-11 s^2\biggr)+256 m^4 M^2 \biggl(m^4+m^2 \Bigl(6 M^2-2 s\Bigr)+5 M^4+2 M^2 s+s^2\biggr)+16 t^2 \biggl(29 m^6+m^4 \Bigl(33 M^2-28 s\Bigr)+m^2 \Bigl(-45 M^4-16 M^2 s+5 s^2\Bigr)+7 M^6-10 M^4 s+7 M^2 s^2\biggr)-32 t \biggl(5 m^8+2 m^6 \Bigl(29 M^2-5 s\Bigr)+m^4 \Bigl(3 M^4-34 M^2 s+5 s^2\Bigr)+4 m^2 M^2 s \Bigl(5 M^2+s\Bigr)+6 M^4 \Bigl(M^2-s\Bigr)^2\biggr)+3 t^5\Biggr)-32 \kappa M^2 \Biggl(t-4 m^2\Biggr)^2 \Biggl(t-4 M^2\Biggr) \Biggl(-2 t \biggl(2 m^2+M^2-2 s\biggr)+2 \biggl(m^2+M^2-s\biggr)^2+t^2\Biggr)+32 M^2 \Biggl(t-4 m^2\Biggr) \Biggl(4 M^2-t\Biggr) \Biggl(t^2 \biggl(s-3 \Bigl(m^2+M^2\Bigr)\biggr)+16 m^2 M^2 \biggl(m^2+M^2-s\biggr)+t^3\Biggr)$
}

\noindent\parbox{\mymathboxwidth}{%
  \raggedright
  \relpenalty=100
  \binoppenalty=100
 \small
  $
M^2 \Biggl(t-4 m^2\Biggr)^2 \Biggl(t-4 M^2\Biggr)^2\,\mathbf{c_{1ab}(2)} = -\Biggl(\biggl(4 m^2-t\biggr) t \biggl(8 \Bigl(t-16 M^2\Bigr) m^6+2 \Bigl(128 \bigl(M^2+s\bigr) M^2+15 t^2-8 \bigl(7 M^2+s\bigr) t\Bigr) m^4-4 \Bigl(2 M^2-2 s-t\Bigr) \Bigl(16 M^4-(16 s+9 t) M^2+(s-t) t\Bigr) m^2+t \Bigl(160 M^6-2 (96 s+73 t) M^4+4 (s+2 t) (8 s+5 t) M^2-t \bigl(2 s^2+6 t s+3 t^2\bigr)\Bigr)+L \Bigl(24 m^6+2 \bigl(8 M^2-24 s-11 t\bigr) m^4+4 \bigl(54 M^4-(12 s+29 t) M^2+6 s^2+5 t^2+9 s t\bigr) m^2+64 M^6+4 M^2 t (11 s+10 t)-2 M^4 (32 s+51 t)-t \bigl(6 s^2+10 t s+5 t^2\bigr)\Bigr)\biggr) \kappa^3\Biggr)-\Biggl(\biggl(5 t^5-2 \Bigl(14 m^2+20 M^2-5 s\Bigr) t^4+2 \Bigl(19 m^4+74 M^2 m^2+51 M^4+3 s^2-6 \bigl(m^2+5 M^2\bigr) s\Bigr) t^3+16 \Bigl(m^6+M^2 m^4+\bigl(-5 M^4-16 s M^2+s^2\bigr) m^2-M^2 \bigl(5 M^4-10 s M^2+s^2\bigr)\Bigr) t^2+32 \Bigl(m^8-2 \bigl(11 M^2+s\bigr) m^6+\bigl(-13 M^4+30 s M^2+s^2\bigr) m^4-4 \bigl(6 M^6-7 s M^4+3 s^2 M^2\bigr) m^2+2 M^4 \bigl(M^2-s\bigr)^2\Bigr) t+256 m^2 M^2 \Bigl(m^2+2 M^2\Bigr) \Bigl(m^2+M^2-s\Bigr)^2\biggr)L+t \biggl(3 t^5+2 \Bigl(2 m^2-20 M^2-9 s\Bigr) t^4+2 \Bigl(-91 m^4+\bigl(94 M^2+70 s\bigr) m^2+45 M^4-11 s^2+38 M^2 s\Bigr) t^3+16 \Bigl(29 m^6+\bigl(33 M^2-28 s\bigr) m^4+\bigl(-45 M^4-16 s M^2+5 s^2\bigr) m^2+7 M^6+7 M^2 s^2-10 M^4 s\Bigr) t^2-32 \Bigl(5 m^8+2 \bigl(29 M^2-5 s\bigr) m^6+\bigl(3 M^4-34 s M^2+5 s^2\bigr) m^4+4 M^2 s \bigl(5 M^2+s\bigr) m^2+6 M^4 \bigl(M^2-s\bigr)^2\Bigr) t+256 m^4 M^2 \Bigl(m^4+\bigl(6 M^2-2 s\bigr) m^2+5 M^4+s^2+2 M^2 s\Bigr)\biggr)\Biggr) \kappa^2+32 M^2 \Biggl(t-4 m^2\Biggr)^2 \Biggl(t-4 M^2\Biggr) \Biggl(2 \biggl(m^2+M^2-s\biggr)^2+t^2-2 \biggl(2 m^2+M^2-2 s\biggr) t\Biggr) \kappa-32 M^2 \Biggl(4 m^2-t\Biggr) \Biggl(t-4 M^2\Biggr) \Biggl(t^3+\biggl(s-3 \Bigl(m^2+M^2\Bigr)\biggr) t^2+16 m^2 M^2 \biggl(m^2+M^2-s\biggr)\Biggr)$
}

\noindent\parbox{\mymathboxwidth}{%
  \raggedright
  \relpenalty=100
  \binoppenalty=100
 \small
  $
2 M^2 \Biggl(t-4 m^2\Biggr)^2 \Biggl(t-4 M^2\Biggr)^2\,\mathbf{c_{1ab}(3)} = \Biggl(4 m^2-t\Biggr) t \Biggl(8 \biggl(t-16 M^2\biggr) m^6+2 \biggl(128 \Bigl(M^2+s\Bigr) M^2+15 t^2-8 \Bigl(7 M^2+s\Bigr) t\biggr) m^4-4 \biggl(2 M^2-2 s-t\biggr) \biggl(16 M^4-(16 s+9 t) M^2+(s-t) t\biggr) m^2+t \biggl(160 M^6-2 (96 s+73 t) M^4+4 (s+2 t) (8 s+5 t) M^2-t \Bigl(2 s^2+6 t s+3 t^2\Bigr)\biggr)+2L \biggl(24 m^6+2 \Bigl(8 M^2-24 s-11 t\Bigr) m^4+4 \Bigl(54 M^4-(12 s+29 t) M^2+6 s^2+5 t^2+9 s t\Bigr) m^2+64 M^6+4 M^2 t (11 s+10 t)-2 M^4 (32 s+51 t)-t \Bigl(6 s^2+10 t s+5 t^2\Bigr)\biggr)\Biggr) \kappa^3+\Biggl(2 \biggl(5 t^5-2 \Bigl(14 m^2+20 M^2-5 s\Bigr) t^4+2 \Bigl(19 m^4+74 M^2 m^2+51 M^4+3 s^2-6 \bigl(m^2+5 M^2\bigr) s\Bigr) t^3+16 \Bigl(m^6+M^2 m^4+\bigl(-5 M^4-16 s M^2+s^2\bigr) m^2-M^2 \bigl(5 M^4-10 s M^2+s^2\bigr)\Bigr) t^2+32 \Bigl(m^8-2 \bigl(11 M^2+s\bigr) m^6+\bigl(-13 M^4+30 s M^2+s^2\bigr) m^4-4 \bigl(6 M^6-7 s M^4+3 s^2 M^2\bigr) m^2+2 M^4 \bigl(M^2-s\bigr)^2\Bigr) t+256 m^2 M^2 \Bigl(m^2+2 M^2\Bigr) \Bigl(m^2+M^2-s\Bigr)^2\biggr)L+t \biggl(3 t^5+2 \Bigl(2 m^2-20 M^2-9 s\Bigr) t^4+2 \Bigl(-91 m^4+\bigl(94 M^2+70 s\bigr) m^2+45 M^4-11 s^2+38 M^2 s\Bigr) t^3+16 \Bigl(29 m^6+\bigl(33 M^2-28 s\bigr) m^4+\bigl(-45 M^4-16 s M^2+5 s^2\bigr) m^2+7 M^6+7 M^2 s^2-10 M^4 s\Bigr) t^2-32 \Bigl(5 m^8+2 \bigl(29 M^2-5 s\bigr) m^6+\bigl(3 M^4-34 s M^2+5 s^2\bigr) m^4+4 M^2 s \bigl(5 M^2+s\bigr) m^2+6 M^4 \bigl(M^2-s\bigr)^2\Bigr) t+256 m^4 M^2 \Bigl(m^4+\bigl(6 M^2-2 s\bigr) m^2+5 M^4+s^2+2 M^2 s\Bigr)\biggr)\Biggr) \kappa^2-32 M^2 \Biggl(t-4 m^2\Biggr)^2 \Biggl(t-4 M^2\Biggr) \Biggl(2 \biggl(m^2+M^2-s\biggr)^2+t^2-2 \biggl(2 m^2+M^2-2 s\biggr) t\Biggr) \kappa+32 M^2 \Biggl(4 m^2-t\Biggr) \Biggl(t-4 M^2\Biggr) \Biggl(t^3+\biggl(s-3 \Bigl(m^2+M^2\Bigr)\biggr) t^2+16 m^2 M^2 \biggl(m^2+M^2-s\biggr)\Biggr)$
}

\noindent\parbox{\mymathboxwidth}{%
  \raggedright
  \relpenalty=100
  \binoppenalty=100
 \small
  $
M^2 \Biggl(t-4 m^2\Biggr)^2\,\mathbf{c_{1ab}(4)} = 4 \kappa^2L m^2 \Biggl(2 (4 \kappa+7) m^4+2 m^2 \biggl((4 \kappa+6) M^2-2 (2 \kappa+5) s-(3 \kappa+5) t\biggr)-2 (\kappa+3) M^2 t+2 (\kappa+4) s t+(\kappa+2) t^2+6 \biggl(M^2-s\biggr)^2\Biggr)$
}

\noindent\parbox{\mymathboxwidth}{%
  \raggedright
  \relpenalty=100
  \binoppenalty=100
 \small
  $
M^2\Biggl(t-4 M^2\Biggr)^2\,\mathbf{c_{1ab}(6)} = 2 \kappa^2L \Biggl(m^4 \biggl(2 (\kappa+3) M^2+\kappa t\biggr)-m^2 \biggl(4 (\kappa-1) M^4+2 M^2 (2 (\kappa+3) s-(\kappa-2) t)+\kappa t (2 s+t)\biggr)+2 (\kappa+3) \biggl(M^3-M s\biggr)^2+t^2 \biggl((\kappa+1) M^2+\kappa s\biggr)+t \biggl(-\Bigl((3 \kappa+4) M^4\Bigr)+\kappa s^2+6 M^2 s\biggr)\Biggr)$
}

\noindent\parbox{\mymathboxwidth}{%
  \raggedright
  \relpenalty=100
  \binoppenalty=100
 \small
  $
M^2 \Biggl(t-4 m^2\Biggr)^2 \Biggl(t-4 M^2\Biggr)^2\,\mathbf{c_{2ab}(1)} = \kappa^2L \Biggl(32 m^8 \biggl(3 \kappa t+8 M^2+t\biggr)+16 m^6 \biggl(4 M^2 ((\kappa-11) t-8 s)+t (-4 (3 \kappa+1) s-7 \kappa t+t)+64 M^4\biggr)+2 m^4 \biggl(8 t^2 \Bigl((1-30 \kappa) M^2+12 \kappa s\Bigr)+16 t \Bigl((27 \kappa-13) M^4-6 (\kappa-5) M^2 s+(3 \kappa+1) s^2\Bigr)+(51 \kappa+19) t^3+128 M^2 \Bigl(5 M^4-6 M^2 s+s^2\Bigr)\biggr)+4 m^2 \biggl(t^3 \Bigl((69 \kappa+37) M^2-(19 \kappa+3) s\Bigr)-4 t^2 \Bigl((39 \kappa+5) M^4+2 (8-7 \kappa) M^2 s+(3 \kappa-1) s^2\Bigr)+32 M^2 t \Bigl(2 (\kappa-3) M^4+(7-2 \kappa) M^2 s-3 s^2\Bigr)-(10 \kappa+7) t^4+128 M^4 \Bigl(M^2-s\Bigr)^2\biggr)+t \biggl(-16 M^6 (4 \kappa t+8 s+5 t)+2 M^4 \Bigl(16 (2 \kappa+5) s t+51 (\kappa+1) t^2+32 s^2\Bigr)-4 M^2 t \Bigl((11 \kappa+15) s t+10 (\kappa+1) t^2+4 s^2\Bigr)+(\kappa+1) t^2 \Bigl(6 s^2+10 s t+5 t^2\Bigr)+64 M^8\biggr)\Biggr)$
}

The remaining $c_{iab}$ coefficient can be obtained as follows:
\begin{align*}
c_{1ab}(4) &= -\,c_{1ab}(5) = -\,c_{2ab}(3) = c_{2ab}(4) = \frac{2}{\sqrt{L}}\,c_{2ab}(7) = -\frac{2}{\sqrt{L}}\,c_{3ab}(1) \\
c_{1ab}(6) &= -\,c_{1ab}(7) = -\,c_{2ab}(5) = c_{2ab}(6) = \frac{2}{\sqrt{L}}\,c_{2ab}(8) = -\frac{2}{\sqrt{L}}\,c_{3ab}(2) \\
c_{2ab}(1) &= -\,c_{2ab}(2) = \frac{2}{\sqrt{L}}\,c_{3ab}(3) \\
c_{1ab}(2) &= -\frac{2}{\sqrt{L}}\,c_{2ab}(9) \\
c_{1ab}(3) &= -\frac{2}{\sqrt{L}}\,c_{2ab}(10) = \frac{4}{L}\,c_{3ab}(4) 
\end{align*}

The coefficients $c_{ic}(j)$ ($i=1,2,3, j=1-11$) of the scalar 3-point functions and their derivatives of Eq.~\ref{eq:ccoeffs} are: 

\noindent\parbox{\mymathboxwidth}{%
  \raggedright
  \relpenalty=100
  \binoppenalty=100
 \small
  $
M^2\,\mathbf{c_{1c}(1)} = 4L \Biggl(\kappa^3 \biggl(m^4-2 m^2 \Bigl(M^2+s\Bigr)+\Bigl(M^2-s\Bigr)^2\biggr)+\kappa^2 \biggl(2 m^4-m^2 \Bigl(M^2+4 s\Bigr)+5 M^4-7 M^2 s+2 s^2\biggr)+4 \kappa M^2 \biggl(m^2+2 M^2-2 s\biggr)+4 M^2 \biggl(m^2+M^2-s\biggr)\Biggr)$
}

\noindent\parbox{\mymathboxwidth}{%
  \raggedright
  \relpenalty=100
  \binoppenalty=100
 \small
  $
M^2 \Biggl(t-4 m^2\Biggr)^2\,\mathbf{c_{1c}(2)} = -2 \Biggl(16 \kappa^2 m^8 t+2 m^6 \biggl(\kappa^2 t (-8 \kappa t+16 s-15 t)+16 M^2 ((\kappa (3 \kappa+4)+2) t-8 s)+128 M^4\biggr)+4 m^4 \biggl(-4 M^4 \Bigl(3 \kappa^2 t+32 s+4 t\Bigr)+M^2 \Bigl(24 \bigl(\kappa^2+2\bigr) s t+(\kappa (5 \kappa+16)+4) t^2+64 s^2\Bigr)+\kappa^2 t \Bigl((3 \kappa+4) t^2-12 s^2-11 s t\Bigr)+64 M^6\biggr)+2 m^2 t \biggl(-t^2 \Bigl((\kappa (\kappa (2 \kappa+17)+28)+12) M^2-2 \kappa^2 (\kappa+5) s\Bigr)+t \Bigl(13 \kappa^2 \bigl(M^2-s\bigr)^2-32 M^2 s\Bigr)-\kappa^2 (\kappa+1) t^3-64 \Bigl(M^3-M s\Bigr)^2\biggr)+t^2 \biggl((\kappa+2) t^2 \Bigl((\kappa+1) (\kappa+2) M^2-\kappa^2 s\Bigr)-2 t \Bigl(\kappa^2 \bigl(M^2-s\bigr)^2-4 M^2 s\Bigr)+16 \Bigl(M^3-M s\Bigr)^2\biggr)\Biggr)$
}

\noindent\parbox{\mymathboxwidth}{%
  \raggedright
  \relpenalty=100
  \binoppenalty=100
 \small
  $
M^2\Biggl(t-4 M^2\Biggr)^2\,\mathbf{c_{1c}(3)} = m^4 \Biggl(\kappa^2 (\kappa+4) t^3+32 (\kappa (3 \kappa+2)+8) M^4 t-2 (\kappa (\kappa (5 \kappa+23)+8)+16) M^2 t^2-512 M^6\Biggr)-m^2 \Biggl(4 M^4 t \biggl(\kappa \Bigl(15 \kappa^2+\kappa-56\Bigr) t+16 (\kappa (3 \kappa+2)+8) s\biggr)+\kappa^2 t^3 (2 (\kappa+4) s+(3 \kappa+4) t)-128 M^6 (-\kappa (2 \kappa+5) t+8 s+t)+2 M^2 t^2 (\kappa (8-3 \kappa (5 \kappa+6)) t-2 (\kappa (\kappa (5 \kappa+23)+8)+16) s)+512 M^8\Biggr)+\kappa^2 s t^3 ((\kappa+4) s+(3 \kappa+4) t)+32 M^8 (16 s-(3 \kappa (\kappa+2)+4) t)-2 M^6 \Biggl(-64 (\kappa-3) s t+(\kappa (\kappa (45 \kappa+127)+104)+16) t^2+256 s^2\Biggr)+M^4 t \Biggl(32 (\kappa (3 \kappa+2)+8) s^2+4 (\kappa (\kappa (17 \kappa+35)+8)+32) s t+(\kappa (\kappa (49 \kappa+160)+160)+48) t^2\Biggr)-M^2 t^2 \Biggl(2 (\kappa (\kappa (5 \kappa+23)+8)+16) s^2+2 (\kappa (\kappa (16 \kappa+27)+8)+8) s t+(\kappa+1) (\kappa (7 \kappa+16)+8) t^2\Biggr)$
}

\noindent\parbox{\mymathboxwidth}{%
  \raggedright
  \relpenalty=100
  \binoppenalty=100
 \small
  $
\frac{M^2}{2} \Biggl(t-4 m^2\Biggr)^2\,\mathbf{c_{1c}(4)} = 2 \Biggl(\kappa^3 \biggl(L-t\biggr) \biggl(t-4 m^2\biggr) \biggl(L m^2 \Bigl(t-2 \bigl(m^2+M^2-s\bigr)\Bigr)+t \Bigl(t \bigl(2 m^2-M^2+s\bigr)-4 m^4\Bigr)\biggr)+\kappa^2 \biggl(2L^2 m^2 \Bigl(7 m^4+m^2 \bigl(6 M^2-5 (2 s+t)\bigr)+3 M^4-3 M^2 (2 s+t)+(s+t) (3 s+t)\Bigr)+L \Bigl(16 m^6 t-2 t^2 \bigl(3 m^4+m^2 \left(4 s-6 M^2\right)+\left(M^2-s\right)^2\bigr)-16 m^4 \bigl(m^2+M^2-s\bigr)^2-M^2 t^3\Bigr)+t \Bigl(16 m^8+m^6 \bigl(96 M^2+32 s-30 t\bigr)+4 m^4 \bigl(t \left(5 M^2-11 s\right)-12 \left(M^2-s\right)^2+4 t^2\bigr)-2 m^2 t \bigl(t \left(17 M^2-10 s\right)-13 \left(M^2-s\right)^2+t^2\bigr)+t^2 \bigl(-2 M^4+M^2 (4 s+5 t)-2 s (s+t)\bigr)\Bigr)\biggr)-4 \kappa M^2 t \biggl(t-4 m^2\biggr)^2 \biggl(L-2 \Bigl(m^2+t\Bigr)\biggr)+4 M^2 \biggl(4 m^2-t\biggr) \biggl(-4 m^2 \Bigl(L-4 M^2+4 s\Bigr) \Bigl(m^2+M^2-s\Bigr)+t^2 \Bigl(L+2 m^2-2 s\Bigr)-2 t \Bigl(L m^2-2 m^4-4 m^2 s+2 \bigl(M^2-s\bigr)^2\Bigr)-t^3\biggr)\Biggr)$
}

\noindent\parbox{\mymathboxwidth}{%
  \raggedright
  \relpenalty=100
  \binoppenalty=100
 \small
  $
\frac{M^2}{2}\Biggl(t-4 M^2\Biggr)^2\,\mathbf{c_{1c}(5)} = \kappa^3 \Biggl(L-t\Biggr) \Biggl(L \biggl(t^2 \Bigl(-m^2+M^2+s\Bigr)+2 M^2 \Bigl(m^4-2 m^2 \bigl(M^2+s\bigr)+\bigl(M^2-s\bigr)^2\Bigr)+t \Bigl(m^4+2 m^2 \bigl(M^2-s\bigr)-3 M^4+s^2\Bigr)\biggr)+t \biggl(2 M^2 \Bigl(2 s \bigl(5 m^2+17 M^2\bigr)-5 \bigl(m^2+3 M^2\bigr)^2-5 s^2\Bigr)+t^2 \Bigl(-3 m^2-7 M^2+3 s\Bigr)+t \Bigl(m^4+m^2 \bigl(30 M^2-2 s\bigr)+49 M^4-32 M^2 s+s^2\Bigr)\biggr)\Biggr)+\kappa^2 \Biggl(L^2 M^2 \biggl(6 m^4+4 m^2 \Bigl(M^2-3 s-t\Bigr)+6 \Bigl(M^2-s\Bigr)^2-4 M^2 t+6 s t+t^2\biggr)+4L \biggl(-8 M^4 \Bigl(m^2+M^2-s\Bigr) \Bigl(m^2+3 M^2-s\Bigr)-t^2 \Bigl(m^4-2 m^2 \bigl(2 M^2+s\bigr)-41 M^4+5 M^2 s+s^2\Bigr)+M^2 t \Bigl(3 m^4-2 m^2 \bigl(M^2+3 s\bigr)-65 M^4+6 M^2 s+3 s^2\Bigr)-6 M^2 t^3\biggr)+t \biggl(m^4 \Bigl(-96 M^4+46 M^2 t-4 t^2\Bigr)+4 m^2 \Bigl(64 M^6+M^4 (48 s+t)-M^2 t (23 s+9 t)+t^2 (2 s+t)\Bigr)+96 M^8+254 M^6 t-4 M^4 \Bigl(24 s^2+35 s t+40 t^2\Bigr)+M^2 t \Bigl(46 s^2+54 s t+23 t^2\Bigr)-4 s t^2 (s+t)\biggr)\Biggr)+8 \kappa M^2 \Biggl(4 M^2-t\Biggr) \Biggl(L \biggl(-2 t \Bigl(m^2+4 M^2-s\Bigr)+2 \Bigl(m^2-3 M^2-s\Bigr) \Bigl(m^2+M^2-s\Bigr)+3 t^2\biggr)+t \biggl(-2 t \Bigl(m^2-4 M^2+s\Bigr)-2 \Bigl(m^4-2 m^2 \bigl(5 M^2+s\bigr)-3 M^4+2 M^2 s+s^2\Bigr)-3 t^2\biggr)\Biggr)-8 M^2 \Biggl(4 M^2-t\Biggr) \Biggl(4 M^2 \biggl(L-4 m^2+4 s\biggr) \biggl(m^2+M^2-s\biggr)+2 t \biggl(L M^2+2 m^4-4 s \Bigl(m^2+M^2\Bigr)-2 M^4+2 s^2\biggr)-t^2 \biggl(L+2 M^2-2 s\biggr)+t^3\Biggr)$
}

\noindent\parbox{\mymathboxwidth}{%
  \raggedright
  \relpenalty=100
  \binoppenalty=100
 \small
  $
M^2 \Biggl(t-4 m^2\Biggr)^2\,\mathbf{c_{1c}(7)} = -2 \Biggl(\kappa^3 t \biggl(4 m^2-t\biggr) \biggl(2L^2 \Bigl(3 m^2+M^2-s-t\Bigr)+2L \Bigl(2 m^2-t\Bigr) \Bigl(m^2-M^2+s\Bigr)+t \Bigl(t \bigl(2 m^2-M^2+s\bigr)-4 m^4\Bigr)\biggr)+\kappa^2 \biggl(16L m^2 \Bigl(L-2 m^2\Bigr) \Bigl(m^2+M^2-s\Bigr)^2+2 t^3 \Bigl(L^2-L M^2+8 m^4+m^2 \bigl(10 s-17 M^2\bigr)-\bigl(M^2-s\bigr)^2\Bigr)-2 t^2 \Bigl(L^2 \bigl(5 m^2+M^2-2 s\bigr)+2L \bigl(3 m^4+m^2 \left(4 s-6 M^2\right)+\left(M^2-s\right)^2\bigr)+15 m^6+m^4 \bigl(22 s-10 M^2\bigr)-13 m^2 \bigl(M^2-s\bigr)^2\Bigr)+2 t \Bigl(L^2 \bigl(5 m^4+2 m^2 \left(s-3 M^2\right)+\left(M^2-s\right)^2\bigr)+16L m^6+8 m^4 \bigl(m^4+2 m^2 \left(3 M^2+s\right)-3 \left(M^2-s\right)^2\bigr)\Bigr)+t^4 \Bigl(-2 m^2+5 M^2-2 s\Bigr)\biggr)-8 \kappa M^2 t \biggl(t-4 m^2\biggr)^2 \biggl(L-m^2-t\biggr)+4 M^2 \biggl(4 m^2-t\biggr) \biggl(-8 m^2 \Bigl(L-2 M^2+2 s\Bigr) \Bigl(m^2+M^2-s\Bigr)+2 t^2 \Bigl(L+m^2-s\Bigr)-4 t \Bigl(L m^2-m^4-2 m^2 s+\bigl(M^2-s\bigr)^2\Bigr)-t^3\biggr)\Biggr)$
}

\noindent\parbox{\mymathboxwidth}{%
  \raggedright
  \relpenalty=100
  \binoppenalty=100
 \small
  $
M^2\Biggl(t-4 M^2\Biggr)^2\,\mathbf{c_{1c}(8)} = \kappa^3 t \Biggl(-\biggl(L^2 \Bigl(6 m^4+4 m^2 \bigl(M^2-3 s-t\bigr)+6 \bigl(M^2-s\bigr)^2-4 M^2 t+6 s t+t^2\Bigr)\biggr)+4L \biggl(6 m^4 M^2+m^2 \Bigl(28 M^4-2 M^2 (6 s+7 t)+t^2\Bigr)+46 M^6-2 M^4 (18 s+13 t)+2 M^2 \Bigl(3 s^2+8 s t+2 t^2\Bigr)-s t^2\biggr)+t \biggl(2 M^2 \Bigl(2 s \bigl(5 m^2+17 M^2\bigr)-5 \bigl(m^2+3 M^2\bigr)^2-5 s^2\Bigr)+t^2 \Bigl(-3 m^2-7 M^2+3 s\Bigr)+t \Bigl(m^4+m^2 \bigl(30 M^2-2 s\bigr)+49 M^4-32 M^2 s+s^2\Bigr)\biggr)\Biggr)-\kappa^2 \Biggl(L^2 \biggl(16 M^2 \Bigl(m^2+M^2-s\Bigr)^2+2 t \Bigl(m^4-2 m^2 \bigl(5 M^2+s\bigr)+M^4+6 M^2 s+s^2\Bigr)+2 t^2 \Bigl(s-2 M^2\Bigr)+t^3\biggr)+8L \biggl(-8 M^4 \Bigl(m^2+M^2-s\Bigr) \Bigl(m^2+3 M^2-s\Bigr)-t^2 \Bigl(m^4-2 m^2 \bigl(2 M^2+s\bigr)-41 M^4+5 M^2 s+s^2\Bigr)+M^2 t \Bigl(3 m^4-2 m^2 \bigl(M^2+3 s\bigr)-65 M^4+6 M^2 s+3 s^2\Bigr)-6 M^2 t^3\biggr)+t \biggl(m^4 \Bigl(-96 M^4+46 M^2 t-4 t^2\Bigr)+4 m^2 \Bigl(64 M^6+M^4 (48 s+t)-M^2 t (23 s+9 t)+t^2 (2 s+t)\Bigr)+96 M^8+254 M^6 t-4 M^4 \Bigl(24 s^2+35 s t+40 t^2\Bigr)+M^2 t \Bigl(46 s^2+54 s t+23 t^2\Bigr)-4 s t^2 (s+t)\biggr)\Biggr)-8 \kappa M^2 \Biggl(4 M^2-t\Biggr) \Biggl(2L \biggl(-2 t \Bigl(m^2+4 M^2-s\Bigr)+2 \Bigl(m^2-3 M^2-s\Bigr) \Bigl(m^2+M^2-s\Bigr)+3 t^2\biggr)+t \biggl(-2 t \Bigl(m^2-4 M^2+s\Bigr)-2 \Bigl(m^4-2 m^2 \bigl(5 M^2+s\bigr)-3 M^4+2 M^2 s+s^2\Bigr)-3 t^2\biggr)\Biggr)+8 M^2 \Biggl(4 M^2-t\Biggr) \Biggl(8 M^2 \biggl(L-2 m^2+2 s\biggr) \biggl(m^2+M^2-s\biggr)+4 t \biggl(L M^2+m^4-2 s \Bigl(m^2+M^2\Bigr)-M^4+s^2\biggr)-2 t^2 \biggl(L+M^2-s\biggr)+t^3\Biggr)$
}

\noindent\parbox{\mymathboxwidth}{%
  \raggedright
  \relpenalty=100
  \binoppenalty=100
 \small
  $
\frac{M^2}{2} \Biggl(t-4 m^2\Biggr)^2\,\mathbf{c_{2c}(2)} = -2L \Biggl(\kappa^3 \biggl(4 m^2-t\biggr) \biggl(2L m^2 \Bigl(2 \bigl(m^2+M^2-s\bigr)-t\Bigr)+t \Bigl(2 m^2-t\Bigr) \Bigl(m^2-M^2+s\Bigr)\biggr)+\kappa^2 \biggl(4L m^2 \Bigl(7 m^4+m^2 \bigl(6 M^2-5 (2 s+t)\bigr)+3 M^4-3 M^2 (2 s+t)+(s+t) (3 s+t)\Bigr)+16 m^6 t-2 t^2 \Bigl(3 m^4+m^2 \bigl(4 s-6 M^2\bigr)+\bigl(M^2-s\bigr)^2\Bigr)-16 m^4 \Bigl(m^2+M^2-s\Bigr)^2-M^2 t^3\biggr)-4 \kappa M^2 t \biggl(t-4 m^2\biggr)^2-4 M^2 \biggl(4 m^2-t\biggr) \biggl(4 m^4+2 m^2 \Bigl(2 M^2-2 s+t\Bigr)-t^2\biggr)\Biggr)$
}

\noindent\parbox{\mymathboxwidth}{%
  \raggedright
  \relpenalty=100
  \binoppenalty=100
 \small
  $
\frac{M^2}{2}\Biggl(t-4 M^2\Biggr)^2\,\mathbf{c_{2c}(3)} = -2L \Biggl(\kappa^3 \biggl(L \Bigl(t^2 \bigl(-m^2+M^2+s\bigr)+2 M^2 \bigl(m^4-2 m^2 \left(M^2+s\right)+\left(M^2-s\right)^2\bigr)+t \bigl(m^4+2 m^2 \left(M^2-s\right)-3 M^4+s^2\bigr)\Bigr)+t \Bigl(-6 m^4 M^2-m^2 \bigl(28 M^4-2 M^2 (6 s+7 t)+t^2\bigr)-46 M^6+M^4 (36 s+26 t)-2 M^2 \bigl(3 s^2+8 s t+2 t^2\bigr)+s t^2\Bigr)\biggr)+\kappa^2 \biggl(L M^2 \Bigl(6 m^4+4 m^2 \bigl(M^2-3 s-t\bigr)+6 \bigl(M^2-s\bigr)^2-4 M^2 t+6 s t+t^2\Bigr)+2 \Bigl(-8 M^4 \bigl(m^2+M^2-s\bigr) \bigl(m^2+3 M^2-s\bigr)-t^2 \bigl(m^4-2 m^2 \left(2 M^2+s\right)-41 M^4+5 M^2 s+s^2\bigr)+M^2 t \bigl(3 m^4-2 m^2 \left(M^2+3 s\right)-65 M^4+6 M^2 s+3 s^2\bigr)-6 M^2 t^3\Bigr)\biggr)-4 \kappa M^2 \biggl(4 M^2-t\biggr) \biggl(-2 m^4+2 m^2 \Bigl(2 \bigl(M^2+s\bigr)+t\Bigr)+6 M^4-4 M^2 (s-2 t)-2 s^2-2 s t-3 t^2\biggr)-4 M^2 \biggl(4 M^2-t\biggr) \biggl(4 M^2 \Bigl(m^2+M^2-s\Bigr)+2 M^2 t-t^2\biggr)\Biggr)$
}

\noindent\parbox{\mymathboxwidth}{%
  \raggedright
  \relpenalty=100
  \binoppenalty=100
 \small
  $
M^2 \Biggl(t-4 m^2\Biggr)^2\,\mathbf{c_{2c}(5)} = 4L \Biggl(\kappa^3 t \biggl(4 m^2-t\biggr) \biggl(2L \Bigl(3 m^2+M^2-s-t\Bigr)+\Bigl(2 m^2-t\Bigr) \Bigl(m^2-M^2+s\Bigr)\biggr)+\kappa^2 \biggl(-2 t^2 \Bigl(L \bigl(5 m^2+M^2-2 s\bigr)+3 m^4+m^2 \bigl(4 s-6 M^2\bigr)+\bigl(M^2-s\bigr)^2\Bigr)+2 t \Bigl(L \bigl(5 m^4+2 m^2 \left(s-3 M^2\right)+\left(M^2-s\right)^2\bigr)+8 m^6\Bigr)+t^3 \Bigl(2L-M^2\Bigr)-16 m^2 (m^2-L) \Bigl(m^2+M^2-s\Bigr)^2\biggr)-4 \kappa M^2 t \biggl(t-4 m^2\biggr)^2-4 M^2 \biggl(4 m^2-t\biggr) \biggl(4 m^4+2 m^2 \Bigl(2 M^2-2 s+t\Bigr)-t^2\biggr)\Biggr)$
}

\noindent\parbox{\mymathboxwidth}{%
  \raggedright
  \relpenalty=100
  \binoppenalty=100
 \small
  $
M^2\Biggl(t-4 M^2\Biggr)^2\,\mathbf{c_{2c}(6)} = 2L \Biggl(\kappa^3 t \biggl(L \Bigl(6 m^4+4 m^2 \bigl(M^2-3 s-t\bigr)+6 \bigl(M^2-s\bigr)^2-4 M^2 t+6 s t+t^2\Bigr)-2 t^2 \Bigl(m^2+4 M^2-s\Bigr)+4 M^2 t \Bigl(7 m^2+13 M^2-8 s\Bigr)-4 M^2 \Bigl(3 m^4-6 s \bigl(m^2+3 M^2\bigr)+14 m^2 M^2+23 M^4+3 s^2\Bigr)\biggr)+\kappa^2 \biggl(L \Bigl(16 M^2 \bigl(m^2+M^2-s\bigr)^2+2 t \bigl(m^4-2 m^2 \left(5 M^2+s\right)+M^4+6 M^2 s+s^2\bigr)+2 t^2 \bigl(s-2 M^2\bigr)+t^3\Bigr)+4 \Bigl(-8 M^4 \bigl(m^2+M^2-s\bigr) \bigl(m^2+3 M^2-s\bigr)-t^2 \bigl(m^4-2 m^2 \left(2 M^2+s\right)-41 M^4+5 M^2 s+s^2\bigr)+M^2 t \bigl(3 m^4-2 m^2 \left(M^2+3 s\right)-65 M^4+6 M^2 s+3 s^2\bigr)-6 M^2 t^3\Bigr)\biggr)-8 \kappa M^2 \biggl(4 M^2-t\biggr) \biggl(-2 m^4+2 m^2 \Bigl(2 \bigl(M^2+s\bigr)+t\Bigr)+6 M^4-4 M^2 (s-2 t)-2 s^2-2 s t-3 t^2\biggr)-8 M^2 \biggl(4 M^2-t\biggr) \biggl(4 M^2 \Bigl(m^2+M^2-s\Bigr)+2 M^2 t-t^2\biggr)\Biggr)$
}

\noindent\parbox{\mymathboxwidth}{%
  \raggedright
  \relpenalty=100
  \binoppenalty=100
 \small
  $
M^2 \Biggl(t-4 m^2\Biggr)^2\,\mathbf{c_{3c}(1)} = 4 \kappa^2L^2 \Biggl((8 \kappa+6) m^6+m^4 \biggl((8 \kappa-4) M^2-4 (2 \kappa+1) s-3 (6 \kappa+5) t\biggr)+m^2 \biggl(6 t \Bigl(\kappa \bigl(s-M^2\bigr)+s\Bigr)+(8 \kappa+7) t^2-2 \Bigl(M^2-s\Bigr)^2\biggr)+t \biggl(M^2 (\kappa t+2 s+t)-(s+t) (\kappa t+s+t)-M^4\biggr)\Biggr)$
}

\noindent\parbox{\mymathboxwidth}{%
  \raggedright
  \relpenalty=100
  \binoppenalty=100
 \small
  $
M^2\Biggl(t-4 M^2\Biggr)^2\,\mathbf{c_{3c}(2)} = \kappa^2L^2 \Biggl(2 m^4 \biggl(2 (\kappa-1) M^2-(2 \kappa+1) t\biggr)+2 m^2 \biggl(-4 (\kappa+3) M^4+M^2 (-4 \kappa s+4 s+6 t)+t (4 \kappa s+\kappa t+2 s)\biggr)+4 (\kappa-1) \biggl(M^3-M s\biggr)^2+2 t^2 \biggl(3 (\kappa+1) M^2-(2 \kappa+1) s\biggr)-2 t \biggl((6 \kappa+5) M^4-6 \kappa M^2 s+(2 \kappa+1) s^2\biggr)-(\kappa+1) t^3\Biggr)$
}

The remaining $c_{ic}$ coefficient can be obtained as follows:
\begin{align*}
c_{1c}(4) & = -\frac{2}{\sqrt{L}}\,c_{2c}(7)  \\
c_{1c}(5) & = -\frac{2}{\sqrt{L}}\,c_{2c}(8)  \\
c_{1c}(1) &= -\,c_{1c}(6) = -\,c_{2c}(1) = c_{2c}(4) = \frac{2}{\sqrt{L}}\,c_{2c}(9) = -\frac{2}{\sqrt{L}}\,c_{3c}(3)  \\
c_{1c}(7) &= -\frac{2}{\sqrt{L}}\,c_{2c}(10) = \frac{4}{L}\,c_{3c}(6) \\
c_{1c}(8) &= -\frac{2}{\sqrt{L}}\,c_{2c}(11) = \frac{4}{L}\,c_{3c}(7) \\
c_{2c}(5) &= -\frac{2}{\sqrt{L}}\,c_{3c}(4)  \\
c_{2c}(6) &= -\frac{2}{\sqrt{L}}\,c_{3c}(5) 
\end{align*}

The coefficients $c_{id}(j)$ ($i=1,2,3, j=1-4$) of the scalar 4-point functions and their derivatives of Eq.~\ref{eq:dcoeffs} are: 

\noindent\parbox{\mymathboxwidth}{%
  \raggedright
  \relpenalty=100
  \binoppenalty=100
 \small
  $
M^2\,\mathbf{c_{1d}(1)} = -2 \Biggl(2 m^6 \biggl(8 M^2-\kappa^2 t\biggr)+m^4 \biggl(\kappa^2 t ((\kappa+2) t+6 s)+2 M^2 (\kappa (\kappa+8) t-24 s)+48 M^4\biggr)+m^2 \biggl(M^2 \Bigl(-\bigl(\kappa \left(2 \kappa^2+\kappa-4\right)-4\bigr) t^2+4 ((\kappa-4) \kappa+2) s t+48 s^2\Bigr)-2 \kappa^2 s t ((\kappa+2) t+3 s)+2 M^4 (\kappa (\kappa+8) t-48 s)+48 M^6\biggr)+\biggl(M^2-s\biggr) \biggl((\kappa+2) t^2 \Bigl((\kappa+1) (\kappa+2) M^2-\kappa^2 s\Bigr)-2 t \Bigl(\kappa^2 \bigl(M^2-s\bigr)^2-4 M^2 s\Bigr)+16 \Bigl(M^3-M s\Bigr)^2\biggr)\Biggr)$
}

\noindent\parbox{\mymathboxwidth}{%
  \raggedright
  \relpenalty=100
  \binoppenalty=100
 \small
  $
M^2\,\mathbf{c_{1d}(2)} = 4 \Biggl(\kappa^3 \biggl(L-t\biggr)^2 \biggl(m^4-2 m^2 \Bigl(M^2+s\Bigr)+\Bigl(M^2-s\Bigr)^2\biggr)+\kappa^2 \biggl(L^2 \Bigl(2 m^4-m^2 \bigl(M^2+4 s\bigr)+5 M^4-7 M^2 s+2 s^2\Bigr)+2L \Bigl(-m^6+m^4 \bigl(M^2+3 s\bigr)+m^2 \bigl(M^4+M^2 (2 s+7 t)-3 s^2\bigr)+M^2 t \bigl(s-M^2\bigr)-\bigl(M^2-s\bigr)^3\Bigr)+t \Bigl(t \bigl(2 m^4-m^2 \left(M^2+4 s\right)+5 M^4-7 M^2 s+2 s^2\bigr)-2 \bigl((m-M)^2-s\bigr) \bigl(m^2+M^2-s\bigr) \bigl((m+M)^2-s\bigr)\Bigr)\biggr)+4 \kappa M^2 \biggl(L^2 \Bigl(m^2+2 M^2-2 s\Bigr)+2L m^2 \Bigl(2 \bigl(m^2+M^2-s\bigr)+t\Bigr)+t \Bigl(4 m^4+m^2 \bigl(4 M^2-4 s+t\bigr)+2 t \bigl(M^2-s\bigr)\Bigr)\biggr)+4 M^2 \biggl(m^2+M^2-s\biggr) \biggl(L^2+2L s+4 \Bigl(m^2+M^2-s\Bigr)^2+2 s t+t^2\biggr)\Biggr)$
}

\noindent\parbox{\mymathboxwidth}{%
  \raggedright
  \relpenalty=100
  \binoppenalty=100
 \small
  $
M^2\,\mathbf{c_{1d}(3)} = -2 \Biggl(\kappa^3 t \biggl(L^2 \Bigl(6 m^2+8 M^2-6 s\Bigr)-4L \Bigl(m^4-2 m^2 \bigl(M^2+s\bigr)+\bigl(M^2-s\bigr)^2\Bigr)+t \Bigl((m-M)^2-s\Bigr) \Bigl((m+M)^2-s\Bigr)\biggr)+\kappa^2 \biggl(-4L \Bigl(m^2+M^2-s\Bigr) \Bigl(L \bigl(-m^2-2 M^2+s\bigr)+m^4-2 m^2 \bigl(M^2+s\bigr)+\bigl(M^2-s\bigr)^2\Bigr)+2 t \Bigl(3L^2 \bigl(m^2+4 M^2-s\bigr)+2L M^2 \bigl(7 m^2-M^2+s\bigr)-\bigl((m-M)^2-s\bigr) \bigl(m^2+M^2-s\bigr) \bigl((m+M)^2-s\bigr)\Bigr)+t^2 \Bigl(2 m^4-m^2 \bigl(M^2+4 s\bigr)+5 M^4-7 M^2 s+2 s^2\Bigr)\biggr)+4 \kappa M^2 \biggl(6L^2 t+4L \Bigl(L+2 m^2\Bigr) \Bigl(m^2+M^2-s\Bigr)+4L m^2 t+t^2 \Bigl(m^2+2 M^2-2 s\Bigr)+4 m^2 t \Bigl(m^2+M^2-s\Bigr)\biggr)+4 M^2 \biggl(2L^2 \Bigl(m^2+M^2-s+t\Bigr)+4L s \Bigl(m^2+M^2-s\Bigr)+\Bigl(m^2+M^2-s\Bigr) \Bigl(4 \bigl(m^2+M^2-s\bigr)^2+2 s t+t^2\Bigr)\biggr)\Biggr)$
}

\noindent\parbox{\mymathboxwidth}{%
  \raggedright
  \relpenalty=100
  \binoppenalty=100
 \small
  $
M^2\,\mathbf{c_{2d}(1)} = -8L \Biggl(\kappa^3 \biggl(L-t\biggr) \biggl(m^4-2 m^2 \Bigl(M^2+s\Bigr)+\Bigl(M^2-s\Bigr)^2\biggr)+\kappa^2 \biggl(L \Bigl(2 m^4-m^2 \bigl(M^2+4 s\bigr)+5 M^4-7 M^2 s+2 s^2\Bigr)-m^6+m^4 \Bigl(M^2+3 s\Bigr)+m^2 \Bigl(M^4+M^2 (2 s+7 t)-3 s^2\Bigr)+M^2 t \Bigl(s-M^2\Bigr)-\Bigl(M^2-s\Bigr)^3\biggr)+4 \kappa M^2 \biggl(L \Bigl(m^2+2 M^2-2 s\Bigr)+m^2 \Bigl(2 \bigl(m^2+M^2-s\bigr)+t\Bigr)\biggr)+4 M^2 \biggl(L+s\biggr) \biggl(m^2+M^2-s\biggr)\Biggr)$
}

\noindent\parbox{\mymathboxwidth}{%
  \raggedright
  \relpenalty=100
  \binoppenalty=100
 \small
  $
M^2\,\mathbf{c_{2d}(2)} = 8L \Biggl(\kappa^3 t \biggl(L \Bigl(3 m^2+4 M^2-3 s\Bigr)-m^4+2 m^2 \Bigl(M^2+s\Bigr)-\Bigl(M^2-s\Bigr)^2\biggr)+\kappa^2 \biggl(t \Bigl(3L \bigl(m^2+4 M^2-s\bigr)+M^2 \bigl(7 m^2-M^2+s\bigr)\Bigr)-\Bigl(m^2+M^2-s\Bigr) \Bigl(-2L \bigl(m^2+2 M^2-s\bigr)+m^4-2 m^2 \bigl(M^2+s\bigr)+\bigl(M^2-s\bigr)^2\Bigr)\biggr)+4 \kappa M^2 \biggl(2 \Bigl(L+m^2\Bigr) \Bigl(m^2+M^2-s\Bigr)+t \Bigl(3L+m^2\Bigr)\biggr)+4 M^2 \biggl(\Bigl(L+s\Bigr) \Bigl(m^2+M^2-s\Bigr)+L t\biggr)\Biggr)$
}

\noindent\parbox{\mymathboxwidth}{%
  \raggedright
  \relpenalty=100
  \binoppenalty=100
 \small
  $
M^2\,\mathbf{c_{3d}(1)} = 4L^2 \Biggl(\kappa^3 \biggl(m^4-m^2 \Bigl(2 \bigl(M^2+s\bigr)+3 t\Bigr)+M^4-2 M^2 (s+2 t)+s (s+3 t)\biggr)+\kappa^2 \biggl(-m^2 \Bigl(7 M^2+3 t\Bigr)+M^4-M^2 (s+12 t)+3 s t\biggr)-4 \kappa M^2 \biggl(m^2+3 t\biggr)-4 M^2 t\Biggr)$
}

The remaining $c_{id}$ coefficients can be obtained as follows:
\begin{align*}
c_{1d}(2) &= -\frac{2}{\sqrt{L}}\,c_{2d}(3) \\
c_{1d}(3) &= -\frac{2}{\sqrt{L}}\,c_{2d}(4) = \frac{4}{L}\,c_{3d}(3) \\
c_{2d}(2) &= -\frac{2}{\sqrt{L}}\,c_{3d}(2)
\end{align*}

\bibliography{refs.bib}

@article{Bloch:1937pw,
    author = "Bloch, F. and Nordsieck, A.",
    title = "{Note on the Radiation Field of the electron}",
    reportNumber = "RX-1199",
    doi = "10.1103/PhysRev.52.54",
    journal = "Phys. Rev.",
    volume = "52",
    pages = "54--59",
    year = "1937"
}

@article{Yennie:1961ad,
    author = "Yennie, D. R. and Frautschi, Steven C. and Suura, H.",
    title = "{The infrared divergence phenomena and high-energy processes}",
    doi = "10.1016/0003-4916(61)90151-8",
    journal = "Annals Phys.",
    volume = "13",
    pages = "379--452",
    year = "1961"
}

@article{ParticleDataGroup:2024cfk,
    author = "Navas, S. and others",
    collaboration = "Particle Data Group",
    title = "{Review of particle physics}",
    doi = "10.1103/PhysRevD.110.030001",
    journal = "Phys. Rev. D",
    volume = "110",
    number = "3",
    pages = "030001",
    year = "2024"
}

@article{Arrington:2003ck,
    author = "Arrington, J.",
    title = "{Evidence for two photon exchange contributions in electron proton and positron proton elastic scattering}",
    eprint = "nucl-ex/0311019",
    archivePrefix = "arXiv",
    doi = "10.1103/PhysRevC.69.032201",
    journal = "Phys. Rev. C",
    volume = "69",
    pages = "032201",
    year = "2004"
}

@article{Arrington:2009qd,
    author = "Arrington, John",
    editor = "Elouadrhiri, Latifa and Grames, Joseph and Melnitchouk, Wally and Forest, Tony A. and Voutier, Eric",
    title = "{Two-photon exchange measurements with positrons and electrons}",
    eprint = "0905.0713",
    archivePrefix = "arXiv",
    primaryClass = "nucl-ex",
    doi = "10.1063/1.3232022",
    journal = "AIP Conf. Proc.",
    volume = "1160",
    number = "1",
    pages = "13--18",
    year = "2009"
}

@article{CLAS:2013mza,
    author = "Moteabbed, M. and others",
    collaboration = "CLAS",
    title = "{Demonstration of a novel technique to measure two-photon exchange effects in elastic $e^\pm p$ scattering}",
    eprint = "1306.2286",
    archivePrefix = "arXiv",
    primaryClass = "nucl-ex",
    reportNumber = "JLAB-PHY-13-1745",
    doi = "10.1103/PhysRevC.88.025210",
    journal = "Phys. Rev. C",
    volume = "88",
    pages = "025210",
    year = "2013"
}

@article{Beenakker:1991ca,
    author = "Beenakker, W. and van der Marck, S. C. and Hollik, W.",
    title = "{e+ e- annihilation into heavy fermion pairs at high-energy colliders}",
    reportNumber = "MPI-PAE-PTH-12-91",
    doi = "10.1016/0550-3213(91)90606-X",
    journal = "Nucl. Phys. B",
    volume = "365",
    pages = "24--78",
    year = "1991"
}

@article{PhysRev.126.2256,
  title = {High-Energy Behavior of Nucleon Electromagnetic Form Factors},
  author = {Sachs, R. G.},
  journal = {Phys. Rev.},
  volume = {126},
  issue = {6},
  pages = {2256--2260},
  numpages = {0},
  year = {1962},
  month = {Jun},
  publisher = {American Physical Society},
  doi = {10.1103/PhysRev.126.2256},
  url = {https://link.aps.org/doi/10.1103/PhysRev.126.2256}
}

@article{Maximon2000,
    title = {Radiative corrections to electron-proton scattering},
    volume = {62},
    ISSN = {1089-490X},
    url = {http://dx.doi.org/10.1103/PhysRevC.62.054320},
    DOI = {10.1103/physrevc.62.054320},
    number = {5},
    journal = {Physical Review C},
    publisher = {American Physical Society (APS)},
    author = {Maximon,  L. C. and Tjon,  J. A.},
    year = {2000},
    month = oct 
}

@article{Mo:1968cg,
    author = "Mo, Luke W. and Tsai, Yung-Su",
    title = "{Radiative Corrections to Elastic and Inelastic e p and mu p Scattering}",
    reportNumber = "SLAC-PUB-0380",
    doi = "10.1103/RevModPhys.41.205",
    journal = "Rev. Mod. Phys.",
    volume = "41",
    pages = "205--235",
    year = "1969"
}

@article{Tsai:1961zz,
    author = "Tsai, Yung-Su",
    title = "{Radiative Corrections to Electron-Proton Scattering}",
    doi = "10.1103/PhysRev.122.1898",
    journal = "Phys. Rev.",
    volume = "122",
    pages = "1898--1907",
    year = "1961"
}

@article{Blunden:2003sp,
    author = "Blunden, P. G. and Melnitchouk, W. and Tjon, J. A.",
    title = "{Two photon exchange and elastic electron proton scattering}",
    eprint = "nucl-th/0306076",
    archivePrefix = "arXiv",
    reportNumber = "JLAB-THY-03-120",
    doi = "10.1103/PhysRevLett.91.142304",
    journal = "Phys. Rev. Lett.",
    volume = "91",
    pages = "142304",
    year = "2003"
}

@misc{Qattan:2024pco,
    author = "Qattan, I. A. and others",
    title = "{High precision measurements of the proton elastic electromagnetic form factors and their ratio at $Q^2$ = 0.50, 2.64, 3.20, and 4.10 GeV$^2$}",
    eprint = "2411.05201",
    archivePrefix = "arXiv",
    primaryClass = "nucl-ex",
    month = "11",
    year = "2024"
}

@article{Gao:2021sml,
    author = "Gao, Haiyan and Vanderhaeghen, Marc",
    title = "{The proton charge radius}",
    eprint = "2105.00571",
    archivePrefix = "arXiv",
    primaryClass = "hep-ph",
    doi = "10.1103/RevModPhys.94.015002",
    journal = "Rev. Mod. Phys.",
    volume = "94",
    number = "1",
    pages = "015002",
    year = "2022"
}

@article{Pohl:2010zza,
    author = "Pohl, Randolf and others",
    title = "{The size of the proton}",
    doi = "10.1038/nature09250",
    journal = "Nature",
    volume = "466",
    pages = "213--216",
    year = "2010"
}

@article{Lee:2015jqa,
    author = "Lee, Gabriel and Arrington, John R. and Hill, Richard J.",
    title = "{Extraction of the proton radius from electron-proton scattering data}",
    eprint = "1505.01489",
    archivePrefix = "arXiv",
    primaryClass = "hep-ph",
    reportNumber = "EFI-PREPRINT-14-35",
    doi = "10.1103/PhysRevD.92.013013",
    journal = "Phys. Rev. D",
    volume = "92",
    number = "1",
    pages = "013013",
    year = "2015"
}

@article{Isaacson:2022cwh,
    author = "Isaacson, Joshua and Jay, William I. and Lovato, Alessandro and Machado, Pedro A. N. and Rocco, Noemi",
    title = "{Introducing a novel event generator for electron-nucleus and neutrino-nucleus scattering}",
    eprint = "2205.06378",
    archivePrefix = "arXiv",
    primaryClass = "hep-ph",
    reportNumber = "FERMILAB-PUB-22-411-T, MIT-CTP/5428",
    doi = "10.1103/PhysRevD.107.033007",
    journal = "Phys. Rev. D",
    volume = "107",
    number = "3",
    pages = "033007",
    year = "2023"
}

@article{Carlson:2007sp,
    author = "Carlson, Carl E. and Vanderhaeghen, Marc",
    title = "{Two-Photon Physics in Hadronic Processes}",
    eprint = "hep-ph/0701272",
    archivePrefix = "arXiv",
    reportNumber = "WM-07-101, JLAB-THY-07-616",
    doi = "10.1146/annurev.nucl.57.090506.123116",
    journal = "Ann. Rev. Nucl. Part. Sci.",
    volume = "57",
    pages = "171--204",
    year = "2007"
}

@article{Shtabovenko_2020,
   title={FeynCalc 9.3: New features and improvements},
   volume={256},
   ISSN={0010-4655},
   url={http://dx.doi.org/10.1016/j.cpc.2020.107478},
   DOI={10.1016/j.cpc.2020.107478},
   journal={Computer Physics Communications},
   publisher={Elsevier BV},
   author={Shtabovenko, Vladyslav and Mertig, Rolf and Orellana, Frederik},
   year={2020},
   month=nov, pages={107478} }

@article{Shtabovenko_2016,
   title={New developments in FeynCalc 9.0},
   volume={207},
   ISSN={0010-4655},
   url={http://dx.doi.org/10.1016/j.cpc.2016.06.008},
   DOI={10.1016/j.cpc.2016.06.008},
   journal={Computer Physics Communications},
   publisher={Elsevier BV},
   author={Shtabovenko, Vladyslav and Mertig, Rolf and Orellana, Frederik},
   year={2016},
   month=oct, pages={432–444} }

@article{MERTIG1991345,
    title = {Feyn Calc - Computer-algebraic calculation of Feynman amplitudes},
    journal = {Computer Physics Communications},
    volume = {64},
    number = {3},
    pages = {345-359},
    year = {1991},
    issn = {0010-4655},
    doi = {https://doi.org/10.1016/0010-4655(91)90130-D},
    url = {https://www.sciencedirect.com/science/article/pii/001046559190130D},
    author = {R. Mertig and M. Böhm and A. Denner}
}

@article{Arrington_2011,
   title={Review of two-photon exchange in electron scattering},
   volume={66},
   ISSN={0146-6410},
   url={http://dx.doi.org/10.1016/j.ppnp.2011.07.003},
   DOI={10.1016/j.ppnp.2011.07.003},
   number={4},
   journal={Progress in Particle and Nuclear Physics},
   publisher={Elsevier BV},
   author={Arrington, J. and Blunden, P.G. and Melnitchouk, W.},
   year={2011},
   month=oct, pages={782–833} 
}

@article{AFANASEV2017245,
    title = {Two-photon exchange in elastic electron–proton scattering},
    journal = {Progress in Particle and Nuclear Physics},
    volume = {95},
    pages = {245-278},
    year = {2017},
    issn = {0146-6410},
    doi = {https://doi.org/10.1016/j.ppnp.2017.03.004},
    url = {https://www.sciencedirect.com/science/article/pii/S0146641017300352},
    author = {A. Afanasev and P.G. Blunden and D. Hasell and B.A. Raue}
}

@article{Lee:2013mka,
    author = "Lee, Roman N.",
    editor = "Wang, Jianxiong",
    title = "{LiteRed 1.4: a powerful tool for reduction of multiloop integrals}",
    eprint = "1310.1145",
    archivePrefix = "arXiv",
    primaryClass = "hep-ph",
    doi = "10.1088/1742-6596/523/1/012059",
    journal = "J. Phys. Conf. Ser.",
    volume = "523",
    pages = "012059",
    year = "2014"
}

@article{tHooft:1978jhc,
    author = "'t Hooft, Gerard and Veltman, M. J. G.",
    title = "{Scalar One Loop Integrals}",
    reportNumber = "PRINT-79-0134 (UTRECHT)",
    doi = "10.1016/0550-3213(79)90605-9",
    journal = "Nucl. Phys. B",
    volume = "153",
    pages = "365--401",
    year = "1979"
}

@article{Patel:2016fam,
    author = "Patel, Hiren H.",
    title = "{Package-X 2.0: A Mathematica package for the analytic calculation of one-loop integrals}",
    eprint = "1612.00009",
    archivePrefix = "arXiv",
    primaryClass = "hep-ph",
    doi = "10.1016/j.cpc.2017.04.015",
    journal = "Comput. Phys. Commun.",
    volume = "218",
    pages = "66--70",
    year = "2017"
}

@article{Denner:2014gla,
    author = "Denner, Ansgar and Dittmaier, Stefan and Hofer, Lars",
    editor = "Mende, Martina",
    title = "{COLLIER - A fortran-library for one-loop integrals}",
    eprint = "1407.0087",
    archivePrefix = "arXiv",
    primaryClass = "hep-ph",
    doi = "10.22323/1.211.0071",
    journal = "PoS",
    volume = "LL2014",
    pages = "071",
    year = "2014"
}

@article{A1:2013fsc,
    author = "Bernauer, J. C. and others",
    collaboration = "A1",
    title = "{Electric and magnetic form factors of the proton}",
    eprint = "1307.6227",
    archivePrefix = "arXiv",
    primaryClass = "nucl-ex",
    doi = "10.1103/PhysRevC.90.015206",
    journal = "Phys. Rev. C",
    volume = "90",
    number = "1",
    pages = "015206",
    year = "2014"
}

@article{Mihovilovic:2016rkr,
    author = "Mihovilovi{\v{c}}, M. and others",
    title = "{First measurement of proton's charge form factor at very low $Q^2$ with initial state radiation}",
    eprint = "1612.06707",
    archivePrefix = "arXiv",
    primaryClass = "nucl-ex",
    doi = "10.1016/j.physletb.2017.05.031",
    journal = "Phys. Lett. B",
    volume = "771",
    pages = "194--198",
    year = "2017"
}

@article{Mihovilovic:2019jiz,
    author = "Mihovilovi{\v{c}}, M. and others",
    title = "{The proton charge radius extracted from the initial-state radiation experiment at MAMI}",
    eprint = "1905.11182",
    archivePrefix = "arXiv",
    primaryClass = "nucl-ex",
    doi = "10.1140/epja/s10050-021-00414-x",
    journal = "Eur. Phys. J. A",
    volume = "57",
    number = "3",
    pages = "107",
    year = "2021"
}

@article{Xiong:2019umf,
    author = "Xiong, W. and others",
    title = "{A small proton charge radius from an electron{\textendash}proton scattering experiment}",
    doi = "10.1038/s41586-019-1721-2",
    journal = "Nature",
    volume = "575",
    number = "7781",
    pages = "147--150",
    year = "2019"
}

@article{Qweak:2018tjf,
    author = "Androi{\'c}, D. and others",
    collaboration = "Qweak",
    title = "{Precision measurement of the weak charge of the proton}",
    eprint = "1905.08283",
    archivePrefix = "arXiv",
    primaryClass = "nucl-ex",
    doi = "10.1038/s41586-018-0096-0",
    journal = "Nature",
    volume = "557",
    number = "7704",
    pages = "207--211",
    year = "2018"
}

@article{Becker:2018ggl,
    author = "Becker, Dominik and others",
    title = "{The P2 experiment}",
    eprint = "1802.04759",
    archivePrefix = "arXiv",
    primaryClass = "nucl-ex",
    doi = "10.1140/epja/i2018-12611-6",
    journal = "Eur. Phys. J. A",
    volume = "54",
    number = "11",
    pages = "208",
    year = "2018"
}

@misc{MUSE:2013uhu,
    author = "Gilman, R. and others",
    collaboration = "MUSE",
    title = "{Studying the Proton ''Radius'' Puzzle with {\textbackslash}mu p Elastic Scattering}",
    eprint = "1303.2160",
    archivePrefix = "arXiv",
    primaryClass = "nucl-ex",
    month = "3",
    year = "2013"
}

@misc{MUSE:2017dod,
    author = "Gilman, R. and others",
    collaboration = "MUSE",
    title = "{Technical Design Report for the Paul Scherrer Institute Experiment R-12-01.1: Studying the Proton ''Radius'' Puzzle with $\mu p$ Elastic Scattering}",
    eprint = "1709.09753",
    archivePrefix = "arXiv",
    primaryClass = "physics.ins-det",
    month = "9",
    year = "2017"
}

@article{Engel:2023arz,
    author = "Engel, T. and Hagelstein, F. and Rocco, M. and Sharkovska, V. and Signer, A. and Ulrich, Y.",
    title = "{Impact of NNLO QED corrections on lepton-proton scattering at MUSE}",
    eprint = "2307.16831",
    archivePrefix = "arXiv",
    primaryClass = "hep-ph",
    reportNumber = "FR-PHENO-2023-08, IPPP/23/39, PSI-PR-23-27, ZU-TH 39/23, MITP-23-040",
    doi = "10.1140/epja/s10050-023-01153-x",
    journal = "Eur. Phys. J. A",
    volume = "59",
    number = "11",
    pages = "253",
    year = "2023"
}

@article{Bonciani:2021okt,
    author = "Bonciani, R. and others",
    title = "{Two-Loop Four-Fermion Scattering Amplitude in QED}",
    eprint = "2106.13179",
    archivePrefix = "arXiv",
    primaryClass = "hep-ph",
    reportNumber = "MPP-2021-84; ZU-TH 29/21",
    doi = "10.1103/PhysRevLett.128.022002",
    journal = "Phys. Rev. Lett.",
    volume = "128",
    number = "2",
    pages = "022002",
    year = "2022"
}

@article{Banerjee:2020rww,
    author = "Banerjee, Pulak and Engel, T. and Signer, A. and Ulrich, Y.",
    title = "{QED at NNLO with McMule}",
    eprint = "2007.01654",
    archivePrefix = "arXiv",
    primaryClass = "hep-ph",
    reportNumber = "PSI-PR-20-09, ZU-TH 23/20",
    doi = "10.21468/SciPostPhys.9.2.027",
    journal = "SciPost Phys.",
    volume = "9",
    pages = "027",
    year = "2020"
}

@article{Bucoveanu:2019hxz,
    author = "Bucoveanu, Razvan-Daniel and Spiesberger, Hubert",
    editor = "Lenisa, Paolo and Ciullo, Giuseppe and Contalbrigo, Marco and Pappalardo, Luciano",
    title = "{QED radiative corrections for Polarized Lepton-Proton Scattering}",
    eprint = "1903.12229",
    archivePrefix = "arXiv",
    primaryClass = "hep-ph",
    reportNumber = "MITP-10-022, MITP/10-022",
    doi = "10.22323/1.346.0115",
    journal = "PoS",
    volume = "SPIN2018",
    pages = "115",
    year = "2019"
}

@article{Kuraev:2013dra,
    author = "Kuraev, E. A. and Ahmadov, A. I. and Bystritskiy, Yu. M. and Tomasi-Gustafsson, E.",
    title = "{Radiative corrections for electron proton elastic scattering taking into account high orders and hard photon emission}",
    eprint = "1311.0370",
    archivePrefix = "arXiv",
    primaryClass = "hep-ph",
    doi = "10.1103/PhysRevC.89.065207",
    journal = "Phys. Rev. C",
    volume = "89",
    number = "6",
    pages = "065207",
    year = "2014"
}

@article{Gerasimov:2015aoa,
    author = "Gerasimov, R. E. and Fadin, V. S.",
    title = "{Analysis of approximations used in calculations of radiative corrections to electron-proton scattering cross section}",
    doi = "10.1134/S1063778815010081",
    journal = "Phys. Atom. Nucl.",
    volume = "78",
    number = "1",
    pages = "69--91",
    year = "2015"
}

@article{Akushevich:2015toa,
    author = "Akushevich, I. and Gao, H. and Ilyichev, A. and Meziane, M.",
    title = "{Radiative corrections beyond the ultra relativistic limit in unpolarized ep elastic and M{\o}ller scatterings for the PRad Experiment at Jefferson Laboratory}",
    doi = "10.1140/epja/i2015-15001-8",
    journal = "Eur. Phys. J. A",
    volume = "51",
    number = "1",
    pages = "1",
    year = "2015"
}

@article{Bystritskiy:2006ju,
    author = "Bystritskiy, Yu. M. and Kuraev, E. A. and Tomasi-Gustafsson, E.",
    title = "{Structure function method applied to polarized and unpolarized electron-proton scattering: A solution of the GE(p)/GM(p) discrepancy}",
    eprint = "hep-ph/0603132",
    archivePrefix = "arXiv",
    reportNumber = "DAPNIA-06-056",
    doi = "10.1103/PhysRevC.75.015207",
    journal = "Phys. Rev. C",
    volume = "75",
    pages = "015207",
    year = "2007"
}

@article{Bohm:1986rj,
    author = "Bohm, M. and Spiesberger, H. and Hollik, W.",
    title = "{On the One Loop Renormalization of the Electroweak Standard Model and Its Application to Leptonic Processes}",
    doi = "10.1002/prop.19860341102",
    journal = "Fortsch. Phys.",
    volume = "34",
    pages = "687--751",
    year = "1986"
}

@article{denner07ewradcorr,
    author = "Denner, Ansgar",
    title = "{Techniques for calculation of electroweak radiative corrections at the one loop level and results for W physics at LEP-200}",
    eprint = "0709.1075",
    archivePrefix = "arXiv",
    primaryClass = "hep-ph",
    reportNumber = "PRINT-91-0349 (WURZBURG)",
    doi = "10.1002/prop.2190410402",
    journal = "Fortsch. Phys.",
    volume = "41",
    pages = "307--420",
    year = "1993"
}

@article{CLAS:2003umf,
    author = "Mecking, B. A. and others",
    collaboration = "CLAS",
    title = "{The CEBAF Large Acceptance Spectrometer (CLAS)}",
    reportNumber = "JLAB-PHY-03-01",
    doi = "10.1016/S0168-9002(03)01001-5",
    journal = "Nucl. Instrum. Meth. A",
    volume = "503",
    pages = "513--553",
    year = "2003"
}

@article{OLYMPUS:2013lem,
    author = "Milner, R. and others",
    collaboration = "OLYMPUS",
    title = "{The OLYMPUS Experiment}",
    eprint = "1312.1730",
    archivePrefix = "arXiv",
    primaryClass = "physics.ins-det",
    doi = "10.1016/j.nima.2013.12.035",
    journal = "Nucl. Instrum. Meth. A",
    volume = "741",
    pages = "1--17",
    year = "2014"
}

@article{Beenakker:1988jr,
    author = "Beenakker, W. and Denner, Ansgar",
    title = "{Infrared Divergent Scalar Box Integrals With Applications in the Electroweak Standard Model}",
    reportNumber = "Print-89-0200 (LEIDEN)",
    doi = "10.1016/0550-3213(90)90636-R",
    journal = "Nucl. Phys. B",
    volume = "338",
    pages = "349--370",
    year = "1990"
}

@article{Dittmaier:2003bc,
    author = "Dittmaier, Stefan",
    title = "{Separation of soft and collinear singularities from one loop N point integrals}",
    eprint = "hep-ph/0308246",
    archivePrefix = "arXiv",
    reportNumber = "MPP-2003-61",
    doi = "10.1016/j.nuclphysb.2003.10.003",
    journal = "Nucl. Phys. B",
    volume = "675",
    pages = "447--466",
    year = "2003"
}

@article{Hill:2016gdf,
    author = "Hill, Richard J.",
    title = "{Effective field theory for large logarithms in radiative corrections to electron proton scattering}",
    eprint = "1605.02613",
    archivePrefix = "arXiv",
    primaryClass = "hep-ph",
    doi = "10.1103/PhysRevD.95.013001",
    journal = "Phys. Rev. D",
    volume = "95",
    number = "1",
    pages = "013001",
    year = "2017"
}

@article{Kivel:2012vs,
    author = "Kivel, N. and Vanderhaeghen, M.",
    title = "{Two-photon exchange corrections to elastic electron-proton scattering at large momentum transfer within the SCET approach}",
    eprint = "1212.0683",
    archivePrefix = "arXiv",
    primaryClass = "hep-ph",
    reportNumber = "HIM-2012-08",
    doi = "10.1007/JHEP04(2013)029",
    journal = "JHEP",
    volume = "04",
    pages = "029",
    year = "2013"
}

@article{PhysRevLett.103.092004,
  title = {Two-Photon Exchange in Elastic Electron-Proton Scattering: A QCD Factorization Approach},
  author = {Kivel, Nikolai and Vanderhaeghen, Marc},
  journal = {Phys. Rev. Lett.},
  volume = {103},
  issue = {9},
  pages = {092004},
  numpages = {4},
  year = {2009},
  month = {Aug},
  publisher = {American Physical Society},
  doi = {10.1103/PhysRevLett.103.092004},
  url = {https://link.aps.org/doi/10.1103/PhysRevLett.103.092004}
}

@article{Talukdar:2019dko,
    author = "Talukdar, Pulak and Shastry, Vanamali C. and Raha, Udit and Myhrer, Fred",
    title = "{Lepton-Proton Two-Photon Exchange in Chiral Perturbation Theory}",
    eprint = "1911.06843",
    archivePrefix = "arXiv",
    primaryClass = "nucl-th",
    doi = "10.1103/PhysRevD.101.013008",
    journal = "Phys. Rev. D",
    volume = "101",
    number = "1",
    pages = "013008",
    year = "2020"
}

@article{PhysRevD.104.053001,
  title = {Radiative and chiral corrections to elastic lepton-proton scattering in chiral perturbation theory},
  author = {Talukdar, Pulak and Shastry, Vanamali C. and Raha, Udit and Myhrer, Fred},
  journal = {Phys. Rev. D},
  volume = {104},
  issue = {5},
  pages = {053001},
  numpages = {43},
  year = {2021},
  month = {Sep},
  publisher = {American Physical Society},
  doi = {10.1103/PhysRevD.104.053001},
  url = {https://link.aps.org/doi/10.1103/PhysRevD.104.053001}
}

@article{Choudhary:2023rsz,
    author = "Choudhary, Poonam and Goswami, Rakshanda and Raha, Udit and Myhrer, Fred and Chakrabarti, Dipankar",
    title = "{Analytical evaluation of elastic lepton-proton two-photon exchange in chiral perturbation theory}",
    eprint = "2306.09454",
    archivePrefix = "arXiv",
    primaryClass = "hep-ph",
    doi = "10.1140/epja/s10050-023-01207-0",
    journal = "Eur. Phys. J. A",
    volume = "60",
    number = "3",
    pages = "69",
    year = "2024",
    note = "[Erratum: Eur.Phys.J.A 62, 31 (2026)]"
}

@article{Dye:2016uep,
    author = "Dye, Steven P. and Gonderinger, Matthew and Paz, Gil",
    title = "{Elements of QED-NRQED effective field theory: NLO scattering at leading power}",
    eprint = "1602.07770",
    archivePrefix = "arXiv",
    primaryClass = "hep-ph",
    reportNumber = "WSU-HEP-1601",
    doi = "10.1103/PhysRevD.94.013006",
    journal = "Phys. Rev. D",
    volume = "94",
    number = "1",
    pages = "013006",
    year = "2016"
}

@article{Dye:2018rgg,
    author = "Dye, Steven P. and Gonderinger, Matthew and Paz, Gil",
    title = "{Elements of QED-NRQED Effective Field Theory: II. Matching of Contact Interactions}",
    eprint = "1812.05056",
    archivePrefix = "arXiv",
    primaryClass = "hep-ph",
    reportNumber = "FERMILAB-PUB-18-675-T, WSU-HEP-1808",
    doi = "10.1103/PhysRevD.100.054010",
    journal = "Phys. Rev. D",
    volume = "100",
    number = "5",
    pages = "054010",
    year = "2019"
}

@article{Blunden:2005ew,
    author = "Blunden, P. G. and Melnitchouk, W. and Tjon, J. A.",
    title = "{Two-photon exchange in elastic electron-nucleon scattering}",
    eprint = "nucl-th/0506039",
    archivePrefix = "arXiv",
    reportNumber = "JLAB-THY-05-372",
    doi = "10.1103/PhysRevC.72.034612",
    journal = "Phys. Rev. C",
    volume = "72",
    pages = "034612",
    year = "2005"
}

@article{Borisyuk:2006fh,
    author = "Borisyuk, Dmitry and Kobushkin, Alexander",
    title = "{Box diagram in the elastic electron-proton scattering}",
    eprint = "nucl-th/0606030",
    archivePrefix = "arXiv",
    doi = "10.1103/PhysRevC.74.065203",
    journal = "Phys. Rev. C",
    volume = "74",
    pages = "065203",
    year = "2006"
}

@article{PhysRevC.78.015205,
  title = {Charge asymmetry for electron (positron)-proton elastic scattering at large angles},
  author = {Kuraev, E. A. and Bytev, V. V. and Bakmaev, S. and Tomasi-Gustafsson, E.},
  journal = {Phys. Rev. C},
  volume = {78},
  issue = {1},
  pages = {015205},
  numpages = {6},
  year = {2008},
  month = {Jul},
  publisher = {American Physical Society},
  doi = {10.1103/PhysRevC.78.015205},
  url = {https://link.aps.org/doi/10.1103/PhysRevC.78.015205}
}

@article{Tomalak:2014dja,
  author  = {Tomalak, O. and Vanderhaeghen, M.},
  title   = {{Two-photon exchange correction to elastic lepton-proton 
             scattering at small momentum transfer}},
  journal = {Phys. Rev. D},
  volume  = {90},
  pages   = {013006},
  year    = {2014},
  doi     = {10.1103/PhysRevD.90.013006},
  eprint  = {1405.1600},
  archivePrefix = {arXiv}
}

@article{Tomalak:2016vbf,
  author  = {Tomalak, O. and Pasquini, B. and Vanderhaeghen, M.},
  title   = {{Two-photon exchange corrections to elastic $e^-$-proton 
             scattering: full dispersive treatment of $\pi N$ states 
             at low momentum transfers}},
  journal = {Phys. Rev. D},
  volume  = {95},
  pages   = {096001},
  year    = {2017},
  doi     = {10.1103/PhysRevD.95.096001},
  eprint  = {1612.07726},
  archivePrefix = {arXiv}
}

@article{Tomalak:2015hva,
  author  = {Tomalak, O. and Vanderhaeghen, M.},
  title   = {{Two-photon exchange correction to muon-proton elastic 
             scattering at low momentum transfer}},
  journal = {Eur. Phys. J. C},
  volume  = {76},
  number  = {3},
  pages   = {125},
  year    = {2016},
  doi     = {10.1140/epjc/s10052-016-3966-3},
  eprint  = {1512.09113},
  archivePrefix = {arXiv}
}

@article{Blunden:2017nby,
  author  = {Blunden, P. G. and Melnitchouk, W.},
  title   = {{Dispersive approach to two-photon exchange in elastic 
             electron-proton scattering}},
  journal = {Phys. Rev. C},
  volume  = {95},
  number  = {6},
  pages   = {065209},
  year    = {2017},
  doi     = {10.1103/PhysRevC.95.065209},
  eprint  = {1703.06181},
  archivePrefix = {arXiv}
}

@article{Ahmed:2020uso,
  author  = {Ahmed, J. and Blunden, P. G. and Melnitchouk, W.},
  title   = {{Two-photon exchange from intermediate state resonances 
             in elastic electron-proton scattering}},
  journal = {Phys. Rev. C},
  volume  = {102},
  number  = {4},
  pages   = {045205},
  year    = {2020},
  doi     = {10.1103/PhysRevC.102.045205},
  eprint  = {2006.12543},
  archivePrefix = {arXiv}
}

@article{Chen:2004tw,
  author  = {Chen, Y.-C. and Afanasev, A. and Brodsky, S. J. 
             and Carlson, C. E. and Vanderhaeghen, M.},
  title   = {{Partonic calculation of the two-photon exchange contribution 
             to elastic electron-proton scattering at large momentum transfer}},
  journal = {Phys. Rev. Lett.},
  volume  = {93},
  pages   = {122301},
  year    = {2004},
  doi     = {10.1103/PhysRevLett.93.122301},
  eprint  = {hep-ph/0403058},
  archivePrefix = {arXiv}
}

@article{Afanasev:2005mp,
  author  = {Afanasev, A. V. and Brodsky, S. J. and Carlson, C. E. 
             and Chen, Y.-C. and Vanderhaeghen, M.},
  title   = {{The Two-photon exchange contribution to elastic 
             electron-nucleon scattering at large momentum transfer}},
  journal = {Phys. Rev. D},
  volume  = {72},
  pages   = {013008},
  year    = {2005},
  doi     = {10.1103/PhysRevD.72.013008},
  eprint  = {hep-ph/0502013},
  archivePrefix = {arXiv}
}

@article{Jamin:1991dp,
    author = "Jamin, Matthias and Lautenbacher, Markus E.",
    title = "{TRACER: Version 1.1: A Mathematica package for gamma algebra in arbitrary dimensions}",
    reportNumber = "TUM-T31-20-91",
    doi = "10.1016/0010-4655(93)90097-V",
    journal = "Comput. Phys. Commun.",
    volume = "74",
    pages = "265--288",
    year = "1993"
}

@article{Kinoshita:1962ur,
    author = "Kinoshita, T.",
    title = "{Mass singularities of Feynman amplitudes}",
    doi = "10.1063/1.1724268",
    journal = "J. Math. Phys.",
    volume = "3",
    pages = "650--677",
    year = "1962"
}

@article{Lee:1964is,
    author = "Lee, T. D. and Nauenberg, M.",
    editor = "Feinberg, G.",
    title = "{Degenerate Systems and Mass Singularities}",
    doi = "10.1103/PhysRev.133.B1549",
    journal = "Phys. Rev.",
    volume = "133",
    pages = "B1549--B1562",
    year = "1964"
}

@article{Afanasev:2023gev,
    author = "Afanasev, Andrei and others",
    title = "{Radiative corrections: from medium to high energy experiments}",
    eprint = "2306.14578",
    archivePrefix = "arXiv",
    primaryClass = "hep-ph",
    doi = "10.1140/epja/s10050-024-01281-y",
    journal = "Eur. Phys. J. A",
    volume = "60",
    number = "4",
    pages = "91",
    year = "2024"
}

@article{Denner:2002ii,
    author = "Denner, Ansgar and Dittmaier, S.",
    title = "{Reduction of one loop tensor five point integrals}",
    eprint = "hep-ph/0212259",
    archivePrefix = "arXiv",
    reportNumber = "MPI-PHT-2002-63, PSI-PR-02-21",
    doi = "10.1016/S0550-3213(03)00184-6",
    journal = "Nucl. Phys. B",
    volume = "658",
    pages = "175--202",
    year = "2003"
}

@article{Denner:2005nn,
    author = "Denner, Ansgar and Dittmaier, S.",
    title = "{Reduction schemes for one-loop tensor integrals}",
    eprint = "hep-ph/0509141",
    archivePrefix = "arXiv",
    reportNumber = "MPP-2005-84, PSI-PR-05-08",
    doi = "10.1016/j.nuclphysb.2005.11.007",
    journal = "Nucl. Phys. B",
    volume = "734",
    pages = "62--115",
    year = "2006"
}

@article{Denner:2010tr,
    author = "Denner, A. and Dittmaier, S.",
    title = "{Scalar one-loop 4-point integrals}",
    eprint = "1005.2076",
    archivePrefix = "arXiv",
    primaryClass = "hep-ph",
    reportNumber = "FR-PHENO-2010-020, PSI-PR-10-10",
    doi = "10.1016/j.nuclphysb.2010.11.002",
    journal = "Nucl. Phys. B",
    volume = "844",
    pages = "199--242",
    year = "2011"
}

@article{Patel:2015tea,
    author = "Patel, Hiren H.",
    title = "{Package-X: A Mathematica package for the analytic calculation of one-loop integrals}",
    eprint = "1503.01469",
    archivePrefix = "arXiv",
    primaryClass = "hep-ph",
    doi = "10.1016/j.cpc.2015.08.017",
    journal = "Comput. Phys. Commun.",
    volume = "197",
    pages = "276--290",
    year = "2015"
}

@article{Denner:1991qq,
    author = "Denner, Ansgar and Nierste, U. and Scharf, R.",
    title = "{A Compact expression for the scalar one loop four point function}",
    reportNumber = "PRINT-91-0280 (WURZBURG)",
    doi = "10.1016/0550-3213(91)90011-L",
    journal = "Nucl. Phys. B",
    volume = "367",
    pages = "637--656",
    year = "1991"
}

@article{Passarino:1978jh,
    author = "Passarino, G. and Veltman, M. J. G.",
    title = "{One Loop Corrections for e+ e- Annihilation Into mu+ mu- in the Weinberg Model}",
    reportNumber = "Print-79-0284 (UTRECHT)",
    doi = "10.1016/0550-3213(79)90234-7",
    journal = "Nucl. Phys. B",
    volume = "160",
    pages = "151--207",
    year = "1979"
}

@misc{Adams:2018pwt,
    author = "Adams, B. and others",
    collaboration = "COMPASS++/AMBER",
    title = "{Letter of Intent: A New QCD facility at the M2 beam line of the CERN SPS (COMPASS++/AMBER)}",
    eprint = "1808.00848",
    archivePrefix = "arXiv",
    primaryClass = "hep-ex",
    reportNumber = "CERN-SPSC-2019-003, SPSC-I-250",
    month = "8",
    year = "2018"
}

@misc{An:2026hqd,
    author = "An, Chenlei and others",
    collaboration = "LUNE",
    title = "{Low-energy Muon-Nucleon scattering experiment: LUNE (White Paper)}",
    eprint = "2608.29509",
    archivePrefix = "arXiv",
    primaryClass = "hep-ex",
    month = "8",
    year = "2026"
}

@article{Borisyuk:2008es,
    author = "Borisyuk, Dmitry and Kobushkin, Alexander",
    title = "{Two-photon exchange in dispersion approach}",
    eprint = "0804.4128",
    archivePrefix = "arXiv",
    primaryClass = "nucl-th",
    doi = "10.1103/PhysRevC.78.025208",
    journal = "Phys. Rev. C",
    volume = "78",
    pages = "025208",
    year = "2008"
}

@article{Tomalak:2018jak,
    author = "Tomalak, Oleksandr and Vanderhaeghen, Marc",
    title = "{Dispersion relation formalism for the two-photon exchange correction to elastic muon-proton scattering: elastic intermediate state}",
    eprint = "1803.05349",
    archivePrefix = "arXiv",
    primaryClass = "hep-ph",
    doi = "10.1140/epjc/s10052-018-5988-5",
    journal = "Eur. Phys. J. C",
    volume = "78",
    number = "6",
    pages = "514",
    year = "2018",
    note = "[Erratum: Eur.Phys.J.C 84, 1208 (2024)]"
}

@article{Brown:1970te,
    author = "Brown, R. W.",
    title = "{Comparison of the scattering of electrons and positrons from protons at small angles}",
    doi = "10.1103/PhysRevD.1.1432",
    journal = "Phys. Rev. D",
    volume = "1",
    pages = "1432--1444",
    year = "1970"
}

@article{Peset:2021iul,
    author = "Peset, Clara and Pineda, Antonio and Tomalak, Oleksandr",
    title = "{The proton radius (puzzle?) and its relatives}",
    eprint = "2106.00695",
    archivePrefix = "arXiv",
    primaryClass = "hep-ph",
    doi = "10.1016/j.ppnp.2021.103901",
    journal = "Prog. Part. Nucl. Phys.",
    volume = "121",
    pages = "103901",
    year = "2021"
}

@article{Burkhardt:2005se,
    author = "Burkhardt, H. and Pietrzyk, B.",
    title = "{Low energy hadronic contribution to the QED vacuum polarization}",
    eprint = "hep-ph/0506323",
    archivePrefix = "arXiv",
    doi = "10.1103/PhysRevD.72.057501",
    journal = "Phys. Rev. D",
    volume = "72",
    pages = "057501",
    year = "2005"
}

@article{Jegerlehner:2008rs,
    author = "Jegerlehner, F.",
    editor = "Bini, Cesare and Venanzoni, Graziano",
    title = "{The Running fine structure constant alpha(E) via the Adler function}",
    eprint = "0807.4206",
    archivePrefix = "arXiv",
    primaryClass = "hep-ph",
    doi = "10.1016/j.nuclphysbps.2008.09.010",
    journal = "Nucl. Phys. B Proc. Suppl.",
    volume = "181-182",
    pages = "135--140",
    year = "2008"
}

@article{Rachek:2014fam,
    author = "Rachek, I. A. and others",
    title = "{Measurement of the two-photon exchange contribution to the elastic $e^{\pm}p$ scattering cross sections at the VEPP-3 storage ring}",
    eprint = "1411.7372",
    archivePrefix = "arXiv",
    primaryClass = "nucl-ex",
    doi = "10.1103/PhysRevLett.114.062005",
    journal = "Phys. Rev. Lett.",
    volume = "114",
    number = "6",
    pages = "062005",
    year = "2015"
}

@article{Tomalak:2014sva,
    author = "Tomalak, O. and Vanderhaeghen, M.",
    title = "{Subtracted dispersion relation formalism for the two-photon exchange correction to elastic electron-proton scattering: comparison with data}",
    eprint = "1408.5330",
    archivePrefix = "arXiv",
    primaryClass = "hep-ph",
    doi = "10.1140/epja/i2015-15024-1",
    journal = "Eur. Phys. J. A",
    volume = "51",
    number = "2",
    pages = "24",
    year = "2015"
}
\end{document}